\documentclass[aps,pra,reprint,superscriptaddress]{revtex4-2}

\usepackage{verbatim} 

\usepackage{graphicx} 
\usepackage{xcolor}

\usepackage{amstext,amsmath,amssymb,amsfonts,amsthm}
\usepackage{mathrsfs}
\usepackage{bm}
\usepackage{todonotes}
\setuptodonotes{inline}
\DeclareMathAlphabet{\mathpzc}{OT1}{pzc}{m}{it} 
\DeclareFontFamily{OT1}{pzc}{} 
\DeclareFontShape{OT1}{pzc}{m}{it}{<-> s * [1.100] pzcmi7t}{} 
\DeclareMathAlphabet{\mathpzc}{OT1}{pzc}{m}{it}

\usepackage{hyperref}

\newcommand{\bra}[1]{\ensuremath{\langle #1 \vert}}
\newcommand{\ket}[1]{\ensuremath{\vert #1  \rangle}}
\newcommand{\braket}[2]{\ensuremath{\langle  #1 \vert #2  \rangle}}
\renewcommand{\b}[1]{\ensuremath{\mathbf{#1}}}
\renewcommand{\d}{\ensuremath{\text{d}}}
\newcommand{\s}{\ensuremath{\text{s}}}
\newcommand{\ee}{\ensuremath{\mathrm{ee}}}

\newcommand{\Hxc}{\ensuremath{\text{Hxc}}}

\newcommand{\x}{\ensuremath{\text{x}}}

\newcommand{\ph}{\ensuremath{\text{ph}}}
\newcommand{\V}{\ensuremath{\text{V}}}
\newcommand{\G}{\ensuremath{\text{G}}}
\newcommand{\KS}{\ensuremath{\text{KS}}}
\newcommand{\GKS}{\ensuremath{\text{GKS}}}
\newcommand{\cor}{\ensuremath{\text{cor}}}
\newcommand{\ea}{\ensuremath{\text{ea}}}
\newcommand{\gs}{\ensuremath{\text{gs}}}

\newcommand{\argstat}[1]{\underset{#1}{\text{argstat}}}
\newcommand{\argmin}[1]{\underset{#1}{\text{argmin}}}
\renewcommand{\l}{\ensuremath{\lambda}}

\newcommand{\eqv}{\equiv}
\newcommand{\eqpsi}{\equiv}

\let\Re\relax

\DeclareMathOperator{\Re}{Re}

\usepackage{color}

\newtheorem{theorem}{Theorem}

\newtheorem{corollary}[theorem]{Corollary}
\newcommand{\R}{\mathbb{R}}
\newcommand{\Z}{\mathbb{Z}}
\newcommand{\C}{\mathbb{C}}

\newcommand{\ii}{\infty}
\usepackage{dsfont}
\newcommand\1{{\ensuremath {\mathds 1} }}

\newcommand{\cT}{\mathcal{T}}
\newcommand{\cQ}{\mathcal{Q}}

\newcommand{\cM}{\mathcal{M}}

\newcommand{\cF}{\mathcal{F}}

\newcommand{\cN}{\mathcal{N}}
\newcommand{\cD}{\mathcal{D}}

\newcommand{\cH}{\mathcal{H}}

\newcommand\pscal[1]{{\ensuremath{\left\langle #1 \right\rangle}}}
\newcommand{\norm}[1]{ \left\| #1 \right\|}
\newcommand{\tr}{{\rm Tr}\,}
\renewcommand{\geq}{\geqslant}
\renewcommand{\leq}{\leqslant}
\renewcommand{\hat}{\widehat}
\renewcommand{\tilde}{\widetilde}
\newcommand{\eps}{\epsilon}
\newcommand{\nn}{\nonumber}
\newcommand{\rd}{\mathrm{d}}

\newcommand{\dr}{\rd \br}

\newcommand{\br}{\mathbf{r}}

\newcommand{\bk}{\mathbf{k}}

\usetikzlibrary{arrows.meta}

\colorlet{regionfill}{black!80}
\colorlet{tubefill}{black!20}
\colorlet{ptcolor}{black}

\newcommand{\bT}{\mathbb{T}}
\newcommand{\abs}[1]{{\left | #1 \right |}}

\begin{document}

\title{Variational and geometric time-dependent density-functional theory\\ of electronic systems}

\author{\'Eric~Canc\`es}
\email{eric.cances@enpc.fr}
\affiliation{CERMICS, CNRS, \'Ecole des Ponts - Institut Polytechnique de Paris and Inria, 6-8 avenue Blaise Pascal, Cité Descartes, 77455 Marne-la-Vallée, France}

\author{Th\'eo~Duez}
\email{theo.duez@enpc.fr}
\affiliation{CERMICS, CNRS, \'Ecole des Ponts - Institut Polytechnique de Paris and Inria, 6-8 avenue Blaise Pascal, Cité Descartes, 77455 Marne-la-Vallée, France}

\author{Jari~van~Gog}
\email{jari.van\_gog@etu.sorbonne-universite.fr}
\affiliation{Laboratoire de Chimie Théorique, Sorbonne Université and CNRS, 4 place Jussieu, 75005 Paris, France}

\author{Asbj{\o}rn~B{\ae}kgaard~Lauritsen}
\email{lauritsen@ceremade.dauphine.fr}
\affiliation{CEREMADE, CNRS, Université Paris-Dauphine, PSL Research University, Place de Lattre de Tassigny, 75016 Paris, France}

\author{Mathieu~Lewin}
\email{Mathieu.Lewin@math.cnrs.fr}
\affiliation{CEREMADE, CNRS, Université Paris-Dauphine, PSL Research University, Place de Lattre de Tassigny, 75016 Paris, France}

\author{Jakob~Oldenburg}
\email{jakob.oldenburg@dauphine.psl.eu}
\affiliation{CEREMADE, CNRS, Université Paris-Dauphine, PSL Research University, Place de Lattre de Tassigny, 75016 Paris, France}

\author{Julien~Toulouse}
\email{julien.toulouse@sorbonne-universite.fr}
\affiliation{Laboratoire de Chimie Théorique, Sorbonne Université and CNRS, 4 place Jussieu, 75005 Paris, France}

\begin{abstract}
A central problem in time-dependent density-functional theory (TDDFT) is to construct a time-dependent many-electron wave function that reproduces a prescribed time-dependent electron density. This problem admits a natural geometric interpretation, which in turn leads to at least two distinct formulations of TDDFT. The first is the usual variational formulation of TDDFT, in which the density constraint is enforced with a real-valued local potential. The second is geometric TDDFT, introduced in recent works, in which the same density constraint is enforced with an imaginary local potential, or equivalently with a Hermitian nonlocal potential. In this work, we develop this unified viewpoint for continuous electronic systems and systematically compare the two formulations. We formulate the corresponding time-dependent Kohn-Sham schemes and derive a number of exact conditions for geometric TDDFT, including conditions associated with causality, time-reversal symmetry, uniform-coordinate scaling, translational invariance, energy and momentum equations of motion, one-electron self-interaction, and the adiabatic limit. The paper also contains a mathematical section where we provide a fully rigorous justification of some parts of the text. We particularly discuss the mathematical properties of two operators underlying the two formulations of TDDFT and establish sufficient conditions for their invertibility. We explain the functional-analytic subtleties underlying the geometric interpretation of TDDFT and the two formulations of TDDFT. This unified perspective clarifies the mathematical and conceptual relation between variational and geometric TDDFT and provides a foundation for the development of approximations within geometric TDDFT.
\end{abstract}

\date{November 16, 2026}
\maketitle

\tableofcontents

\section{Introduction}

Time-dependent density-functional theory (TDDFT) is one of the major computational approaches for describing the quantum electron dynamics of electronic systems subject to a time-dependent external potential (see, e.g., Refs.~\onlinecite{Lee-IJMPB-01,MarUllNogRubBurGro-06,Ullrichs-11,MarMaiNogGroRub-12,ProIsb-IJQC-16,Mai-JCP-16,Ull-APLCP-25}). The usual formulation of TDDFT is based on the idea that a given one-electron density trajectory can be generated from a fictitious Hamiltonian with an effective time-dependent external potential, whose uniqueness (modulo time-dependent constants) for a given initial state is established by the Runge-Gross theorem~\cite{RunGro-84} under certain regularity conditions~\cite{FouLamLewSor-16}. In the time-dependent Kohn-Sham (TDKS) scheme~\cite{RunGro-84,vanLeeuwen-99}, this is exploited to obtain, in principle exactly, the physical density trajectory of an interacting electronic system by solving a time-dependent Schrödinger equation (TDSE) with an auxiliary noninteracting Hamiltonian with an effective time-dependent external potential. The main practical difficulty of TDDFT lies in constructing accurate density-functional approximations to this effective time-dependent external potential. In particular, most practical TDDFT calculations rely on the so-called adiabatic approximation which, although remarkably successful in many applications, is known to have important limitations in describing strongly nonadiabatic electron dynamics (see, e.g., Ref.~\onlinecite{LacMai-NPJCM-23}).

Recently, some of the present authors~\cite{MAQUI_TDDFT-26,MAQUI_finite-26} introduced an abstract formulation of TDDFT in terms of the geometry of the space of many-body states having a given density at a given time. The density constraint fixes the component of the wave-function velocity normal to the set of states with the prescribed instantaneous density, while leaving the tangential component free. This general formulation leads to different ways of constructing fictitious Hamiltonians leading to the same density trajectory. The usual TDDFT formulation in which the density trajectory is imposed using an effective time-dependent real-valued local potential is only one possible construction. This usual TDDFT formulation can be understood as resulting from imposing the time-dependent variational principle (or stationarity of the action), and for this reason we will refer to this usual TDDFT formulation as \textit{variational TDDFT}. It turns out that the geometric interpretation of the abstract TDDFT problem leads to another very natural construction in which the density trajectory is imposed using an effective time-dependent \textit{imaginary} local potential, or equivalently a time-dependent Hermitian \textit{nonlocal} potential, whose uniqueness can be established by an extension of the Runge-Gross theorem referred to as the geometric Runge-Gross theorem~\cite{MAQUI_TDDFT-26}. This leads to a new formulation of TDDFT that we refer to as \textit{geometric TDDFT}. 

One can then set up a geometric TDKS-like scheme in which the physical density trajectory of an interacting electronic system is obtained, in principle exactly, by solving a TDSE with a non-interacting fictitious Hamiltonian including an effective time-dependent imaginary local potential or, equivalently, a time-dependent Hermitian nonlocal potential. Of course, the practical utility of this geometric TDDFT approach will depend on whether accurate approximations can be developed for the effective imaginary local potential or Hermitian nonlocal potential. Preliminary studies on toy-model systems~\cite{MAQUI_TDDFT-26,MAQUI_finite-26} suggest that, for strongly nonadiabatic electron dynamics, the effective imaginary potential of geometric TDDFT might be easier to approximate than the effective real-valued potential of usual variational TDDFT.

In Ref.~\onlinecite{MAQUI_finite-26}, a fully mathematically rigorous account of this abstract geometric formulation of TDDFT was given for the case of finite-lattice systems, and the new geometric TDDFT approach was numerically illustrated on the Hubbard dimer. In Ref.~\onlinecite{MAQUI_TDDFT-26}, this approach was extended to electronic systems in continuous space, with states in an infinite-dimensional Hilbert space, and numerically illustrated on two-electron systems on the real line. The case of an infinite-dimensional Hilbert space is considerably more subtle than the finite-dimensional lattice case, and many aspects of the theory were not made explicit in Ref.~\onlinecite{MAQUI_TDDFT-26}. In the present paper, we give a more complete account of the abstract geometric formulation of TDDFT for real electronic systems. With this framework, we systematically review the usual variational TDDFT and compare it with our new geometric TDDFT. In particular, we review the main exact conditions on the effective potential known in variational TDDFT and show how to extend them to geometric TDDFT. We also provide a detailed mathematical study of the main objects on which variational and geometric TDDFT rely. The complete mathematical well-posedness of variational and geometric TDDFT at short times was proved recently for particles evolving on the torus (i.e., a cubic box with periodic boundary conditions) in Refs.~\onlinecite{LauLewOld-V-26_ppt,LauLewOld-V-short-26_ppt,LauLewOld-W-26_ppt}.

The paper is organized as follows. In Section~\ref{sec:var_geom_TDDFT} we present the abstract geometric formulation of TDDFT, and the resulting variational and geometric TDDFT schemes. In Section~\ref{sec:exact_constraints} we derive and discuss some exact conditions on the effective potentials in variational and geometric TDDFT. These are useful for constructing and assessing approximations (see, e.g., Ref.~\onlinecite{KurHea-PRA-09}). Finally, in Section~\ref{sec:math} we provide a detailed mathematical study of the geometric interpretation of TDDFT, and of variational and geometric TDDFT. This section is fully mathematically rigorous and thus written in a more technical way, but we have nevertheless tried to make it accessible to all readers. At the end of Section~\ref{sec:math}, we give a summary of the mathematical results and their consequences for TDDFT. Section~\ref{sec:conclusion} contains our conclusions.
Several appendices contain complementary comments, results and proofs. Hartree atomic units are used throughout this work.

\section{Variational and geometric TDDFT}
\label{sec:var_geom_TDDFT}

In this section, we review the usual variational TDDFT and the recently proposed geometric TDDFT~\cite{MAQUI_TDDFT-26,MAQUI_finite-26} for electronic systems in continuous space, from the point of view of the abstract geometric formulation of TDDFT.

\subsection{Abstract formulation of TDDFT}

We start by giving an abstract formulation of TDDFT for $N$ electrons in dimension $d \in \{1,2,3 \}$ with states in the antisymmetric Hilbert space $${\cal H} := \bigwedge^N  L^2(\mathbb{R}^d \times \{ \uparrow,\downarrow \};\mathbb{C})$$ with inner product denoted by $\braket{\,\cdot\,}{\,\cdot\,}$. We consider the physical (ph) $N$-electron time-dependent Hamiltonian $\hat H^\ph : t \mapsto \hat{H}^\ph(t)$ with
\begin{equation}
\hat{H}^\ph(t) := \hat{T} + \hat{V}_\ee + \hat{V}^\ph(t),
\end{equation}
where $\hat{T}$ is the kinetic-energy operator,
\begin{equation}
\hat{T} := -\frac{1}{2} \int_{\mathbb{R}^d} [\nabla^2_\b{r} \hat{\gamma}(\b{r},\b{r}')]_{\b{r}'=\b{r}} \d \b{r},
\label{eq:T}
\end{equation}
$\hat{V}_\ee$ is the operator associated with the electron-electron interaction potential $v_\ee(\b{r}_1-\b{r}_2)$,
\begin{equation}
\hat{V}_\ee := \frac{1}{2} \int_{\mathbb{R}^d\times\mathbb{R}^d} v_\ee(\b{r}_1-\b{r}_2) \, \hat{\rho}_2(\b{r}_1,\b{r}_2) \d \b{r}_1  \d \b{r}_2,
\label{eq:Vee}
\end{equation}
and $\hat{V}^\ph(t)$ is the operator associated with the physical time-dependent external one-electron local potential $v^\ph(\b{r},t)$,
\begin{equation}
\hat{V}^\ph(t) := \int_{\mathbb{R}^d} v^\ph(\b{r},t) \hat{\rho}(\b{r}) \d \b{r},
\label{eq:Vext}
\end{equation}
where we have used the one-electron density-matrix operator $\hat{\gamma}(\b{r},\b{r}'):= \sum_{\sigma \in \{ \uparrow,\downarrow \}} \hat{\psi}^\dagger_\sigma(\b{r}') \hat{\psi}_\sigma(\b{r})$ written with the creation and annihilation field operators $\hat{\psi}_\sigma^\dagger(\b{r})$ and $\hat{\psi}_\sigma(\b{r})$, the density operator $\hat{\rho}(\b{r}) := \hat{\gamma}(\b{r},\b{r})$, and the pair-density operator $\hat{\rho}_2(\b{r}_1,\b{r}_2):= \hat{\rho}(\b{r}_1) \hat{\rho}(\b{r}_2) - \hat{\rho}(\b{r}_1)  \delta(\b{r}_1-\b{r}_2)$.
Starting from a physical initial wave function $\Psi_0^\ph$, the physical time-dependent wave function $\Psi^\ph(t)$ is the solution to the time-dependent Schr\"odinger equation (TDSE),
\begin{eqnarray}
\left\{ 
\begin{aligned}
i\, \partial_t \Psi^\ph(t) &= \hat{H}^\ph(t) \Psi^\ph(t),\\[0.3cm]
\Psi^\ph(0) &= \Psi_0^\ph,
\end{aligned}
\right.
\label{TDSE}
\end{eqnarray}
and leads to the physical time-dependent density $\rho^\ph(\b{r},t) := \rho_{\Psi^\ph(t)}(\b{r})$, where $\rho_\Psi(\b{r}) := \bra{\Psi} \hat{\rho}(\b{r}) \ket{\Psi}$ designates the density of the wave function $\Psi$. 

The goal of TDDFT is to find the physical density $\rho^\ph(\b{r},t)$ without calculating the many-electron wave function $\Psi^\ph(t)$. We can attempt to achieve this by first considering the following auxiliary inverse problem. For a given density trajectory $\rho : t \mapsto \rho(t)$, a given time-dependent reference Hamiltonian $\hat{H}: t \mapsto \hat{H}(t)$ and a given initial wave function $\Psi_0$ such that $\rho_{\Psi_0}(\b{r}) = \rho(\b{r},0)$, we try to find a time-dependent operator $\hat{F}: t \mapsto \hat{F}(t)$ in a given class of operators (e.g., one-electron local potential operators) such that the solution to the modified TDSE,
\begin{eqnarray}
\left\{ 
\begin{aligned}
i\, \partial_t \Psi(t) &= \left( \hat{H}(t) + \hat{F}(t) \right) \Psi(t),\\[0.3cm]
\Psi(0) &= \Psi_0,
\end{aligned}
\right.
\label{TDSEwithF}
\end{eqnarray}
gives the prescribed density trajectory, i.e. $\rho_{\Psi(t)}(\b{r}) = \rho(\b{r},t)$ for all times $t$. If this density-enforcing correction $\hat{F}$ exists, and if we have a Runge-Gross-type uniqueness theorem for this class of operators for a given $\Psi_0$ and $\hat{H}$, it gives us a unique way (possibly modulo additive time-dependent constants) of defining $\hat{F}$ as a functional of $\rho$, $\Psi_0$, and $\hat{H}$, i.e. $\hat{F} = \hat{F}[\rho,\Psi_0,\hat{H}]$. For $\Psi_0 = \Psi_0^\ph$ and $\hat{H} = \hat{H}^\ph$, this gives us a fundamental TDDFT equation for the physical density trajectory $\rho^\ph$:
\begin{eqnarray}
\hat{F}[\rho,\Psi_0^\ph,\hat{H}^\ph]=0 \, \Longleftrightarrow \, \rho=\rho^\ph,
\label{eq:F=0}
\end{eqnarray}
i.e., $\rho^\ph$ is the unique solution to the equation $\hat{F}[\rho,\Psi_0^\ph,\hat{H}^\ph]=0$. Indeed, if $\hat{F}[\rho,\Psi_0^\ph,\hat{H}^\ph]=0$, Eq.~\eqref{TDSEwithF} reduces to Eq.~\eqref{TDSE} and thus $\rho=\rho^\ph$. Conversely, if $\rho=\rho^\ph$, we must have $\hat{F}[\rho,\Psi_0^\ph,\hat{H}^\ph]=0$ since $\hat{F}$ is unique. Of course, if we choose for $\hat{H}(t)$ a Hamiltonian different from the physical one (for instance a fictitious non-interacting Hamiltonian), then $\hat{F}[\rho^{\rm ph},\Psi_0^{\rm ph},\hat{H}]$ will generally not vanish for the sought-after physical density $\rho^{\rm ph}$.

\subsection{Geometric interpretation of the abstract formulation of TDDFT}
\label{sec:geometric_interpretation}

We give now a geometric interpretation of the abstract formulation of TDDFT. In the modified TDSE~\eqref{TDSEwithF}, at a given time $t$, fixing the operator $\hat{F}(t)$ amounts to fixing the wave-function velocity $\partial_t \Psi(t)$. If all we want is to reproduce a given density trajectory, there is a lot of freedom in choosing the wave-function velocity $\partial_t \Psi(t)$ at any given time $t$. This can be understood geometrically as follows.

Suppose that we have a wave-function trajectory $(\Psi(s))_{0\leq s \leq t}$ giving the prescribed density trajectory $(\rho(s))_{0\leq s \leq t}$ until time $t$. In particular, at time $t$, $\Psi(t)$ gives the prescribed density $\rho(t)$ at that time. We want to choose a wave-function velocity $\partial_t \Psi(t)$ at time $t$ in order to extend the wave-function trajectory, while still continuing to reproduce the prescribed density trajectory. To understand how to do that, it is useful to introduce the set of states $\Psi \in {\cal H}$ constrained to have a fixed density $\rho:\b{r}\mapsto \rho(\b{r})$,
\begin{eqnarray}
{\cal M}_{\rho} \! :=\Big\{ \Psi \in {\cal H} \; |\; \bra{\Psi}\hat{\rho}(\b{r})\ket{\Psi}=\rho(\b{r}) \mbox{ for a.e. } \b{r} \in \mathbb{R}^d \Big\},
\end{eqnarray}
where a.e. stands for ``almost every'' in the Lebesgue sense (i.e., everywhere except on sets of measure zero). The corresponding ``tangent space'' associated with the linearization of the density constraint at $\Psi \in {\cal M}_{\rho}$ is then
\begin{multline}
{\cal T}_{\Psi} {\cal M}_{\rho}\\
:=\Big\{  \delta \Psi_\parallel \in {\cal H} \; |\; \Re\bra{\delta \Psi_\parallel}\hat{\rho}(\b{r})\ket{\Psi} \!=\!0 \mbox{ for a.e. } \b{r} \in \mathbb{R}^d\Big\}.
\label{eq:TM}
\end{multline}
It is the set of Hilbert-space vectors $\delta \Psi_\parallel$ corresponding to wave-function variations which do not change the density of $\Psi$ to first order. The real part in Eq.~\eqref{eq:TM} implies that ${\cal T}_{\Psi} {\cal M}_{\rho}$ is not a complex vector space but a real vector space, i.e. only real-valued coefficients are allowed when forming linear combinations of tangent vectors. We write ``tangent space" in quotation marks because it is rather involved to define $\cM_\rho$ as a smooth manifold in the rigorous mathematical sense, which is required for $\cT_\Psi\cM_\rho$ to really be the tangent space of $\cM_\rho$ in the sense of differential geometry. We will come back to this below.

We also define the corresponding ``normal space'' as the orthogonal complement of ${\cal T}_{\Psi} {\cal M}_{\rho}$ in ${\cal H}$ with respect to the real inner product $\Re\braket{\,\cdot\,}{\,\cdot\,}$,
\begin{multline}
{\cal N}_{\Psi} {\cal M}_{\rho} \! := \left( {\cal T}_{\Psi} {\cal M}_{\rho} \right)^{\perp_\R} \\
= \Big\{ \delta \Psi_\perp \in {\cal H} \; |\; \forall \delta \Psi_\parallel \in {\cal T}_{\Psi}{\cal M}_{\rho},\; \Re\braket{\delta \Psi_\parallel}{\delta \Psi_\perp} \!=\!0 \Big\},
\label{eq:NMdef}
\end{multline}
which is the set of Hilbert-space vectors $\delta \Psi_\perp$ orthogonal to linearized density-preserving variations at $\Psi \in {\cal M}_{\rho}$. Formally, the normal space is spanned by the continuous family of states $\{\hat{\rho}(\b{r})\Psi\}_{\b{r}\in \mathbb{R}^d}$, which have the ``overlap matrix''
\begin{eqnarray}
S_\Psi(\b{r},\b{r}') &:=& \bra{\Psi} \hat{\rho}(\b{r}') \hat{\rho}(\b{r}) \ket{\Psi}
\label{eq:SPsirrp}
\end{eqnarray}
which should really be viewed as the kernel of an operator $S_\Psi$ acting on an appropriate function space. We will always assume that the operator $S_\Psi$ is (boundedly) invertible (see Section~\ref{sec:S}). The normal space can then be more explicitly written as 
\begin{eqnarray}
{\cal N}_{\Psi} {\cal M}_{\rho} 
=\Big\{\scalebox{1.2}{$\int$}_{\!\mathbb{R}^d} \d \b{r} \, w(\b{r}) \hat{\rho}(\b{r})\,\Psi, \;\;w: \mathbb{R}^d \to \mathbb{R}
\Big\},
\label{eq:NM}
\end{eqnarray}
for suitable real-valued potentials $w$. The precise function space of suitable potentials $w$ is specified in Section~\ref{sec:manifold}. Thus, each normal vector is generated by the action of a one-electron real-valued local potential operator $\scalebox{1.2}{$\int$}_{\!\mathbb{R}^d} \d \b{r} \, w(\b{r}) \hat{\rho}(\b{r})$ on the wave function $\Psi$. Since the potential $w$ is real-valued, ${\cal N}_{\Psi} {\cal M}_{\rho}$ is also a real vector space. 

When the Hilbert space ${\cal H}$ has finite dimension, ${\cal M}_{\rho}$ is generically a smooth submanifold of ${\cal H}$, and ${\cal T}_{\Psi} {\cal M}_{\rho}$ and ${\cal N}_{\Psi} {\cal M}_{\rho}$ are then the usual tangent and normal spaces of ${\cal M}_{\rho}$ in the meaning of differential geometry~\cite{MAQUI_finite-26}. For the present infinite-dimensional Hilbert space ${\cal H}$, the situation is somewhat more complicated, because one has to specify a topology and it turns out that the one of the Hilbert space $\cH$ is not sufficient (see Section~\ref{sec:manifold} and Appendix~\ref{sec:not_a_manifold}). This mirrors the similar topological difficulties of ground-state DFT~\cite{Garrigue-21,CorLae-25,OpenAI-KS}. We will come back to all these mathematical technicalities in Section~\ref{sec:manifold}. We nevertheless use the suggestive notation ${\cal T}_{\Psi}{\cal M}_{\rho}$ and ${\cal N}_{\Psi}{\cal M}_{\rho}$ for the spaces defined in Eqs.~\eqref{eq:TM} and~\eqref{eq:NMdef}, and refer to them as the tangent and normal spaces, with the understanding that they are defined through the linearized density constraint rather than through a smooth-manifold structure.

\begin{figure}[t]
\includegraphics[scale=1,angle=0]{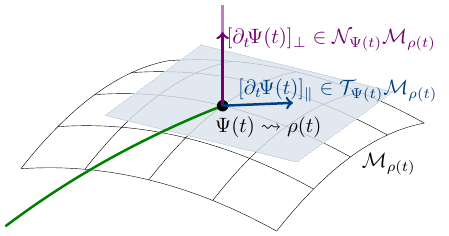}
        \caption{Geometric interpretation of the abstract formulation of TDDFT. We consider wave-function trajectories constrained to have a given density trajectory. At a given time $t$, the wave function $\Psi(t)$ gives the density $\rho(t)$. The set of states constrained to have the density $\rho(t)$ is denoted as ${\cal M}_{\rho(t)}$. Its corresponding linearized tangent space and normal space at state $\Psi(t)$ are denoted by ${\cal T}_{\Psi(t)} {\cal M}_{\rho(t)}$ and ${\cal N}_{\Psi(t)} {\cal M}_{\rho(t)}$. The wave-function velocity $\partial_t \Psi(t)$ can then be decomposed into a tangent component $[\partial_t \Psi(t)]_\parallel \in {\cal T}_{\Psi(t)} {\cal M}_{\rho(t)}$ and a normal component $[\partial_t \!\Psi(t)]_\perp \in {\cal N}_{\Psi(t)} {\cal M}_{\rho(t)}$. The normal component $[\partial_t \!\Psi(t)]_\perp$ is fixed by the density velocity $\partial_t \rho(t)$, while the tangent component $[\partial_t \!\Psi(t)]_\parallel$ is completely free.}
\label{fig:manifold}
\end{figure}

Along a wave-function trajectory, at each time $t$, the wave function $\Psi(t)$ must belong to the fixed-density set $\cM_{\rho(t)}$ for the prescribed density $\rho(t)$ at that time. Introduction of the tangent and normal spaces at the wave function $\Psi(t)$ at time $t$ allows us to decompose the wave-function velocity $\partial_t \Psi(t)$ into a tangent component $[\partial_t \!\Psi(t)]_\parallel \in {\cal T}_{\Psi(t)} {\cal M}_{\rho(t)}$ and a normal component $[\partial_t \!\Psi(t)]_\perp \in {\cal N}_{\Psi(t)} {\cal M}_{\rho(t)}$ (see Fig.~\ref{fig:manifold}),
\begin{eqnarray}
\partial_t \Psi(t) = [\partial_t \!\Psi(t)]_\parallel + [\partial_t \!\Psi(t)]_\perp,
\label{eq:dtPsidecomp}
\end{eqnarray}
where the tangent component $[\partial_t \!\Psi(t)]_\parallel$ is completely free and only the normal component $[\partial_t \!\Psi(t)]_\perp$ is fixed by the density constraint $\bra{\Psi(t)}\hat{\rho}(\b{r})\ket{\Psi(t)} = \rho(\b{r},t)$. To see this, we take the time derivative of the density constraint,
\begin{eqnarray}
2\Re\bra{\partial_t \Psi(t)}\hat{\rho}(\b{r})\ket{\Psi(t)} = \partial_t \rho(\b{r},t),
\end{eqnarray}
and, since by definition the tangent velocity has no contribution, i.e. $\Re\bra{[\partial_t \Psi(t)]_\parallel}\hat{\rho}(\b{r})\ket{\Psi(t)}  = 0$, it remains
\begin{eqnarray}
2\Re\bra{[\partial_t \Psi(t)]_\perp} \hat{\rho}(\b{r})\ket{\Psi(t)} = \partial_t \rho(\b{r},t).
\label{eq:RedtPsitperp}
\end{eqnarray}
Since $[\partial_t \Psi(t)]_\perp$ is in the normal space, it can be written as $[\partial_t \Psi(t)]_\perp = \scalebox{1.2}{$\int$}_{\!\mathbb{R}^d} \d \b{r} \, w_\perp(\b{r},t) \hat{\rho}(\b{r})\,\Psi(t)$ for some suitable function $w_\perp(\cdot,t): \mathbb{R}^d \to \mathbb{R}$. Equation~\eqref{eq:RedtPsitperp} then leads to the linear equation for $w_\perp$
\begin{eqnarray}
2\int_{\mathbb{R}^d} S_{\Psi(t)}(\b{r},\b{r}') w_\perp(\b{r}',t) \d \b{r}' = \partial_t \rho(\b{r},t),
\label{eq:Swperp}
\end{eqnarray}
where we have used that $\Re \bra{\Psi} \hat{\rho}(\b{r}') \hat{\rho}(\b{r}) \ket{\Psi} =  \bra{\Psi} \hat{\rho}(\b{r}') \hat{\rho}(\b{r}) \ket{\Psi} = S_\Psi(\b{r},\b{r}')$. Since we assume that the operator $S_{\Psi(t)}$ is invertible, Eq.~\eqref{eq:Swperp} gives a unique solution for $w_\perp(\b{r},t)$, and thus for $[\partial_t \Psi(t)]_\perp$.

Thus, the density constraint imposes only the normal component of the wave-function velocity $[\partial_t \!\Psi(t)]_\perp$. The tangent component of the wave-function velocity $[\partial_t \!\Psi(t)]_\parallel$ is completely arbitrary. At a given time $t$, a choice of the operator $\hat{F}(t)$ in Eq.~\eqref{TDSEwithF} determines a choice of $[\partial_t \!\Psi(t)]_\parallel$, namely
\begin{eqnarray}
[\partial_t \!\Psi(t)]_\parallel = -i \left( \hat{H}(t) + \hat{F}(t) \right) \Psi(t) - [\partial_t \!\Psi(t)]_\perp.
\label{}
\end{eqnarray}
We thus see that, for a given reference Hamiltonian $\hat{H}(t)$, there are many possible wave-function trajectories constrained to have a given density trajectory, and thus potentially many possible TDDFT schemes. To define a particular TDDFT scheme, we need to choose $\hat{F}(t)$ or $[\partial_t \!\Psi(t)]_\parallel$ according to some criterion.

\subsection{Variational TDDFT}
\label{sec:standardTDDFT}

Standard TDDFT can be interpreted as choosing the operator $\hat{F}(t)$ in the modified TDSE~\eqref{TDSEwithF} according to the time-dependent variational principle~\cite{KraSar-81,KucMeyCed-88,MarBur-20,LasSu-22}, which can be formulated as follows. Let us consider a wave-function trajectory $\Psi:[0,T] \to {\cal H}$ from initial time $0$ to any arbitrary final time $T>0$. This wave-function trajectory $\Psi$ satisfies the time-dependent variational principle under the constraint of yielding a given density trajectory $\rho$ if $\Psi$ is a stationary point of the action functional with respect to variations over all trial continuously differentiable wave-function trajectories $\tilde{\Psi}$ with fixed endpoints, $\tilde{\Psi}(0) = \Psi(0)$ and $\tilde{\Psi}(T) = \Psi(T)$, yielding the density trajectory $\rho$. This can be written as
\begin{eqnarray}
\Psi \in \!\!\! \argstat{\substack{\tilde{\Psi}\rightsquigarrow \rho\\ \tilde{\Psi}(0)=\Psi(0),\; \tilde{\Psi}(T)=\Psi(T)}} \!\!\! \int_0^T \!\!\! \d t\, \bra{\tilde{\Psi}(t)} i \partial_t - \hat{H}(t) \ket{\tilde{\Psi}(t)},
\label{eq:statPsirho}
\end{eqnarray}
where argstat designates the set of stationary points and $\tilde{\Psi} \rightsquigarrow \rho$ means that the variations of $\tilde{\Psi}$ are taken under the constraint that $\tilde{\Psi}$ yields the given density trajectory $\rho$. Notice that in Eq.~\eqref{eq:statPsirho} the variational space itself depends on the stationary trajectory $\Psi$ through its endpoints $\Psi(0)$ and $\Psi(T)$. Thus, Eq.~\eqref{eq:statPsirho} should be understood as an implicit stationarity condition rather than as a standard variational problem with a prescribed variational space. This unusual feature has historically caused a great deal of confusion in the literature about causality in TDDFT (see Appendix~\ref{app:Lagrange}).
Equation~\eqref{eq:statPsirho} is a global-in-time condition, but by making wave-function variations in Eq.~\eqref{eq:statPsirho} localized around any given time $t$, it leads to the local-in-time condition~\cite{MAQUI_finite-26,LasSu-22} 
\begin{eqnarray}
\Re\bra{\delta \Psi_\parallel} i \partial_t - \hat{H}(t) \ket{\Psi(t)}=0, \;\; \forall\, \delta \Psi_\parallel \in  {\cal T}_{\Psi(t)} {\cal M}_{\rho(t)},
\label{eq:statPsirhoTM}
\end{eqnarray}
i.e., $( i \partial_t - \hat{H}(t)) \Psi(t)$ must be orthogonal to all vectors of the tangent space. Since $( i \partial_t - \hat{H}(t)) \Psi(t) = \hat{F}(t) \Psi(t)$, this means that $\hat{F}(t) \Psi(t)$ must be in the normal space, 
\begin{eqnarray}
\hat{F}(t) \Psi(t) \in {\cal N}_{\Psi(t)} {\cal M}_{\rho(t)}.
\label{eq:FPsiinN}
\end{eqnarray}
Using the expression of the normal space in Eq.~\eqref{eq:NM}, we arrive at the conclusion that, in the variational (V) principle, $\hat{F}(t)$ in the modified TDSE~\eqref{TDSEwithF} can be chosen of the form of a one-electron local potential operator
\begin{eqnarray}
\hat{F}^\text{V}(t) := \int_{\mathbb{R}^d} \d\b{r} \, v(\b{r},t) \hat{\rho}(\b{r}),
\label{eq:FVP}
\end{eqnarray}
with a real-valued potential $v(\cdot,t): \mathbb{R}^d \to \mathbb{R}$ which must be such that the solution to the modified TDSE~\eqref{TDSEwithF} with the operator $\hat{F}^\text{V}(t)$ gives the prescribed density trajectory $\rho$. The potential $v(\b{r},t)$ can be interpreted as the Lagrange multiplier for the density constraint $\bra{\Psi(t)} \hat{\rho}(\b{r}) \ket{\Psi(t)}= \rho(\b{r},t)$ (see Appendix~\ref{app:Lagrange}). 

If we choose $\hat{H} := \hat{T} + \hat{V}_\ee$ as the reference Hamiltonian in Eq.~\eqref{TDSEwithF}, we recover the usual TDDFT approach that we will refer to here as variational TDDFT. Note that we do not discuss yet the TDKS scheme, which will be discussed in Section~\ref{sec:variationalTDKS}.
For a given density trajectory $\rho$ and initial wave function $\Psi_0$, if the potential $v$ in Eq.~\eqref{eq:FVP} exists, then the Runge-Gross theorem~\cite{RunGro-84} (see, also, Ref.~\onlinecite{FouLamLewSor-16} for a mathematically rigorous proof) states that, for spatially smooth $v_\ee$, $\Psi_0$, and $v$, and time-analytic $v$, the potential $v$ is unique up to an arbitrary additive time-dependent constant $C(t)$. This means that $v$ is a functional of $\rho$ and $\Psi_0$, i.e. $v = v[\rho,\Psi_0]$, modulo $C(t)$. Very recently, conditions for both short-time existence and uniqueness of the potential $v$ were given for the case of the torus~\cite{LauLewOld-V-26_ppt}. For variational TDDFT, the equation~\eqref{eq:F=0} giving the physical density $\rho^\ph$ then takes the form
\begin{eqnarray}
v[\rho,\Psi_0^\ph] \eqv v^\ph  \, \Longleftrightarrow \, \rho=\rho^\ph,
\label{eq:v=0_universal}
\end{eqnarray}
where $\eqv$ designates here the equality between potentials modulo additive time-dependent constants $C(t)$.

Thus, in variational TDDFT, a prescribed density trajectory $\rho$ is reproduced from a time-dependent wave function $\Psi^\V(t)$ satisfying an effective TDSE,
\begin{eqnarray}
\left\{
\begin{aligned}
i \partial_t \Psi^\V(t) &= \left( \hat{T} + \hat{V}_\text{ee} + \hat{V}(t)\right) \Psi^\V(t), \\
\Psi^\V(0) &= \Psi_0,
\end{aligned}
\right.
\label{eq:TDSE-TDDFT}
\end{eqnarray}
with the universal one-electron local potential operator $\hat{V}(t) := \int_{\mathbb{R}^d} \d\b{r} \, v[\rho,\Psi_0](\b{r},t) \, \hat{\rho}(\b{r})$ and an initial wave function $\Psi_0$ having the prescribed initial density, i.e. $\rho_{\Psi_0}(\b{r}) = \rho(\b{r},0)$. Since the potential $v[\rho,\Psi_0]$, imposing the density $\rho$, is a functional of $\rho$ and $\Psi_0$, the wave function $\Psi^\V(t)$ in Eq.~\eqref{eq:TDSE-TDDFT} is also a functional of $\rho$ and $\Psi_0$, i.e. $\Psi^\V(t) = \Psi^\V[\rho,\Psi_0](t)$ (modulo multiplicative time-dependent global phase factors).

The continuity equation associated with Eq.~\eqref{eq:TDSE-TDDFT} always takes the standard form
\begin{eqnarray}
\partial_t \rho(\b{r},t) + \nabla \cdot \b{j}_{\Psi^\V(t)}(\b{r}) = 0,
\label{eq:conteq}
\end{eqnarray}
where $\b{j}_{\Psi}(\b{r}):= \bra{\Psi} \, \hat{\b{j}}(\b{r}) \ket{\Psi} $ is the current density of the wave function $\Psi$, expressed with the current density operator 
$\hat{\b{j}}(\b{r}) := 1/(2i) [(\nabla_{\b{r}} - \nabla_{\b{r}'}) \hat{\gamma}(\b{r},\b{r}')]_{\b{r}'=\b{r}}$. 
In particular, Eq.~\eqref{eq:conteq} at $t=0$ implies an additional compatibility condition between the initial wave function and the initial density velocity: $\nabla \cdot \b{j}_{\Psi_0}(\b{r}) = - \partial_t \rho(\b{r},t)|_{t=0}$. The continuity equation~\eqref{eq:conteq} cannot be directly used to extract the universal potential $v$ reproducing a given density trajectory $\rho$, since $v$ does not explicitly appear in it. To extract $v$, we must write down the second-order time-derivative of the density~\cite{vanLeeuwen-99}
\begin{eqnarray}
\partial_t^2 \rho(\b{r},t) &=& -2\int_{\mathbb{R}^d} \! {\cal K}_{\Psi^\V(t)}(\b{r},\b{r}') v(\b{r}',t) \d \b{r}' 
+ q_{\Psi^\V(t)}(\b{r}),
\nonumber\\
\label{eq:divlocforceKPsi}
\end{eqnarray}
where $q_\Psi(\b{r}) := - \bra{\Psi} [\hat{T}+\hat{V}_\ee,[\hat{T},\hat{\rho}(\b{r})]] \ket{\Psi}$ and ${\cal K}_{\Psi}(\b{r},\b{r}')$ is the operator kernel defined by
\begin{eqnarray}
{\cal K}_{\Psi}(\b{r},\b{r}') &:=& 
\frac{1}{2} \bra{\Psi} [\hat{\rho}(\b{r}'),[\hat{T},\hat{\rho}(\b{r})]] \ket{\Psi} 
\nonumber\\
&=& 
\frac{1}{2} \nabla_{\b{r}} \cdot \left( \rho_\Psi(\b{r}) \,  \nabla_{\b{r}'} \delta(\b{r}' - \b{r}) \right).
\label{eq:KPsi}
\end{eqnarray}
Since ${\cal K}_{\Psi}(\b{r},\b{r}')$ depends in fact only on the density of $\Psi$, it is more convenient and transparent to define the operator kernel as parametrized by $\rho$ (and without the factor $1/2$),
\begin{eqnarray}
K_{\rho}(\b{r},\b{r}'):= \nabla_{\b{r}} \cdot \left( \rho(\b{r}) \,  \nabla_{\b{r}'} \delta(\b{r}' - \b{r}) \right),
\label{eq:K}
\end{eqnarray}
i.e., $\int_{\mathbb{R}^d} K_{\rho}(\b{r},\b{r}') v(\b{r}') \d \b{r}' =  - \nabla \cdot (\rho(\b{r}) \, \nabla v(\b{r}) )$. Equation~\eqref{eq:divlocforceKPsi} is then rewritten as
\begin{eqnarray}
\partial_t^2 \rho(\b{r},t) &=& \nabla \cdot (\rho(\b{r},t) \, \nabla v(\b{r},t) ) + q_{\Psi^\V(t)}(\b{r}), \;\;\;\;\;
\label{eq:divlocforce}
\end{eqnarray}
which is known as the force-balance equation (or divergence of the local force equation)~\cite{vanLeeuwen-99,RugPenLee-15}. It is the fundamental equation of variational TDDFT which gives a link between the potential $v(\b{r},t)$ and the second-order time-derivative of the density $\partial_t^2 \rho(\b{r},t)$. The operator $K_\rho$ is central to variational TDDFT.  In particular, the proof of the Runge-Gross theorem in Ref.~\onlinecite{FouLamLewSor-16} requires the invertibility of $K_{\rho_{\Psi_0}}$ (modulo constants). The mathematical properties of the operator $K_{\rho}$ are described in Section~\ref{sec:K}. The existence of $v$ for a given density trajectory $\rho(t)$ was recently proved in Refs.~\onlinecite{LauLewOld-V-26_ppt,LauLewOld-V-short-26_ppt} for particles evolving on the torus, using the force-balance equation.

\subsection{Geometric TDDFT}
\label{sec:geometricTDDFT}

\begin{figure}[t]
\includegraphics[scale=1,angle=0]{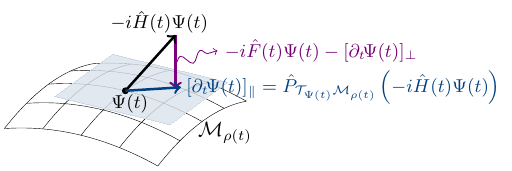}
        \caption{Geometric principle in TDDFT. For a given time $t$ and wave function $\Psi(t)$ with density $\rho(t)$, the tangent component of the wave-function velocity $[\partial_t \!\Psi(t)]_\parallel$ is chosen as the orthogonal projection of the Schrödinger wave-function velocity $- i \hat{H}(t) \Psi(t)$ onto the linearized tangent space ${\cal T}_{\Psi(t)} {\cal M}_{\rho(t)}$ associated with the density constraint at state $\Psi(t)$. The normal projection vector is $-i \hat{F}(t) \Psi(t) - [\partial_t \!\Psi(t)]_\perp$.}
\label{fig:geometric_principle}
\end{figure}

In geometric TDDFT~\cite{MAQUI_TDDFT-26,MAQUI_finite-26}, we choose the operator $\hat{F}(t)$ in the modified TDSE~\eqref{TDSEwithF} according to the geometric principle (known as McLachlan or Dirac-Frenkel principle in other contexts~\cite{Dirac-30,Frenkel-34,McLachlan-64,LasSu-22}). Let us consider again a wave-function trajectory $\Psi:[0,T] \to {\cal H}$ from initial time $0$ to any arbitrary final time $T>0$. This wave-function trajectory $\Psi$ satisfies the time-dependent geometric principle under the constraint of yielding a given density trajectory $\rho$ if, at every time $t$, the wave-function velocity $\partial_t \Psi(t)$ minimizes the $L^2$ norm of the residual of the TDSE over all trial wave-function velocities $\Xi \in \cH$ yielding a given density velocity $\partial_t \rho(t)$,
\begin{eqnarray}
\partial_t \Psi(t) = \argmin{\substack{{\Xi}\rightsquigarrow  \partial_t \rho(t)}} \left\| i \, {\Xi} -\hat{H}(t) \Psi(t) \right\|^2_{{\cal H}},
\label{eq:mindtPsi}
\end{eqnarray}
where $\Xi \rightsquigarrow  \partial_t \rho(t)$ means that the variations of $\Xi$ are taken under the constraint $2\Re\bra{\Xi}\hat{\rho}(\b{r})\ket{\Psi(t)} = \partial_t \rho(\b{r},t)$. This is the appropriate constraint since, for any time-dependent wave function $\Psi(t)$, we have $2\Re\bra{\partial_t \Psi(t)}\hat{\rho}(\b{r})\ket{\Psi(t)} = \partial_t \rho_{\Psi(t)}(\b{r})$. In other words, we choose the wave-function velocity $\partial_t \Psi(t)$ to be the closest to the unconstrained Schrödinger velocity $-i \hat{H}(t) \Psi(t)$ among those reproducing the prescribed density velocity.
Since fixing $\partial_t \rho(t)$ is equivalent to fixing the normal component of the wave-function velocity $[\partial_t \!\Psi(t)]_\perp$ according to Eq.~\eqref{eq:Swperp}, the constrained minimization in Eq.~\eqref{eq:mindtPsi} can be reformulated as the unconstrained minimization over the tangent component of the wave-function velocity $[\partial_t \!\Psi(t)]_\parallel$. The minimum is achieved when $[\partial_t \!\Psi(t)]_\parallel$ is the orthogonal projection of the Schrödinger wave-function velocity $- i \hat{H}(t) \Psi(t)$ onto the tangent space,
\begin{eqnarray}
[\partial_t \!\Psi(t)]_\parallel = \hat{P}_{{\cal T}_{\Psi(t)} {\cal M}_{\rho(t)}} \left( - i \hat{H}(t) \Psi(t)\right),
\end{eqnarray}
where $\hat{P}_{{\cal T}_{\Psi(t)} {\cal M}_{\rho(t)}}$ is the orthogonal projection operator onto ${\cal T}_{\Psi(t)} {\cal M}_{\rho(t)}$ for the real inner product $\Re\braket{\,\cdot\,}{\,\cdot\,}$. 
This implies that the difference $[\partial_t \!\Psi(t)]_\parallel - (- i \hat{H}(t) \Psi(t))$ must be a vector in the normal space, but since $[\partial_t \!\Psi(t)]_\parallel - (- i \hat{H}(t) \Psi(t)) = -i \hat{F}(t) \Psi(t) - [\partial_t \!\Psi(t)]_\perp$, we obtain that $-i \hat{F}(t) \Psi(t)$ must be in the normal space (see Fig.~\ref{fig:geometric_principle}),
\begin{eqnarray}
-i \hat{F}(t) \Psi(t) \in {\cal N}_{\Psi(t)} {\cal M}_{\rho(t)}.
\label{eq:iFPsiinN}
\end{eqnarray}
Note that the above discussion starting from Eq.~\eqref{eq:mindtPsi} can also be reinterpreted as the fact that $\hat{F}(t) \Psi(t)$ must be chosen with minimal $L^2$-norm under the density constraint. 

The condition in Eq.~\eqref{eq:iFPsiinN} looks quite similar to the condition in Eq.~\eqref{eq:FPsiinN} for the variational principle, but since ${\cal N}_{\Psi(t)} {\cal M}_{\rho(t)}$ is not a complex vector space but only a real vector space, the two conditions are in fact different. Using the expression of the normal space in Eq.~\eqref{eq:NM}, we arrive at the conclusion that, in the geometric (G) principle, $\hat{F}(t)$ can be chosen of the form of an imaginary one-electron local potential operator
\begin{eqnarray}
\hat{F}^\text{G}(t) := i \int_{\mathbb{R}^d} \d\b{r} \, w(\b{r},t) \hat{\rho}(\b{r}),
\label{eq:FGP}
\end{eqnarray}
with a real-valued function $w(.,t): \mathbb{R}^d \to \mathbb{R}$ which must be such that the solution to the modified TDSE~\eqref{TDSEwithF} with the operator $\hat{F}^\text{G}(t)$ gives the prescribed density $\rho$. We will refer to $w(\b{r},t)$ as the geometric potential. The geometric potential $w(\b{r},t)$ can be interpreted as the Lagrange multiplier for the density-velocity constraint, $2\Re\bra{\partial_t \Psi(t)}\hat{\rho}(\b{r})\ket{\Psi(t)} = \partial_t \rho(\b{r},t)$, in the minimization in Eq.~\eqref{eq:mindtPsi} (see Appendix~\ref{app:Lagrange}).

If we choose $\hat{H}(t) := \hat{T} + \hat{V}_\ee + \hat{{\cal V}}(t)$ with some local external potential operator $\hat{{\cal V}}(t) := \int_{\mathbb{R}^d} \d\b{r} \, \nu(\b{r},t) \, \hat{\rho}(\b{r})$ as the reference Hamiltonian in Eq.~\eqref{TDSEwithF}, we recover geometric TDDFT~\cite{MAQUI_TDDFT-26,MAQUI_finite-26}. For a given density trajectory $\rho$, initial wave function $\Psi_0$, and external potential $\nu$, if the geometric potential $w$ in Eq.~\eqref{eq:FGP} exists, then the geometric extension of the Runge-Gross theorem~\cite{MAQUI_TDDFT-26} states that, for spatially smooth $v_\ee$, $\Psi_0$, $\nu$, and $w$, and time-analytic $w$, the function $w$ is unique. This means that $w$ is a functional of $\rho$, $\Psi_0$, and $\nu$, i.e. $w = w[\rho,\Psi_0,\nu]$. For geometric TDDFT, the equation~\eqref{eq:F=0} giving the physical density $\rho^\ph$ thus takes the form
\begin{eqnarray}
w[\rho,\Psi_0^\ph,v^\ph] = 0 \, \Longleftrightarrow \, \rho=\rho^\ph.
\label{eq:w=0}
\end{eqnarray}
Similarly to variational TDDFT, we could also define a universal interacting geometric potential functional $w[\rho,\Psi_0]:=w[\rho,\Psi_0,0]$, but we will not make use of it.

Thus, in geometric TDDFT, a prescribed density trajectory $\rho$ is reproduced from a time-dependent wave function $\Psi^\G(t)$ satisfying an effective TDSE,
\begin{eqnarray}
\left\{
\begin{aligned}
i \partial_t \Psi^\G(t) &= \left( \hat{T} + \hat{V}_\ee + \hat{{\cal V}}(t) + i \hat{W}(t)\right) \Psi^\G(t), \\
\Psi^\G(0) &= \Psi_0,
\end{aligned}
\right.
\label{eq:TDSE-TDDFTgeo}
\end{eqnarray}
with $\hat{W}(t) := \int_{\mathbb{R}^d} \d\b{r} \, w[\rho,\Psi_0,\nu](\b{r},t) \, \hat{\rho}(\b{r})$ and an initial wave function $\Psi_0$ having the prescribed initial density, i.e. $\rho_{\Psi_0}(\b{r}) = \rho(\b{r},0)$. Since the geometric potential $w[\rho,\Psi_0,\nu]$, imposing the density $\rho$, is a functional of $\rho$, $\Psi_0$ and $\nu$, the wave function $\Psi^\G(t)$ in Eq.~\eqref{eq:TDSE-TDDFTgeo} is also a functional of $\rho$, $\Psi_0$, and $\nu$, i.e. $\Psi^\G(t) = \Psi^\G[\rho,\Psi_0,\nu](t)$. This wave function $\Psi^\G(t)$ will generally differ from $\Psi^\V(t)$ in Eq.~\eqref{eq:TDSE-TDDFT}. At first sight, one may be worried by the fact that Eq.~\eqref{eq:TDSE-TDDFTgeo} involves a non-Hermitian Hamiltonian. However, $\hat{W}(t)$ is such that it reproduces the time-dependent density $\rho(t)$, so it preserves the norm of $\Psi^\G(t)$, leading to the condition
\begin{eqnarray}
\bra{\Psi^\G(t)} \hat{W}(t) \ket{\Psi^\G(t)}=0.
\label{eq:PsiWPsi=0}
\end{eqnarray}
This can be used to rewrite Eq.~\eqref{eq:TDSE-TDDFTgeo} in a Hermitian form,
\begin{eqnarray}
\left\{
\begin{aligned}
i \partial_t \Psi^\G(t) &= \left( \hat{T} + \hat{V}_\ee + \hat{{\cal V}}(t) + \hat{\Sigma}(t)\right) \Psi^\G(t),  \\
\Psi^\G(0) &= \Psi_0,
\end{aligned}
\right.
\label{eq:TDSE-TDDFTgeocomm}
\end{eqnarray}
where $\hat{\Sigma}(t)$ is an $N$-electron state-dependent nonlocal Hermitian operator written as a commutator
\begin{eqnarray}
\hat{\Sigma}(t) :=  i \left[\hat{W}(t),\ket{\Psi^\G(t)}\bra{\Psi^\G(t)}\right],
\label{eq:Sigma}
\end{eqnarray}
where $\ket{\Psi^\G(t)}\bra{\Psi^\G(t)}$ is the $N$-electron density matrix associated with $\Psi^\G(t)$. Note that, when using the commutator form in Eqs.~\eqref{eq:TDSE-TDDFTgeocomm}-\eqref{eq:Sigma}, any arbitrary time-dependent constant $C(t)$ can be added to the geometric potential $w(\b{r},t)$, whereas when using the imaginary-potential form in Eq.~\eqref{eq:TDSE-TDDFTgeo}, the time-dependent constant is fixed by the norm-preserving condition in Eq.~\eqref{eq:PsiWPsi=0}.

Equation~\eqref{eq:TDSE-TDDFTgeo} (or Eq.~\eqref{eq:TDSE-TDDFTgeocomm} with the condition~\eqref{eq:PsiWPsi=0}) leads to a modified form of the continuity equation
\begin{eqnarray}
\partial_t \rho(\b{r},t) + \nabla \cdot \b{j}_{\Psi^\G(t)}(\b{r}) = 2\int_{\mathbb{R}^d} S_{\Psi^\G(t)}(\b{r},\b{r}') w(\b{r}',t) \d \b{r}',
\nonumber\\
\label{eq:modconteq}
\end{eqnarray}
where $S_{\Psi}(\b{r},\b{r}')$ is the operator kernel defined in Eq.~\eqref{eq:SPsirrp}, which can also be written as
\begin{eqnarray}
S_{\Psi}(\b{r},\b{r}') = \rho_\Psi(\b{r}) \delta(\b{r}-\b{r}') + \rho_{2,\Psi}(\b{r},\b{r}'),
\end{eqnarray}
where $\rho_{2,\Psi}(\b{r},\b{r}'):= \bra{\Psi} \hat{\rho}_2(\b{r},\b{r}') \ket{\Psi}$ is the pair density of $\Psi$.
Thus, by contrast with variational TDDFT, in geometric TDDFT the geometric potential $w$ reproducing a given density trajectory $\rho$ can be directly extracted from the modified continuity equation~\eqref{eq:modconteq}. The presence of $w(\b{r},t)$ in Eq.~\eqref{eq:modconteq} also means that, contrary to variational TDDFT, there is no additional compatibility condition on $\nabla \cdot \b{j}_{\Psi_0}(\b{r})$ in geometric TDDFT, since the initial function $w(\b{r},0)$ will automatically be such that Eq.~\eqref{eq:modconteq} is satisfied at $t=0$, whatever the values of $\partial_t \rho(\b{r},t) |_{t=0}$ and $\nabla \cdot \b{j}_{\Psi_0}(\b{r})$. Equation~\eqref{eq:modconteq} is the fundamental equation of geometric TDDFT which gives a link between the geometric potential $w(\b{r},t)$ and the first-order time-derivative of the density $\partial_t \rho(\b{r},t)$. The operator $S_{\Psi}$ appearing in this equation is central to geometric TDDFT. In particular, the proof of the geometric Runge-Gross theorem in Ref.~\onlinecite{MAQUI_TDDFT-26} requires the invertibility of $S_{\Psi_0}$. The mathematical properties of the operator $S_{\Psi}$ are described in Section~\ref{sec:S}.

According to the modified continuity equation~\eqref{eq:modconteq}, the geometric potential $w(\b{r},t)$ generates effective sources and sinks of density (similar to the situation in Refs.~\onlinecite{Ern-JCP-06,ZhoErn-JCP-12} for ground-state DFT of open systems). Whereas in variational TDDFT the potential $v(\b{r},t)$ can be thought of as generating a force moving the density across space and time (according to Eq.~\eqref{eq:divlocforce}), in geometric TDDFT the geometric potential $w(\b{r},t)$ directly creates or absorbs the density where and when this is needed. This can be considered a more direct way of imposing a prescribed density. Note that the total number of electrons $N$ remains conserved, since the source/sink term in Eq.~\eqref{eq:modconteq} integrates to zero,
\begin{eqnarray}
\int_{\mathbb{R}^d} \int_{\mathbb{R}^d} S_{\Psi^\G(t)}(\b{r},\b{r}') w(\b{r}',t) \d \b{r} \d \b{r}' \phantom{xxxxxxxxx}
\nonumber\\
=  N \bra{\Psi^\G(t)} \hat{W}(t) \ket{\Psi^\G(t)}=0,
\label{}
\end{eqnarray}
where we have used Eqs.~\eqref{eq:SPsirrp} and~\eqref{eq:PsiWPsi=0}.

\subsection{Variational and geometric TDKS}

We now describe several TDKS schemes based on variational or geometric TDDFT.

\subsubsection{Variational TDKS}
\label{sec:variationalTDKS}
Repeating the reasoning in Section~\ref{sec:standardTDDFT} with the non-interacting reference Hamiltonian $\hat{H} := \hat{T}$ in Eq.~\eqref{TDSEwithF}, we obtain that a prescribed density trajectory $\rho$ can be reproduced from a single-determinant wave function $\Phi^\V(t)$ satisfying an effective non-interacting TDSE,~\cite{RunGro-84}
\begin{eqnarray}
\left\{
\begin{aligned}
i \partial_t \Phi^\V(t) &= \left( \hat{T} + \hat{V}_\s(t)\right) \Phi^\V(t), \\
\Phi^\V(0) &= \Phi_0,
\end{aligned}
\right.
\label{eq:TDDFT-TDKS}
\end{eqnarray}
with a one-electron local potential operator $\hat{V}_\s(t) := \int_{\mathbb{R}^d} \d\b{r} \, v_\s[\rho,\Phi_0](\b{r},t) \, \hat{\rho}(\b{r})$ and an initial single-determinant wave function $\Phi_0$ having the prescribed initial density, i.e. $\rho_{\Phi_0}(\b{r}) = \rho(\b{r},0)$, and a compatible current-density divergence, i.e. $\nabla \cdot \b{j}_{\Phi_0}(\b{r}) = -\partial_t \rho(\b{r},t)|_{t=0}$. The potential $v_\s[\rho,\Phi_0](\b{r},t)$, imposing the density trajectory $\rho$, is called the KS potential and is a functional of $\rho$ and $\Phi_0$. The single-determinant wave function $\Phi^\V(t)$ is also a functional of $\rho$ and $\Phi_0$, i.e. $\Phi^\V(t) = \Phi^\V[\rho,\Phi_0](t)$. The uniqueness (up to an arbitrary additive time-dependent constant) of the KS potential $v_\s[\rho,\Phi_0](\b{r},t)$ is guaranteed by the Runge-Gross theorem applied to the non-interacting system. The recent work of Ref.~\onlinecite{LauLewOld-V-26_ppt} gives conditions for both short-time existence and uniqueness of the KS potential $v_\s$ on the torus.

The continuity equation associated with Eq.~\eqref{eq:TDDFT-TDKS} takes the form
\begin{eqnarray}
\partial_t \rho(\b{r},t) + \nabla \cdot \b{j}_{\Phi^\V(t)}(\b{r}) = 0,
\label{eq:conteqTDKS}
\end{eqnarray}
which means that, for a given density, the non-interacting current density $\b{j}_{\Phi^\V(t)}(\b{r})$ and the interacting current density $\b{j}_{\Psi^\V(t)}(\b{r})$ appearing in Eq.~\eqref{eq:conteq} have the same longitudinal component. Finally, the force-balance equation associated with Eq.~\eqref{eq:TDDFT-TDKS} is
\begin{eqnarray}
\partial_t^2 \rho(\b{r},t) = \nabla \cdot (\rho(\b{r},t) \, \nabla v_\s(\b{r},t) ) + q_{\s,\Phi^\V(t)}(\b{r}), \;\;\;
\label{eq:divlocforceTDKS}
\end{eqnarray}
where $q_{\s,\Phi}:= - \bra{\Phi} [\hat{T},[\hat{T},\hat{\rho}(\b{r})]] \ket{\Phi}$. Equation~\eqref{eq:divlocforceTDKS} can in principle be used to extract $v_\s(\b{r},t)$. 

If we insert the physical density $\rho^\ph$ of the physical interacting system inside the KS potential, i.e. $v_\s[\rho^\ph,\Phi_0](\b{r},t)$, then by construction the solution $\Phi^\V(t)$ to Eq.~\eqref{eq:TDDFT-TDKS} gives back the physical density, i.e. $\rho_{\Phi^\V[\rho^\ph,\Phi_0](t)}(\b{r}) = \rho^\ph(\b{r},t)$. However, Eq.~\eqref{eq:TDDFT-TDKS} cannot be used to predict $\rho^\ph$ since $v_\s[\rho,\Phi_0]$ does not contain any information about the electron-electron interaction $v_\ee$ and the external potential $v^\ph$ of the physical system. To obtain a predictive scheme, a connection must be made with the physical system. This is achieved by first introducing the Hartree-exchange-correlation (Hxc) potential~\cite{RugPenLee-15}
\begin{eqnarray}
v_\Hxc[\rho,\Phi_0,\Psi_0] := v_\s[\rho,\Phi_0] - v[\rho,\Psi_0],
\label{eq:vHxc}
\end{eqnarray}
where $v[\rho,\Psi_0]$ is the universal interacting functional potential in Eq.~\eqref{eq:TDSE-TDDFT}. We then obtain the usual variational TDKS scheme as an effective non-interacting non-linear TDSE,
\begin{eqnarray}
\left\{
\begin{aligned}
i \partial_t \Phi^\KS(t) &= \left( \hat{T} + \hat{V}^\ph(t) + \hat{V}_\Hxc(t)\right) \Phi^\KS(t), \\
\Phi^\KS(0) &= \Phi_0,
\end{aligned}
\right.
\label{eq:TDKS}
\end{eqnarray}
with the Hxc potential operator $\hat{V}_\Hxc(t) := \int_{\mathbb{R}^d} \d\b{r} \, v_\Hxc[\rho_{\Phi^\KS},\Phi_0,\Psi_0^\ph](\b{r},t) \, \hat{\rho}(\b{r})$. Note that $v_\Hxc$ is evaluated at the physical initial wave function $\Psi_0^\ph$ and at the density $\rho_{\Phi^\KS} := (\rho_{\Phi^\KS(s)})_{0\leq s \leq t}$ which designates the density trajectory calculated from the single-determinant wave function $\Phi^\KS$ until time $t$. This makes Eq.~\eqref{eq:TDKS} a differential equation highly nonlinear in $\Phi^\KS(t)$. By the Runge-Gross theorem, $v^\ph + v_\Hxc[\rho_{\Phi^\KS},\Phi_0,\Psi_0^\ph] \eqv v_\s[\rho_{\Phi^\KS},\Phi_0]$, and then by Eqs.~\eqref{eq:vHxc} and~\eqref{eq:v=0_universal}, we recover the physical density, i.e. $\rho_{\Phi^\KS(t)}(\b{r}) = \rho^\ph(\b{r},t)$. The KS wave function $\Phi^\KS(t)$ in Eq.~\eqref{eq:TDKS} thus corresponds to the wave function in Eq.~\eqref{eq:TDDFT-TDKS} for the physical density, i.e. $\Phi^\KS(t) \eqpsi \Phi^\V[\rho^\ph,\Phi_0](t)$ where $\eqpsi$ designates here the equality between wave functions modulo multiplicative time-dependent phase factors.

\subsubsection{Geometric TDKS}
\label{sec:geometricTDKS}

Applying the reasoning in Section~\ref{sec:geometricTDDFT} with the non-interacting reference Hamiltonian $\hat{H}(t) := \hat{T} + \hat{{\cal V}}(t)$ in Eq.~\eqref{TDSEwithF} with some local external potential operator $\hat{{\cal V}}(t) := \int_{\mathbb{R}^d} \d\b{r} \, \nu(\b{r},t) \, \hat{\rho}(\b{r})$, we obtain that a prescribed density trajectory $\rho$ can be reproduced from a single-determinant wave function $\Phi^\G(t)$ satisfying an effective non-interacting TDSE,~\cite{MAQUI_TDDFT-26,MAQUI_finite-26}
\begin{eqnarray}
\left\{
\begin{aligned}
i \partial_t \Phi^\G(t) &= \left( \hat{T} + \hat{{\cal V}}(t) + i \hat{W}_\s(t) \right) \Phi^\G(t), \\
\Phi^\G(0) &= \Phi_0,
\end{aligned}
\right.
\label{eq:TDDFT-TDGKS}
\end{eqnarray}
with an imaginary one-electron local potential operator $i \hat{W}_\s(t) := i \int_{\mathbb{R}^d} \d\b{r} \, w_\s[\rho,\Phi_0,\nu](\b{r},t) \, \hat{\rho}(\b{r})$ and an initial single-determinant wave function $\Phi_0$ having the prescribed initial density, i.e. $\rho_{\Phi_0}(\b{r}) = \rho(\b{r},0)$. The function $w_\s[\rho,\Phi_0,\nu](\b{r},t)$, imposing the density trajectory $\rho$, will be referred to as the KS geometric potential, and is a functional of $\rho$, $\Phi_0$, and $\nu$. The single-determinant wave function $\Phi^\G(t)$ is also a functional of $\rho$, $\Phi_0$, and $\nu$, i.e. $\Phi^\G(t) = \Phi^\G[\rho,\Phi_0,\nu](t)$. The uniqueness of the KS geometric potential $w_\s[\rho,\Phi_0,\nu](\b{r},t)$ is guaranteed by the geometric Runge-Gross theorem~\cite{MAQUI_TDDFT-26} applied to the non-interacting system. The recent work of Ref.~\onlinecite{LauLewOld-W-26_ppt} gives conditions for both short-time existence and uniqueness of the KS geometric potential $w_\s$ on the torus.

Like in the interacting case, since $\hat{W}_\s(t)$ reproduces the time-dependent density $\rho(t)$, it must preserve the norm of $\Phi^\G(t)$, leading to the condition $\bra{\Phi^\G(t)} \hat{W}_\s(t) \ket{\Phi^\G(t)}=0$. This can be used to rewrite Eq.~\eqref{eq:TDDFT-TDGKS} in a Hermitian form,
\begin{eqnarray}
\left\{
\begin{aligned}
i \partial_t \Phi^\G(t) &= \left( \hat{T} + \hat{{\cal V}}(t) + \hat{\Sigma}_\s(t) \right) \Phi^\G(t), \\
\Phi^\G(0) &= \Phi_0,
\end{aligned}
\right.
\label{eq:TDDFT-TDGKScomm}
\end{eqnarray}
where $\hat{\Sigma}_\s(t):=  i \left[\hat{W}_\s(t),\ket{\Phi^\G(t)}\bra{\Phi^\G(t)}\right]$. Since $\Phi^\G(t)$ is a single-determinant wave function and $\hat{W}_\s(t)$ is a one-electron operator, the action of the $N$-electron operator $\hat{\Sigma}_\s(t)$ on $\Phi^\G(t)$ can in fact be represented exactly by the action of a one-electron nonlocal operator,
\begin{eqnarray}
\hat{\Sigma}_\s(t)  \Phi^\G(t) = \phantom{xxxxxxxxxxxxxxxxxxxxxxxxx}
\nonumber\\
\Biggl( \sum_{\sigma \in \{\uparrow,\downarrow\}} \int_{\mathbb{R}^d} \!\int_{\mathbb{R}^d} \!\! {\scriptstyle \Sigma}_\s^\sigma(\b{r},\b{r}',t) \hat{\gamma}^\sigma(\b{r}',\b{r}) \d \b{r} \d \b{r}' \Biggl) \! \Phi^\G(t),
\end{eqnarray}
with the spin-dependent nonlocal potential
\begin{eqnarray}
{\scriptstyle \Sigma}_\text{s}^\sigma(\b{r},\b{r}',t) := i \Bigl( w_\s(\b{r},t) - w_\s(\b{r}',t) \Bigr) \gamma_{\Phi^\G(t)}^\sigma(\b{r},\b{r}'),
\label{}
\end{eqnarray}
and $\gamma_{\Phi}^\sigma(\b{r},\b{r}'):=\bra{\Phi} \hat{\gamma}^\sigma(\b{r},\b{r}') \ket{\Phi}$ where $\hat{\gamma}^\sigma(\b{r},\b{r}'):= \hat{\psi}^\dagger_\sigma(\b{r}') \hat{\psi}_\sigma(\b{r})$ is the spin-resolved density-matrix operator. Note that we have assumed that $\Phi^\G(t)$ is a collinear-spin Slater determinant, i.e. an eigenstate of the spin-projection operator $\hat{S}_z$.

The modified continuity equation associated with Eq.~\eqref{eq:TDDFT-TDGKS} (or Eq.~\eqref{eq:TDDFT-TDGKScomm}) takes the form
\begin{eqnarray}
\partial_t \rho(\b{r},t) + \nabla \cdot \b{j}_{\Phi^\G(t)}(\b{r}) = 2\int_{\mathbb{R}^d} S_{\Phi^\G(t)}(\b{r},\b{r}') w_\s(\b{r}',t) \d \b{r}',
\nonumber\\
\label{eq:modconteqTDGKS}
\end{eqnarray}
where $S_{\Phi}(\b{r},\b{r}') = \rho_\Phi(\b{r}) \delta(\b{r} - \b{r}') +  \rho_{2,\Phi}(\b{r},\b{r}')$ and, for a single-determinant wave function $\Phi$, $\rho_{2,\Phi}(\b{r},\b{r}')=\rho(\b{r}) \rho(\b{r}') + \rho_{2,\x,\Phi}(\b{r},\b{r}')$ contains Hartree and exchange contributions only. The exchange contribution to the pair density is $\rho_{2,\x,\Phi}(\b{r},\b{r}') := - \sum_{\sigma \in \{\uparrow,\downarrow\}} |\gamma_{\Phi}^\sigma(\b{r},\b{r}')|^2$.

If we insert the physical density $\rho^\ph$ inside the KS geometric potential, i.e. $w_\s[\rho^\ph,\Phi_0,\nu](\b{r},t)$, then by construction the solution $\Phi^\G(t)$ to Eq.~\eqref{eq:TDDFT-TDGKS} gives back the physical density, i.e. $\rho_{\Phi^\G[\rho^\ph,\Phi_0,\nu](t)}(\b{r}) = \rho^\ph(\b{r},t)$. However, similarly to Eq.~\eqref{eq:TDDFT-TDKS}, Eq.~\eqref{eq:TDDFT-TDGKS} cannot be used to predict $\rho^\ph$ since $w_\s[\rho,\Phi_0,\nu]$ does not contain any information about the electron-electron interaction $v_\ee$ of the physical system. To obtain a predictive scheme, we can define the following universal Hxc geometric potential~\cite{MAQUI_TDDFT-26}
\begin{eqnarray}
w_\Hxc[\rho,\Phi_0,\Psi_0] := w_\s[\rho,\Phi_0,v[\rho,\Psi_0]],
\label{eq:wHxc}
\end{eqnarray}
where $v[\rho,\Psi_0]$ is again the universal interacting functional potential in Eq.~\eqref{eq:TDSE-TDDFT}. In other words, the quantity $w_\Hxc[\rho,\Phi_0,\Psi_0]$ is the geometric potential needed to produce the prescribed density $\rho$ in an auxiliary non-interacting system with external potential $v[\rho,\Psi_0]$. Note that the use of $v[\rho,\Psi_0]$ implies that we need again the additional compatibility condition of variational TDDFT between the initial wave function $\Psi_0$ and the initial density velocity: $\nabla \cdot \b{j}_{\Psi_0}(\b{r}) = - \partial_t \rho(\b{r},t)|_{t=0}$. Even though $v[\rho,\Psi_0]$ is defined only up to additive time-dependent constants, $w_\Hxc[\rho,\Phi_0,\Psi_0]$ is uniquely defined. We then obtain the geometric TDKS scheme (GKS) as an effective non-interacting nonlinear TDSE,~\cite{MAQUI_TDDFT-26}
\begin{eqnarray}
\left\{
\begin{aligned}
i \partial_t \Phi^\GKS(t) &= \left( \hat{T} + \hat{V}^\ph(t) + i\hat{W}_\Hxc(t)\right) \Phi^\GKS(t), \\
\Phi^\GKS(0) &= \Phi_0,
\end{aligned}
\right.
\;\;
\label{eq:TDGKS}
\end{eqnarray}
with $\hat{W}_\Hxc(t) := \int_{\mathbb{R}^d} \d\b{r} \, w_\Hxc[\rho_{\Phi^\GKS},\Phi_0,\Psi_0^\ph](\b{r},t) \, \hat{\rho}(\b{r})$.  By the geometric Runge-Gross theorem, $w_\Hxc[\rho_{\Phi^\GKS},\Phi_0,\Psi_0^\ph] = w_\s[\rho_{\Phi^\GKS},\Phi_0,v^\ph]$, and then by Eqs.~\eqref{eq:wHxc} and~\eqref{eq:v=0_universal}, we recover the physical density, i.e. $\rho_{\Phi^\GKS(t)}(\b{r}) = \rho^\ph(\b{r},t)$. The GKS wave function $\Phi^\GKS(t)$ in Eq.~\eqref{eq:TDGKS} thus corresponds to the wave function in Eq.~\eqref{eq:TDDFT-TDGKS} for the physical density and the physical external potential, i.e. $\Phi^\GKS(t) = \Phi^\G[\rho^\ph,\Phi_0,v^\ph](t)$.

We can also define the geometric potential correction $w^\text{cor}[\rho,\Phi_0,\Psi_0]$ to a given Hxc potential approximation $v_\Hxc^\text{app}[\rho,\Phi_0,\Psi_0]$,~\cite{MAQUI_TDDFT-26}
\begin{eqnarray}
w^\text{cor}[\rho,\Phi_0,\Psi_0] := w_\s[\rho,\Phi_0,v[\rho,\Psi_0] +v_\Hxc^\text{app}[\rho,\Phi_0,\Psi_0]],
\nonumber\\
\label{eq:wcor}
\end{eqnarray}
leading to the TDKS scheme with geometric correction (KS+G)~\cite{MAQUI_TDDFT-26,MAQUI_finite-26}
\begin{eqnarray}
\left\{
\begin{aligned}
i \partial_t \Phi^\text{KS+G}(t) &= \Bigl( \hat{T} + \hat{V}^\ph(t) + \hat{V}_\Hxc^\text{app}(t) \\
&\phantom{xxxxxxx}+ i\hat{W}^\text{cor}(t)\Bigl) \Phi^\text{KS+G}(t), \\
\Phi^\text{KS+G}(0) &= \Phi_0,
\end{aligned}
\right.
\label{eq:TDKS+G}
\end{eqnarray}
with $\hat{V}_\Hxc^\text{app}(t) := \int_{\mathbb{R}^d} \d\b{r} \, v_\Hxc^\text{app}[\rho_{\Phi^\text{KS+G}},\Phi_0,\Psi_0^\ph](\b{r},t) \, \hat{\rho}(\b{r})$ and $\hat{W}^\text{cor}(t) := \int_{\mathbb{R}^d} \d\b{r} \, w^\text{cor}[\rho_{\Phi^\text{KS+G}},\Phi_0,\Psi_0^\ph](\b{r},t) \, \hat{\rho}(\b{r})$. The physical density is also recovered with this scheme, i.e. $\rho_{\Phi^\text{KS+G}(t)}(\b{r}) = \rho^\ph(\b{r},t)$.

In the special case where $v_\Hxc^\text{app}$ is chosen as the exact adiabatic Hxc potential~\cite{ThiGroKum-PRL-08},
\begin{eqnarray}
v_\Hxc^\text{ea}[\rho](\b{r},t) := v_\Hxc^\text{gs}[\rho(t)](\b{r}),
\label{eq:vea}
\end{eqnarray}
where $v_\Hxc^\text{gs}[\rho](\b{r})$ is the exact Hxc potential of static ground-state DFT, the geometric potential correction $w^\text{cor}$ will be called the non-adiabatic (na) geometric potential correction,
\begin{eqnarray}
w^\text{na}[\rho,\Phi_0,\Psi_0] := w_\s[\rho,\Phi_0,v[\rho,\Psi_0] +v_\Hxc^\text{ea}[\rho]].
\label{eq:wna}
\end{eqnarray}

\section{Exact conditions in variational and geometric TDDFT}
\label{sec:exact_constraints}

In this section, we review some exact conditions known in variational TDDFT and extend them to geometric TDDFT. For reviews of exact conditions in variational TDDFT, see Refs.~\onlinecite{Bur-INC-06,Ullrichs-11,WagYanBur-INC-12}.

\subsection{Causality and initial-state/density-history consistency}
\label{sec:causality}

Perhaps the most basic exact conditions on the potential $v[\rho,\Psi_0](\b{r},t)$ of variational TDDFT are causality (see, e.g., Ref.~\onlinecite{Ullrichs-11}) and initial-state/density-history consistency~\cite{MaiBurWoo-PRL-02}, which directly stem from the Runge-Gross theorem. These conditions can be explicitly shown as follows.

Consider a time interval $[0,T]$, a given density trajectory $\rho$ defined on this time interval, and an initial wave function $\Psi_0$. The wave function $\Psi^\V =  \Psi^\V[\rho,\Psi_0]$ of variational TDDFT is also defined on the time interval $[0,T]$ and satisfies the effective TDSE in Eq.~\eqref{eq:TDSE-TDDFT}. The restriction of $\Psi^\V$ on the time interval $[t_1,t] \subseteq [0,T]$ satisfies the same effective TDSE for $t'\in[t_1,t]$ but with the initial condition at $t_1$,
\begin{eqnarray}
\left\{
\begin{aligned}
\! i \partial_{t'} \! \Psi^\V{|_{[t_1,t]}}(t') \! &= \! \left( \! \hat{T} + \hat{V}_\text{ee} + \hat{V}{|_{[t_1,t]}}(t') \! \right) \Psi^\V{|_{[t_1,t]}}(t'), \\
\Psi^\V{|_{[t_1,t]}}(t_1) &= \Psi^\V(t_1),
\end{aligned}
\right.
\;\;
\label{eq:TDSE-TDDFT-R}
\end{eqnarray}
with $\hat{V}{|_{[t_1,t]}}(t') := \int_{\mathbb{R}^d} \d\b{r} \, v[\rho,\Psi_0]{|_{[t_1,t]}}(\b{r},t') \, \hat{\rho}(\b{r})$ where $v[\rho,\Psi_0]{|_{[t_1,t]}}(\b{r},t')$ is the potential restricted to the time interval $[t_1,t]$. Thus, the restricted wave function $\Psi^\V{|_{[t_1,t]}}$ satisfies the effective TDSE in Eq.~\eqref{eq:TDSE-TDDFT-R} starting from the initial wave function $\Psi^\V[\rho,\Psi_0](t_1)$ and with the restricted potential $v[\rho,\Psi_0]{|_{[t_1,t]}}(\b{r},t')$, and yields the density restricted to the time interval $[t_1,t]$, i.e. $\rho{|_{[t_1,t]}}$. According to the Runge-Gross theorem, this implies that, for $t'\in[t_1,t]$,
\begin{eqnarray}
\Psi^\V [\rho,\Psi_0]{|_{[t_1,t]}}(t') \eqpsi \Psi^\V[\rho{|_{[t_1,t]}},\Psi^\V [\rho,\Psi_0](t_1)](t'),
\label{}
\end{eqnarray}
and
\begin{eqnarray}
v[\rho,\Psi_0]{|_{[t_1,t]}}(\b{r},t') \eqv v[\rho{|_{[t_1,t]}},\Psi^\V [\rho,\Psi_0](t_1)](\b{r},t').
\label{eq:vrhoPsi0restrict}
\end{eqnarray}
Proceeding exactly in the same way for the non-interacting TDSE in Eq.~\eqref{eq:TDDFT-TDKS}, we find the same condition for the KS potential,
\begin{eqnarray}
v_\s[\rho,\Phi_0]{|_{[t_1,t]}}(\b{r},t') \eqv v_\s[\rho{|_{[t_1,t]}},\Phi^\V [\rho,\Phi_0](t_1)](\b{r},t'),
\label{}
\end{eqnarray}
and therefore, using Eq.~\eqref{eq:vHxc}, also for the Hxc potential,
\begin{eqnarray}
v_\Hxc[\rho,\Phi_0,\Psi_0]{|_{[t_1,t]}}(\b{r},t') \eqv \phantom{xxxxxxxxxxxxxxx}
\nonumber\\
v_\Hxc[\rho{|_{[t_1,t]}},\Phi^\V [\rho,\Phi_0](t_1),\Psi^\V [\rho,\Psi_0](t_1)](\b{r},t').
\label{eq:vHxcrestrict}
\end{eqnarray}
The condition in Eq.~\eqref{eq:vHxcrestrict} contains both the properties of causality and initial-state/density-history consistency. Indeed, for $t_1=0$ and $t'=t$, Eq.~\eqref{eq:vHxcrestrict} gives the causality property
\begin{eqnarray}
v_\Hxc[\rho,\Phi_0,\Psi_0](\b{r},t) \eqv v_\Hxc[\rho{|_{[0,t]}},\Phi_0,\Psi_0](\b{r},t),
\label{}
\end{eqnarray}
i.e., $v_\Hxc[\rho,\Phi_0,\Psi_0](\b{r},t)$ is a causal functional of $\rho$ (it depends only on the densities $\rho(t')$ for $t'\leq t$). Note that causality does not forbid dependence on instantaneous density time-derivatives like $\partial_t \rho(t)$ or $\partial_t^2 \rho(t)$. Besides, for $t_1 \not= 0$ and $t'=t$, Eq.~\eqref{eq:vHxcrestrict} gives the initial-state/density-history consistency property~\cite{MaiBurWoo-PRL-02}
\begin{eqnarray}
v_\Hxc[\rho,\Phi_0,\Psi_0](\b{r},t) \eqv \phantom{xxxxxxxxxxxxxxxxxxx}
\nonumber\\
v_\Hxc[\rho{|_{[t_1,t]}},\Phi^\V [\rho,\Phi_0](t_1),\Psi^\V [\rho,\Psi_0](t_1)](\b{r},t),
\label{}
\end{eqnarray}
i.e., the density history on the time interval $[0,t_1]$ can be absorbed into the new initial states $\Phi^\V [\rho,\Phi_0](t_1)$ and $\Psi^\V [\rho,\Psi_0](t_1)$.

For geometric TDDFT, the situation is very similar. For a given density trajectory $\rho$ and external potential $\nu$ defined on the time interval $[0,T]$, and a given initial wave function $\Psi_0$, the wave function $\Psi^\G =  \Psi^\G[\rho,\Psi_0,\nu]$ of geometric TDDFT satisfies the effective TDSE in Eq.~\eqref{eq:TDSE-TDDFTgeo}. The restriction of $\Psi^\G$ on the time interval $[t_1,t]$ satisfies the following equation for $t'\in[t_1,t]$,
\begin{eqnarray}
\left\{
\begin{aligned}
i \partial_{t'} \Psi^\G{|_{[t_1,t]}}(t') &= \Bigl( \hat{T} + \hat{V}_\text{ee} + \hat{{\cal V}}{|_{[t_1,t]}}(t') \\
& \quad + i \hat{W}{|_{[t_1,t]}}(t') \Big) \Psi^\G{|_{[t_1,t]}}(t'), \\
\Psi^\G{|_{[t_1,t]}}(t_1) &= \Psi^\G(t_1),
\end{aligned}
\right.
\label{eq:TDSE-TDDFTgeo-R}
\end{eqnarray}
with $\hat{{\cal V}}{|_{[t_1,t]}}(t') := \int_{\mathbb{R}^d} \d\b{r} \, \nu{|_{[t_1,t]}}(\b{r},t') \, \hat{\rho}(\b{r})$ and $\hat{W}{|_{[t_1,t]}}(t') := \int_{\mathbb{R}^d} \d\b{r} \, w[\rho,\Psi_0,\nu]{|_{[t_1,t]}}(\b{r},t') \, \hat{\rho}(\b{r})$. Thus, the restricted wave function $\Psi^\G{|_{[t_1,t]}}$ satisfies the effective TDSE in Eq.~\eqref{eq:TDSE-TDDFTgeo-R} starting from the initial wave function $\Psi^\G[\rho,\Psi_0,\nu](t_1)$, with the restricted external potential $\nu{|_{[t_1,t]}}(\b{r},t')$ and the restricted geometric potential $w[\rho,\Psi_0,\nu]{|_{[t_1,t]}}(\b{r},t')$, and yields the restricted density $\rho{|_{[t_1,t]}}$. According to the geometric Runge-Gross theorem, this implies that, for $t'\in[t_1,t]$,
\begin{eqnarray}
\Psi^\G [\rho,\Psi_0,\nu]{|_{[t_1,t]}}(t') = \phantom{xxxxxxxxxxxxxx}
\nonumber\\
\Psi^\G[\rho{|_{[t_1,t]}},\Psi^\G[\rho,\Psi_0,\nu](t_1),\nu{|_{[t_1,t]}}](t'),
\label{}
\end{eqnarray}
and
\begin{eqnarray}
w[\rho,\Psi_0,\nu]{|_{[t_1,t]}}(\b{r},t') = \phantom{xxxxxxxxxxxxxx}
\nonumber\\
w[\rho{|_{[t_1,t]}},\Psi^\G[\rho,\Psi_0,\nu](t_1),\nu{|_{[t_1,t]}}](\b{r},t').
\label{}
\end{eqnarray}
Proceeding exactly in the same way for the non-interacting TDSE in Eq.~\eqref{eq:TDDFT-TDGKS}, we find the same condition for the KS geometric potential,
\begin{eqnarray}
w_\s[\rho,\Phi_0,\nu]{|_{[t_1,t]}}(\b{r},t') = \phantom{xxxxxxxxxxxxxx}
\nonumber\\
w_\s[\rho{|_{[t_1,t]}},\Phi^\G[\rho,\Phi_0,\nu](t_1),\nu{|_{[t_1,t]}}](\b{r},t'),
\label{}
\end{eqnarray}
and therefore, using Eqs.~\eqref{eq:wHxc} and~\eqref{eq:vrhoPsi0restrict}, also for the Hxc geometric potential,
\begin{eqnarray}
w_\Hxc[\rho,\Phi_0,\Psi_0]{|_{[t_1,t]}}(\b{r},t') = \phantom{xxxxxxxxxxxxxxxxxxxx}
\nonumber\\
w_\Hxc[\rho{|_{[t_1,t]}},\Phi^\G[\rho,\Phi_0,v[\rho,\Psi_0]](t_1),\Psi^\V[\rho,\Psi_0](t_1)](\b{r},t').
\nonumber\\
\label{eq:wHxcrestrict}
\end{eqnarray}
Equation~\eqref{eq:wHxcrestrict} encapsulates both the causality and initial-state/density-history consistency of $w_\Hxc[\rho,\Phi_0,\Psi_0](\b{r},t)$. An analogous condition holds for the geometric potential correction $w^\text{cor}[\rho,\Phi_0,\Psi_0]$ in Eq.~\eqref{eq:wcor}, provided that the corresponding approximate Hxc potential $v_\Hxc^\text{app}[\rho,\Phi_0,\Psi_0]$ is causal and satisfies initial-state/density-history consistency along the corresponding geometrically corrected evolution.

\subsection{Time reversal}

The effect of time reversal on the Hxc potential is usually not discussed in variational TDDFT, but it is interesting to discuss it for comparison with geometric TDDFT. 

In this subsection, we will work on a symmetric time interval $[-T,T]$. For any time-dependent wave function $\Psi(t)$, we define the time-reversed wave function,
\begin{eqnarray}
\Psi_{-}(t) := \Psi^*(-t).
\label{}
\end{eqnarray}
Note that the physical time-reversal operator also transforms the spin coordinates, but for deriving the transformation of the local spin-independent potentials, we can ignore spin coordinates. If $\Psi(t)$ gives the time-dependent density $\rho(\b{r},t)$, then $\Psi_{-}(t)$ gives the time-reversed density,
\begin{eqnarray}
\rho_{-}(\b{r},t) := \rho(\b{r},-t).
\label{}
\end{eqnarray}
We have seen that in variational TDDFT, for a given density trajectory $\rho$ and initial wave function $\Psi_0$, there is a time-dependent wave function $\Psi^\V(t) = \Psi^\V[\rho,\Psi_0](t)$ satisfying the effective TDSE in Eq.~\eqref{eq:TDSE-TDDFT} and giving the density $\rho$. If we reverse time, $t \to - t$, and take the complex conjugate in Eq.~\eqref{eq:TDSE-TDDFT}, we obtain 
\begin{eqnarray}
\left\{
\begin{aligned}
i \partial_t \Psi^\V_{-}(t) &= \left( \hat{T} + \hat{V}_\text{ee} + \hat{V}(-t)\right) \Psi^\V_{-}(t), \\
\Psi^\V_{-}(0) &= \Psi_0^*,
\end{aligned}
\right.
\label{eq:TDSE-TDDFT-TR}
\end{eqnarray}
where $\hat{V}(-t) = \int_{\mathbb{R}^d} \d\b{r} \, v[\rho,\Psi_0](\b{r},-t) \, \hat{\rho}(\b{r})$. Thus, the time-reversed wave function $\Psi^\V_{-}[\rho,\Psi_0](t)$ satisfies the effective TDSE in Eq.~\eqref{eq:TDSE-TDDFT-TR} with the time-reversed potential $v[\rho,\Psi_0](\b{r},-t)$ and initial wave function $\Psi_0^*$, and yields the time-reversed density $\rho_{-}$. According to the Runge-Gross theorem, this implies that
\begin{eqnarray}
\Psi^\V[\rho_{-},\Psi_0^*] \eqpsi \Psi^\V_{-}[\rho,\Psi_0],
\end{eqnarray}
and
\begin{eqnarray}
v[\rho_{-},\Psi_0^*](\b{r},t) \eqv  v[\rho,\Psi_0](\b{r},-t).
\label{eq:v-TR}
\end{eqnarray}
Proceeding in exactly the same way for the non-interacting TDSE in Eq.~\eqref{eq:TDDFT-TDKS}, we obtain the same time-reversal transformation for the KS potential,
\begin{eqnarray}
v_\s[\rho_{-},\Phi_0^*](\b{r},t) \eqv  v_\s[\rho,\Phi_0](\b{r},-t),
\end{eqnarray}
and, using Eq.~\eqref{eq:vHxc}, for the Hxc potential,
\begin{eqnarray}
v_\Hxc[\rho_{-},\Phi_0^*,\Psi_0^*](\b{r},t) \eqv v_\Hxc[\rho,\Phi_0,\Psi_0](\b{r},-t),
\label{eq:vHxcTR}
\end{eqnarray}
i.e. $v_\Hxc$ is time-even under time reversal.

Similarly, in geometric TDDFT, for a given density trajectory $\rho$, initial wave function $\Psi_0$, and external potential $\nu$, there is a time-dependent wave function $\Psi^\G(t) = \Psi^\G[\rho,\Psi_0,\nu](t)$ satisfying the effective TDSE in Eq.~\eqref{eq:TDSE-TDDFTgeo} and giving the density $\rho$. If we reverse time, $t \to - t$, and take the complex conjugate in Eq.~\eqref{eq:TDSE-TDDFTgeo}, we obtain
\begin{eqnarray}
\left\{
\begin{aligned}
i \partial_t \Psi^\G_{-}(t) &= \left( \hat{T} + \hat{V}_\ee + \hat{{\cal V}}(-t) - i \hat{W}(-t)\right) \Psi^\G_{-}(t), \\
\Psi^\G_{-}(0) &= \Psi_0^*,
\end{aligned}
\right.
\;\;
\label{eq:TDSE-TDDFTgeo-TR}
\end{eqnarray}
with $\hat{W}(-t) = \int_{\mathbb{R}^d} \d\b{r} \, w[\rho,\Psi_0,\nu](\b{r},-t) \, \hat{\rho}(\b{r})$. Thus, the time-reversed wave function $\Psi^\G_{-}[\rho,\Psi_0,\nu](t)$ satisfies the effective TDSE in Eq.~\eqref{eq:TDSE-TDDFTgeo-TR} with the time-reversed external potential $\nu_{-}(\b{r},t):= \nu(\b{r},-t)$, the time-reversed geometric potential $-w[\rho,\Psi_0,\nu](\b{r},-t)$ and initial wave function $\Psi_0^*$, and yields the time-reversed density $\rho_{-}$. According to the geometric Runge-Gross theorem, this implies that
\begin{eqnarray}
\Psi^\G[\rho_{-},\Psi_0^*,\nu_{-}] = \Psi^\G_{-}[\rho,\Psi_0,\nu],
\end{eqnarray}
and
\begin{eqnarray}
w[\rho_{-},\Psi_0^*,\nu_{-}](\b{r},t) =  - w[\rho,\Psi_0,\nu](\b{r},-t).
\end{eqnarray}
Proceeding in exactly the same way for the non-interacting TDSE in Eq.~\eqref{eq:TDDFT-TDGKS}, we obtain the same time-reversal transformation for the KS geometric potential,
\begin{eqnarray}
w_\s[\rho_{-},\Phi_0^*,\nu_{-}](\b{r},t) =  -w_\s[\rho,\Phi_0,\nu](\b{r},-t).
\label{eq:ws-TR}
\end{eqnarray}
Using then Eqs.~\eqref{eq:wHxc},~\eqref{eq:v-TR}, and~\eqref{eq:ws-TR}, we obtain the time-reversal transformation for the Hxc geometric potential,
\begin{eqnarray}
w_\Hxc[\rho_{-},\Phi_0^*,\Psi_0^*](\b{r},t) = -w_\Hxc[\rho,\Phi_0,\Psi_0](\b{r},-t),
\label{eq:wHxcTR}
\end{eqnarray}
i.e. $w_\Hxc$ is time-odd under time reversal. The condition in Eq.~\eqref{eq:wHxcTR} also applies to the geometric potential correction $w^\text{cor}[\rho,\Phi_0,\Psi_0]$ in Eq.~\eqref{eq:wcor}, provided that the corresponding Hxc potential approximation $v_\Hxc^\text{app}[\rho,\Phi_0,\Psi_0]$ fulfils the condition in Eq.~\eqref{eq:vHxcTR}.

\subsection{Adiabatic connection and coordinate scaling}

\subsubsection{Variational TDDFT}

We start by reviewing the adiabatic connection in standard variational TDDFT~\cite{Gor-PRA-97,Gor-INC-06} and the derivation of the coordinate scaling relations (see Refs.~\onlinecite{HesParBur-PRL-99} and~\onlinecite{Ullrichs-11}). 

In the adiabatic connection for TDDFT, the interaction strength is varied while the prescribed density trajectory is held fixed by adjusting the external potential.
For a given density trajectory $\rho$ and an initial wave function $\tilde{\Psi}_0$ (with compatible density $\rho_{\tilde{\Psi}_0}(\b{r}) = \rho(\b{r},0)$ and current-density divergence $\nabla \cdot \b{j}_{\tilde{\Psi}_0}(\b{r}) = - \partial_t \rho(\b{r},t)|_{t=0}$), we consider the TDSE with a coupling constant $\l \in \mathbb{R}$,
\begin{eqnarray}
\left\{ 
\begin{aligned}
i\, \partial_t \Psi^{\V,\l}(t) &= \Bigl( \hat{T} + \l \hat{V}_\text{ee} + \hat{V}^\l(t) \Bigl) \Psi^{\V,\l}(t),\\[0.3cm]
\Psi^{\V,\l}(0) &= \tilde{\Psi}_0,
\end{aligned}
\right.
\label{eq:TDSEAC}
\end{eqnarray}
where $\hat{V}^\l(t) := \int_{\mathbb{R}^d} \d \b{r} \, v^\l[\rho,\tilde{\Psi}_0](\b{r},t) \hat{\rho}(\b{r})$ is the external potential operator such that $\Psi^{\V,\l}(t)$ gives the density $\rho$ for all $\l$, i.e. $\rho_{\Psi^{\V,\l}(t)}(\b{r}) = \rho(\b{r},t)$. The wave function $\Psi^{\V,\l}(t)$ is a functional of $\rho$ and $\tilde{\Psi}_0$, i.e. $\Psi^{\V,\l}(t) = \Psi^{\V,\l}[\rho,\tilde{\Psi}_0](t)$. At $\l=1$, if we choose the same initial wave function as in Eq.~\eqref{eq:TDSE-TDDFT}, i.e. $\tilde{\Psi}_0=\Psi_0$, we obtain the same potential $v^{\l=1}[\rho,\Psi_0](\b{r},t) \eqv v[\rho,\Psi_0](\b{r},t)$ and wave function $\Psi^{\V,\l=1}[\rho,\Psi_0](t) \eqpsi \Psi^\V[\rho,\Psi_0](t)$ as in Eq.~\eqref{eq:TDSE-TDDFT}. At $\l=0$, if we choose the same initial wave function as in Eq.~\eqref{eq:TDDFT-TDKS}, i.e. $\tilde{\Psi}_0= \Phi_0$, we obtain the same potential $v^{\l=0}[\rho,\Phi_0](\b{r},t) \eqv v_\s[\rho,\Phi_0](\b{r},t)$ and wave function $\Psi^{\V,\l=0}[\rho,\Phi_0](t) \eqpsi \Phi^\V[\rho,\Phi_0](t)$ as in Eq.~\eqref{eq:TDDFT-TDKS}.

Let us derive the coordinate scaling relation for the potential $v^\l[\rho,\tilde{\Psi}_0](\b{r}, t)$. For any time-dependent wave function $\Psi$, we define the norm-preserving coordinate-scaled wave function with the scaling factor $\gamma>0$ as
\begin{eqnarray}
\Psi_{\gamma}(\b{r}_1,...,\b{r}_N,t) := \gamma^{dN/2} \Psi(\gamma\b{r}_1,...,\gamma\b{r}_N,\gamma^2 t),
\label{}
\end{eqnarray}
where we have omitted the spin variables that are left untouched by the scaling. As we will see below, the time variable is scaled by $\gamma^2$ in order to cancel out the fact the kinetic-energy operator scales by $\gamma^{-2}$ when the spatial coordinates are scaled by $\gamma$. If the wave function $\Psi$ yields the density $\rho(\b{r},t)$, then the scaled wave function $\Psi_{\gamma}$ yields the scaled density
\begin{eqnarray}
\rho_{\gamma}(\b{r},t) := \gamma^d \rho(\gamma\b{r},\gamma^2 t),
\label{}
\end{eqnarray}
which preserves the electron number, i.e. $\int_{\mathbb{R}^d} \rho_{\gamma}(\b{r},t) \d \b{r} = \int_{\mathbb{R}^d} \rho(\b{r},\gamma^2 t) \d \b{r} = N$. We see that $\gamma \to \infty$ corresponds to a high and rapidly-varying density limit, and $\gamma \to 0$ corresponds to a low and slowly-varying density limit. 

If we scale the space and time coordinates in Eq.~\eqref{eq:TDSEAC}, $\b{r}_i \to \gamma \b{r}_i$ and $t \to \gamma^2 t$, and assuming that the two-electron interaction scales as
\begin{eqnarray}
v_\text{ee}(\gamma \br) =\gamma^{-1} v_\text{ee}(\br) 
\end{eqnarray}
i.e. like the Coulomb interaction $v_\text{ee}(\br)=1/|\br|$, we obtain
\begin{eqnarray}
\left\{ 
\begin{aligned}
i \partial_t \Psi^{\V,\l}_{\gamma}(t) &= \Bigl( \hat{T} + \l \gamma \hat{V}_\text{ee} + \hat{V}^\l_\gamma(t) \Bigl) \Psi^{\V,\l}_{\gamma}(t), \\[0.3cm]
\Psi^{\V,\l}_{\gamma}(0) &= \tilde{\Psi}_{0,\gamma},
\end{aligned}
\right.
\label{eq:TDSEACscaled}
\end{eqnarray}
where we have introduced the scaled potential operator $\hat{V}_\gamma^\l(t) := \int_{\mathbb{R}^d} \d \b{r} \, v_{\gamma}^\l[\rho,\tilde{\Psi}_0](\b{r},t) \hat{\rho}(\b{r})$ with $ v_{\gamma}^\l[\rho,\tilde{\Psi}_0](\b{r},t) := \gamma^2  v^\l[\rho,\tilde{\Psi}_0](\gamma\b{r},\gamma^2 t)$, and the scaled initial wave function $\tilde{\Psi}_{0,\gamma}(\b{r}_1,...,\b{r}_N):=\gamma^{dN/2} \tilde{\Psi}_{0}(\gamma \b{r}_1, ...,\gamma \b{r}_N)$. Thus, the scaled time-dependent wave function $\Psi^{\V,\l}_{\gamma}[\rho,\tilde{\Psi}_0](t)$ satisfies the TDSE with coupling constant $\l \gamma$ in Eq.~\eqref{eq:TDSEACscaled} starting from the scaled initial wave function $\tilde{\Psi}_{0,\gamma}$, and yields the scaled density $\rho_{\gamma}(\b{r},t)$. According to the Runge-Gross theorem, this implies that $\Psi^{\V,\l}_{\gamma}[\rho,\tilde{\Psi}_0]$ is identical to the wave function for the coupling constant $\l \gamma$ evaluated at the scaled quantities $\rho_{\gamma}$ and $\tilde{\Psi}_{0,\gamma}$,~\cite{HesParBur-PRL-99}
\begin{eqnarray}
\Psi^{\V,\l}_{\gamma}[\rho,\tilde{\Psi}_0] \eqpsi \Psi^{\V,\l \gamma}[\rho_{\gamma},\tilde{\Psi}_{0,\gamma}],
\label{}
\end{eqnarray}
and that the external potential appearing in Eq.~\eqref{eq:TDSEACscaled} is identical to the external potential for the coupling constant $\l \gamma$ evaluated at the scaled quantities $\rho_{\gamma}$ and $\tilde{\Psi}_{0,\gamma}$
\begin{eqnarray}
\gamma^2 v^\l[\rho,\tilde{\Psi}_0](\gamma\b{r}, \gamma^2 t) \eqv v^{\l \gamma}[\rho_{\gamma},\tilde{\Psi}_{0,\gamma}](\b{r},t),
\label{eq:vlscaling0}
\end{eqnarray}
or, equivalently,
\begin{eqnarray}
v^{\l}[\rho_{\gamma},\tilde{\Psi}_{0,\gamma}](\b{r},t) \eqv \gamma^2 v^{\l/\gamma}[\rho,\tilde{\Psi}_0](\gamma\b{r}, \gamma^2 t).
\label{eq:vlscaling}
\end{eqnarray}
In particular, for $\l=0$ and $\tilde{\Psi}_0 = \Phi_0$, we obtain the scaling relation of the single-determinant wave function in Eq.~\eqref{eq:TDDFT-TDKS}
\begin{eqnarray}
\Phi^\V_{\gamma}[\rho,\Phi_0] \eqpsi \Phi^\V[\rho_{\gamma},\Phi_{0,\gamma}],
\label{}
\end{eqnarray}
and for the KS potential
\begin{eqnarray}
v_\text{s}[\rho_\gamma,\Phi_{0,\gamma}](\b{r}, t) \eqv \gamma^2 v_\text{s}[\rho,\Phi_{0}](\gamma\b{r},\gamma^2 t).
\label{eq:vsscaling}
\end{eqnarray}

Let us now define the $\l$-dependent Hxc potential associated with the interaction $\l v_\text{ee}$, similarly as in Eq.~\eqref{eq:vHxc},
\begin{eqnarray}
v_\Hxc^{\l}[\rho,\Phi_0,\tilde{\Psi}_0] := v_\s[\rho,\Phi_0] - v^\l[\rho,\tilde{\Psi}_0],
\label{eq:vHxcl}
\end{eqnarray}
and we have $v_\Hxc^{\l=0}[\rho,\Phi_0,\Phi_0]=0$ and $v_\Hxc^{\l=1}[\rho,\Phi_0,\Psi_0]=v_\Hxc[\rho,\Phi_0,\Psi_0]$. 
From Eqs.~\eqref{eq:vlscaling},~\eqref{eq:vsscaling} and~\eqref{eq:vHxcl}, we find the scaling relation for the Hxc potential at $\l=1$ and $\tilde{\Psi}_0 = \Psi_0$,~\cite{HesParBur-PRL-99}
\begin{eqnarray}
v_\text{Hxc}[\rho_\gamma,\Phi_{0,\gamma},\Psi_{0,\gamma}](\b{r}, t) \eqv \gamma^2 v_\text{Hxc}^{1/\gamma}[\rho,\Phi_{0},\Psi_{0}](\gamma\b{r},\gamma^2 t). \;\;
\nonumber\\
\label{eq:vHxcscaling}
\end{eqnarray}

\subsubsection{Geometric TDDFT}

We now formulate the adiabatic connection for geometric TDDFT and derive the corresponding coordinate scaling relations. 

For a given density trajectory $\rho$, initial wave function $\tilde{\Psi}_0$ (with compatible density $\rho_{\tilde{\Psi}_0}(\b{r}) = \rho(\b{r},0)$), and external potential $\nu$, we consider the geometric TDSE with coupling constant $\l$,
\begin{eqnarray}
\left\{ 
\begin{aligned}
\! i \partial_t \Psi^{\G,\l}(t) &= \Bigl( \hat{T} + \l \hat{V}_\text{ee} + \hat{{\cal V}}(t) + i \hat{W}^\l(t) \Bigl) \Psi^{\G,\l}(t),\\[0.3cm]
\! \Psi^{\G,\l}(0) &= \tilde{\Psi}_0,
\end{aligned}
\right.
\label{eq:TDDFTgeoAC}
\end{eqnarray}
with $\hat{{\cal V}}(t) := \int_{\mathbb{R}^d} \d \b{r} \, \nu(\b{r},t) \hat{\rho}(\b{r})$ and $\hat{W}^\l(t) :=  \int_{\mathbb{R}^d} \d \b{r} \, w^\l[\rho,\tilde{\Psi}_0,\nu](\b{r},t) \hat{\rho}(\b{r})$ where $w^\l$ is such that $\Psi^{\G,\l}(t)$ gives the density $\rho$ for all $\l$, i.e. $\rho_{\Psi^{\G,\l}(t)}(\b{r}) = \rho(\b{r},t)$. The wave function $\Psi^{\G,\l}(t)$ is a functional of $\rho$, $\tilde{\Psi}_0$ and $\nu$, i.e. $\Psi^{\G,\l}(t) = \Psi^{\G,\l}[\rho,\tilde{\Psi}_0,\nu](t)$. 

If we scale the space and time coordinates in Eq.~\eqref{eq:TDDFTgeoAC}, $\b{r}_i \to \gamma \b{r}_i$ and $t \to \gamma^2 t$, assuming again that $v_\text{ee}(\gamma \br) =\gamma^{-1} v_\text{ee}(\br)$, we obtain
\begin{eqnarray}
\!\!\left\{ 
\begin{aligned}
\! i \partial_t \Psi^{\G,\l}_{\gamma}(t) &= \! \! \Bigl( \hat{T} + \l \gamma \hat{V}_\text{ee} + \hat{{\cal V}}_\gamma(t) + i\hat{W}^\l_\gamma(t) \Bigl) 
\Psi^{\G,\l}_{\gamma}(t), \\[0.3cm]
\! \Psi^{\G,\l}_{\gamma}(0) &= \tilde{\Psi}_{0,\gamma},
\end{aligned}
\right.
\label{eq:TDDFTgeoACscaled}
\end{eqnarray}
with $\hat{{\cal V}}_\gamma(t) := \int_{\mathbb{R}^d} \d \b{r} \, \nu_{\gamma}(\b{r},t) \hat{\rho}(\b{r})$ and $\hat{W}^\l_\gamma(t) :=  \int_{\mathbb{R}^d} \d \b{r} \, w^\l_\gamma[\rho,\tilde{\Psi}_0,\nu](\b{r},t) \hat{\rho}(\b{r})$ where $\nu_{\gamma}(\b{r},t) := \gamma^2  \nu(\gamma\b{r},\gamma^2 t)$ and $w^\l_{\gamma}[\rho,\tilde{\Psi}_0,\nu](\b{r},t) := \gamma^2  w^\l[\rho,\tilde{\Psi}_0,\nu](\gamma\b{r},\gamma^2 t)$. Thus, the scaled time-dependent wave function $\Psi^{\G,\l}_{\gamma}[\rho,\tilde{\Psi}_0,\nu](t)$ satisfies the TDSE with coupling constant $\l \gamma$ in Eq.~\eqref{eq:TDDFTgeoACscaled} with external potential $\nu_{\gamma}$ starting from the scaled initial wave function $\tilde{\Psi}_{0,\gamma}$, and yields the scaled density $\rho_{\gamma}(\b{r},t)$. According to the geometric Runge-Gross theorem, this implies that $\Psi^{\G,\l}_{\gamma}[\rho,\tilde{\Psi}_0,\nu]$ is identical to the wave function for the coupling constant $\l \gamma$ evaluated at the scaled quantities $\rho_{\gamma}$, $\tilde{\Psi}_{0,\gamma}$ and $\nu_\gamma$,
\begin{eqnarray}
\Psi^{\G,\l}_{\gamma}[\rho,\tilde{\Psi}_0,\nu] = \Psi^{\G,\l \gamma}[\rho_{\gamma}, \tilde{\Psi}_{0,\gamma},\nu_{\gamma}],
\end{eqnarray}
and
\begin{eqnarray}
\gamma^2 w^\l[\rho,\tilde{\Psi}_0,\nu](\gamma \b{r},\gamma^2 t) = w^{\l \gamma}[\rho_{\gamma}, \tilde{\Psi}_{0,\gamma},\nu_{\gamma}](\b{r},t),
\end{eqnarray}
or, equivalently,
\begin{eqnarray}
w^{\l}[\rho_{\gamma}, \tilde{\Psi}_{0,\gamma},\nu_{\gamma}](\b{r},t)
= \gamma^2 w^{\l/\gamma}[\rho,\tilde{\Psi}_0,\nu](\gamma \b{r},\gamma^2 t).
\end{eqnarray}
In particular, for $\l=0$ and $\tilde{\Psi}_0 = \Phi_0$, we obtain the scaling relation for the single-determinant wave function in Eq.~\eqref{eq:TDDFT-TDGKS}
\begin{eqnarray}
\Phi_{\gamma}^\G[\rho,\Phi_0,\nu] = \Phi^\G[\rho_{\gamma}, \Phi_{0,\gamma},\nu_{\gamma}],
\end{eqnarray}
and for the geometric KS potential
\begin{eqnarray}
w_\text{s}[\rho_{\gamma}, \Phi_{0,\gamma},\nu_{\gamma}](\b{r},t) = \gamma^2 w_\text{s}[\rho,\Phi_0,\nu](\gamma\b{r},\gamma^2t). 
\label{eq:wsrhogamma}
\end{eqnarray}

Let us now define the  $\l$-dependent Hxc geometric potential, similarly as in Eq.~\eqref{eq:wHxc}, as
\begin{eqnarray}
w^{\l}_\Hxc[\rho,\Phi_0,\tilde{\Psi}_0] := w_\text{s}[\rho,\Phi_0,v^\l[\rho,\tilde{\Psi}_0]],
\label{eq:wlHxc}
\end{eqnarray}
and we have  $w^{\l=0}_\Hxc[\rho,\Phi_0,\Phi_0]=0$ (since $v^{\l=0}[\rho,\Phi_0]=v_\s[\rho,\Phi_0]$ is the exact KS potential which does not need any geometric correction) and $w^{\l=1}_\Hxc[\rho,\Phi_0,\Psi_0]=w_\Hxc[\rho,\Phi_0,\Psi_0]$. The use of $v^\l[\rho,\tilde{\Psi}_0]$ requires the additional compatibility condition $\nabla \cdot \b{j}_{\tilde{\Psi}_0}(\b{r}) = - \partial_t \rho(\b{r},t)|_{t=0}$.
From Eqs.~\eqref{eq:vlscaling},~\eqref{eq:wsrhogamma} and~\eqref{eq:wlHxc}, we find the scaling relation for the Hxc geometric potential at $\l=1$ and $\tilde{\Psi}_0 =\Psi_0$,
\begin{eqnarray}
w_\Hxc[\rho_\gamma,\Phi_{0,\gamma},\Psi_{0,\gamma}](\b{r},t) &=& 
\gamma^2 w^{1/\gamma}_\Hxc[\rho,\Phi_0,\Psi_0](\gamma\b{r},\gamma^2t),
\nonumber\\
\end{eqnarray}
i.e. it has the same scaling relation as $v_\Hxc[\rho,\Phi_0,\Psi_0]$ in Eq.~\eqref{eq:vHxcscaling}. 

We may also extend the geometric correction functional in Eq.~\eqref{eq:wcor} to a $\l$-dependent one as
\begin{eqnarray}
w^{\cor,\l}[\rho,\Phi_0,\tilde{\Psi}_0] := \phantom{xxxxxxxxxxxxxxxxxxxx}
\nonumber\\
w_\text{s}[\rho,\Phi_0,v^\l[\rho,\tilde{\Psi}_0]+v^{\text{app},\l}_\text{Hxc}[\rho,\Phi_0,\tilde{\Psi}_0]],
\label{}
\end{eqnarray}
where $v^{\text{app},\l}_\text{Hxc}[\rho,\Phi_0,\tilde{\Psi}_0](\b{r},t):= \l^2 v^{\text{app}}_\text{Hxc}[\rho_{1/\l},\Phi_{0,1/\l},\tilde{\Psi}_{0,1/\l}](\l\b{r},\l^2 t)$ for $\l>0$. We obtain again the scaling relation at $\l=1$ and $\tilde{\Psi}_0 =\Psi_0$,
\begin{eqnarray}
w^{\cor}[\rho_\gamma,\Phi_{0,\gamma},\Psi_{0,\gamma}](\b{r},t) &=& \gamma^2 w^{\cor,1/\gamma}[\rho,\Phi_0,\Psi_0](\gamma\b{r},\gamma^2t).
\nonumber\\
\end{eqnarray}

\subsection{Generalized translational invariance}

An important exact condition in variational TDDFT is the so-called generalized translational invariance~\cite{Vig-PRL-95} (see also Ref.~\onlinecite{Ullrichs-11}), which can be shown as follows.

We consider a transformation to an arbitrary accelerated reference frame, $\b{r} \to \b{r} + \b{x}(t)$, where $\b{x}(t) \in \mathbb{R}^d$ is an arbitrary sufficiently regular time-dependent vector.
For a given time-dependent wave function $\Psi(t)$, the corresponding accelerated-frame wave function is 
\begin{eqnarray}
\Psi_{\b{x}}(\b{r}_1,...,\b{r}_N,t) := \phantom{xxxxxxxxxxxxxxx}
\nonumber\\
e^{-i \dot{\b{x}}(t)\cdot \hat{\b{R}}} \; \Psi(\b{r}_1 + \b{x}(t),...,\b{r}_N+\b{x}(t),t)
\end{eqnarray}
where $\dot{\b{x}}$ designates the time derivative of $\b{x}$ and $\hat{\b{R}} := \int_{\mathbb{R}^d} \b{r} \; \hat{\rho}(\b{r}) \d \b{r}$ is the total position operator. If $\Psi(t)$ yields the time-dependent density $\rho(\b{r},t)$, then $\Psi_\b{x}(t)$ yields the accelerated-frame density
\begin{eqnarray}
\rho_\b{x}(\b{r},t) := \rho(\b{r}+\b{x}(t),t).
\end{eqnarray}
For a given density trajectory $\rho$ and initial wave function $\Psi_0$, the wave function $\Psi^\V(t) = \Psi^\V[\rho,\Psi_0](t)$ of variational TDDFT satisfies the effective TDSE in Eq.~\eqref{eq:TDSE-TDDFT}. The corresponding accelerated-frame wave function $\Psi^\V_\b{x}(t)$ then satisfies the following effective TDSE
\begin{eqnarray}
\left\{
\begin{aligned}
i \partial_{t} \Psi^\V_\b{x}(t)  &= \left( \hat{T} + \hat{V}_\text{ee} + \hat{V}_\b{x}(t)  \right) \Psi^\V_\b{x}(t), \\
\Psi^\V_\b{x}(0) &= \Psi_{0,\b{x}},
\end{aligned}
\right.
\label{eq:TDSE-TDDFT-A}
\end{eqnarray}
with the accelerated-frame initial wave function $\Psi_{0,\b{x}}:=e^{-i \dot{\b{x}}(0)\cdot \hat{\b{R}}} \; \Psi_0(\b{r}_1 + \b{x}(0),...,\b{r}_N+\b{x}(0))$
and $\hat{V}_\b{x}(t) := \int_{\mathbb{R}^d} \d\b{r} \, v_\b{x}[\rho,\Psi_0](\b{r},t) \, \hat{\rho}(\b{r})$ where $v_\b{x}[\rho,\Psi_0](\b{r},t)$ is the accelerated-frame potential,
\begin{equation}
v_\b{x}[\rho,\Psi_0](\b{r},t) = v[\rho,\Psi_0](\b{r}+\b{x}(t),t) + \ddot{\b{x}}(t) \cdot \b{r} - \frac{1}{2} \dot{\b{x}}(t)^2,
\label{eq:varhoPsi0}
\end{equation}
where $\ddot{\b{x}}(t) \cdot \b{r}$ is the reference-frame inertial potential and $-\dot{\b{x}}(t)^2/2$ is an irrelevant time-dependent constant. Thus, the wave function $\Psi^\V_\b{x}[\rho,\Psi_0](t)$ satisfies the effective TDSE in Eq.~\eqref{eq:TDSE-TDDFT-A} with potential $v_\b{x}[\rho,\Psi_0](\b{r},t)$ and initial wave function $\Psi_{0,\b{x}}$, and yields the accelerated-frame density $\rho_\b{x}$. According to the Runge-Gross theorem, this implies that
\begin{eqnarray}
\Psi^\V[\rho_\b{x},\Psi_{0,\b{x}}](t) \eqpsi \Psi^\V_\b{x}[\rho,\Psi_0](t),
\end{eqnarray}
and
\begin{eqnarray}
v[\rho_\b{x},\Psi_{0,\b{x}}](\b{r},t) \eqv v_\b{x}[\rho,\Psi_0](\b{r},t).
\label{eq:varhoPsi0cond}
\end{eqnarray}
Proceeding in exactly the same way for the non-interacting TDSE in Eq.~\eqref{eq:TDDFT-TDKS}, we find the same condition for the KS potential,
\begin{eqnarray}
v_{\s}[\rho_\b{x},\Phi_{0,\b{x}}](\b{r},t) \eqv v_{\s,\b{x}}[\rho,\Phi_0](\b{r},t),
\end{eqnarray}
where
\begin{equation}
v_{\s,\b{x}}[\rho,\Phi_0](\b{r},t) = v_\s[\rho,\Phi_0](\b{r}+\b{x}(t),t) + \ddot{\b{x}}(t) \cdot \b{r} - \frac{1}{2} \dot{\b{x}}(t)^2,
\label{eq:vsarhoPhi0}
\end{equation}
and $\Phi_{0,\b{x}}:=e^{-i \dot{\b{x}}(0)\cdot \hat{\b{R}}} \; \Phi_0(\b{r}_1 + \b{x}(0),...,\b{r}_N+\b{x}(0))$ is the accelerated-frame initial Slater-determinant wave function. Using Eq.~\eqref{eq:vHxc}, we find the generalized translational invariance of the Hxc potential,~\cite{Vig-PRL-95}
\begin{eqnarray}
v_{\Hxc}[\rho_\b{x},\Phi_{0,\b{x}},\Psi_{0,\b{x}}](\b{r},t)  \eqv v_{\Hxc}[\rho,\Phi_0,\Psi_0](\b{r}+\b{x}(t),t), \;
\nonumber\\
\label{eq:vHxca}
\end{eqnarray}
where we have used the fact that the last two terms in Eqs.~\eqref{eq:varhoPsi0} and~\eqref{eq:vsarhoPhi0} cancel out in the difference between $v_\b{x}[\rho,\Psi_0](\b{r},t)$ and $v_{\s,\b{x}}[\rho,\Phi_0](\b{r},t)$.

For geometric TDDFT, the situation is similar. For a given density trajectory $\rho$, initial wave function $\Psi_0$, and external potential $\nu$, the wave function $\Psi^\G(t) = \Psi^\G[\rho,\Psi_0,\nu](t)$ satisfies the effective TDSE in Eq.~\eqref{eq:TDSE-TDDFTgeo}. The corresponding accelerated-frame wave function $\Psi^\G_\b{x}(t)$ then satisfies the following effective TDSE
\begin{eqnarray}
\left\{
\begin{aligned}
i \partial_{t} \Psi^\G_\b{x}(t)  &= \left( \hat{T} + \hat{V}_\text{ee} + \hat{{\cal V}}_{\b{x}}(t) + i \hat{W}_\b{x}(t) \right) \Psi^\G_\b{x}(t), \\
\Psi^\G_\b{x}(0) &= \Psi_{0,\b{x}},
\end{aligned}
\right.
\;\;\;\:
\label{eq:TDSE-TDDFTgeo-A}
\end{eqnarray}
with $\hat{{\cal V}}_{\b{x}}(t) := \int_{\mathbb{R}^d} \d\b{r} \, \nu_{\b{x}}(\b{r},t) \, \hat{\rho}(\b{r})$ where 
\begin{eqnarray}
\nu_{\b{x}}(\b{r},t) := \nu(\b{r}+\b{x}(t),t) + \ddot{\b{x}}(t) \cdot \b{r} - \frac{1}{2} \dot{\b{x}}(t)^2,
\end{eqnarray}
and $\hat{W}_\b{x}(t) := \int_{\mathbb{R}^d} \d\b{r} \, w_\b{x}[\rho,\Psi_0,\nu](\b{r},t) \, \hat{\rho}(\b{r})$ where
\begin{eqnarray}
w_\b{x}[\rho,\Psi_0,\nu](\b{r},t) := w[\rho,\Psi_0,\nu](\b{r}+\b{x}(t),t).
\end{eqnarray}
Thus, the wave function $\Psi^\G_\b{x}[\rho,\Psi_0,\nu](t)$ satisfies the effective TDSE in Eq.~\eqref{eq:TDSE-TDDFTgeo-A} with external potential $\nu_{\b{x}}(\b{r},t)$, geometric potential $w_\b{x}[\rho,\Psi_0,\nu](\b{r},t)$ and initial wave function $\Psi_{0,\b{x}}$, and yields the accelerated-frame density $\rho_\b{x}$. According to the geometric Runge-Gross theorem, this implies that
\begin{eqnarray}
\Psi^\G[\rho_\b{x},\Psi_{0,\b{x}},\nu_{\b{x}}](t) = \Psi^\G_\b{x}[\rho,\Psi_0,\nu](t),
\end{eqnarray}
and
\begin{eqnarray}
w[\rho_\b{x},\Psi_{0,\b{x}},\nu_{\b{x}}](\b{r},t) = w[\rho,\Psi_0,\nu](\b{r}+\b{x}(t),t).
\end{eqnarray}
Proceeding in exactly the same way for the non-interacting TDSE in Eq.~\eqref{eq:TDDFT-TDGKS}, we find the same condition for the KS geometric potential,
\begin{eqnarray}
w_{\s}[\rho_\b{x},\Phi_{0,\b{x}},\nu_{\b{x}}](\b{r},t) = w_{\s}[\rho,\Phi_0,\nu](\b{r}+\b{x}(t),t),\;
\end{eqnarray}
and, using Eqs.~\eqref{eq:wHxc} and~\eqref{eq:varhoPsi0cond}, we find the generalized translational invariance of the Hxc geometric potential,
\begin{equation}  
w_{\Hxc}[\rho_\b{x},\Phi_{0,\b{x}},\Psi_{0,\b{x}}](\b{r},t)  = w_{\Hxc}[\rho,\Phi_0,\Psi_0](\b{r}+\b{x}(t),t). 
\label{eq:wHxca}
\end{equation}
The condition in Eq.~\eqref{eq:wHxca} also applies to the geometric potential correction $w^\text{cor}[\rho,\Phi_0,\Psi_0]$ in Eq.~\eqref{eq:wcor}, provided that the corresponding Hxc potential approximation $v_\Hxc^\text{app}[\rho,\Phi_0,\Psi_0]$ fulfils the condition in Eq.~\eqref{eq:vHxca}.

\subsection{Time-dependent total energy}

\subsubsection{Variational TDDFT}

Even though the total energy is generally not conserved along the dynamics, it is nevertheless interesting to consider it to find an exact condition linking the energy and the potential~\cite{HesParBur-PRL-99,HesMaiBur-JCP-02}. The time-dependent total energy associated with the variational TDDFT dynamics in Eq.~\eqref{eq:TDSE-TDDFT} is
\begin{eqnarray}
E^\V(t) &:=&  \bra{\Psi^\V(t)}\hat{H}^\V(t)\ket{\Psi^\V(t)},
\end{eqnarray}
with the Hamiltonian $\hat{H}^\V(t):= \hat{T} + \hat{V}_\text{ee} + \hat{V}(t)$ where $\hat{V}(t) := \int_{\mathbb{R}^d} \d \b{r} \, v[\rho,\Psi_0](\b{r},t) \hat{\rho}(\b{r})$. Since $\Psi^\V(t)$ is a functional of the density trajectory $\rho$ and of the initial wave function $\Psi_0$, i.e. $\Psi^\V(t)=\Psi^\V[\rho,\Psi_0](t)$, the time-dependent total energy is also a functional of $\rho$ and $\Psi_0$, i.e. $E^\V(t) = E^\V[\rho,\Psi_0](t)$. 
Note that since $v[\rho,\Psi_0](\b{r},t) $ is defined up to an arbitrary additive time-dependent constant, so is the total energy $E^\V(t)$. We can write the total energy as the sum of its different contributions,
\begin{eqnarray}
E^\V(t) =  T^\V(t) + V_\text{ee}^\V(t) + V(t),
\label{eq:EVtcontrib}
\end{eqnarray}
where $T^\V(t) := \bra{\Psi^\V(t)}\hat{T}\ket{\Psi^\V(t)}$, $V_\text{ee}^\V(t) :=\bra{\Psi^\V(t)}\hat{V}_\text{ee}\ket{\Psi^\V(t)}$, and $V(t) := \bra{\Psi^\V(t)}\hat{V}(t)\ket{\Psi^\V(t)}=\int_{\mathbb{R}^d} \d \b{r} \, v(\b{r},t) \rho(\b{r},t)$. 
Using the Ehrenfest equation of motion for an operator $\hat{A}(t)$ and Hamiltonian $\hat{H}^\V(t)$,
\begin{eqnarray}
\frac{\d}{\d t} \bra{\Psi^\V(t)}\hat{A}(t)\ket{\Psi^\V(t)} = \bra{\Psi^\V(t)} \frac{\d}{\d t} \hat{A}(t) \ket{\Psi^\V(t)} 
\nonumber\\
+ i \bra{\Psi^\V(t)} [\hat{H}^\V(t),\hat{A}(t)]\ket{\Psi^\V(t)},
\end{eqnarray}
in the special case where $\hat{A}(t)$ is the Hamiltonian $\hat{H}^\V(t)$, we obtain
\begin{eqnarray}
\frac{\d}{\d t} E^\V(t) = \int_{\mathbb{R}^d} (\partial_t v(\b{r},t)) \rho(\b{r},t) \d \b{r}.
\end{eqnarray}
Combining this equation with the time-derivative of Eq.~\eqref{eq:EVtcontrib},
\begin{equation}  
\frac{\d}{\d t} E^\V(t) = \frac{\d}{\d t} T^\V(t) + \frac{\d}{\d t} V_\text{ee}^\V(t) +  \int_{\mathbb{R}^d} \partial_t [ v(\b{r},t) \rho(\b{r},t)] \d \b{r},
\end{equation}
we arrive at
\begin{eqnarray}
\frac{\d}{\d t} T^\V(t) + \frac{\d}{\d t} V_\text{ee}^\V(t) =  -\int_{\mathbb{R}^d} v(\b{r},t) \partial_t  \rho(\b{r},t) \d \b{r}.
\label{dTdtdVeedt}
\end{eqnarray}
Note that Eq.~\eqref{dTdtdVeedt} is insensitive to an additive time-dependent constant in $v(\b{r},t)$ since $\int_{\mathbb{R}^d} \partial_t  \rho(\b{r},t) \d \b{r}=0$.

Similarly, for the non-interacting variational TDDFT dynamics in Eq.~\eqref{eq:TDDFT-TDKS}, the time-dependent total energy is
\begin{eqnarray}
E_\s^\V(t) &:=&  \bra{\Phi^\V(t)}\hat{H}_\s^\V(t)\ket{\Phi^\V(t)},
\end{eqnarray}
with the Hamiltonian $\hat{H}_\s^\V(t):= \hat{T} + \hat{V}_\s(t)$ where $\hat{V}_\s(t) := \int_{\mathbb{R}^d} \d \b{r} \, v_\s[\rho,\Phi_0](\b{r},t) \hat{\rho}(\b{r})$. Since the single-determinant wave function $\Phi^\V(t)=\Phi^\V[\rho,\Phi_0](t)$ is a functional of the density $\rho$ and of the initial single-determinant wave function $\Phi_0$, $E_\s^\V(t)=E_\s^\V[\rho,\Phi_0](t)$ is also a functional of $\rho$ and $\Phi_0$ (defined up to an arbitrary additive time-dependent constant). Repeating the same steps as above, we find
\begin{eqnarray}
\frac{\d}{\d t} T_\text{s}^\V(t)  =  -\int_{\mathbb{R}^d} v_\text{s}(\b{r},t) \partial_t \rho(\b{r},t) \d \b{r},
\label{dTsdt}
\end{eqnarray}
where $T_\text{s}^\V(t) :=  \bra{\Phi^\V(t)} \hat{T} \ket{\Phi^\V(t)}$ is the time-dependent variational non-interacting kinetic energy.

As in ground-state KS DFT, the total energy $E^\V(t)$ can be decomposed as
\begin{eqnarray}
E^\V(t) = T_\text{s}^\V(t) + V(t) + E_\text{Hxc}^\V(t),
\label{eq:EVtdecomp}
\end{eqnarray}
where $E_\text{Hxc}^\V(t) := T^\V(t) + V_\text{ee}^\V(t) -  T_\text{s}^\V(t)$ is the time-dependent variational Hxc energy, which is a functional of $\rho$, $\Phi_0$, and $\Psi_0$, i.e. $E_\text{Hxc}^\V(t)=E_\text{Hxc}^\V [\rho,\Phi_0,\Psi_0](t)$. Note that the arbitrary additive time-dependent constant in $E^\V(t)$ and $V(t)$ cancels out so that $E_\text{Hxc}^\V(t)$ does not depend on this constant. Using Eqs.~\eqref{dTdtdVeedt} and~\eqref{dTsdt}, we then obtain the relation between $E_\text{Hxc}^\V(t)$ and $v_\text{Hxc}(\b{r},t)$,~\cite{HesParBur-PRL-99}
\begin{eqnarray}
\frac{\d}{\d t} E_\text{Hxc}^\V(t)  =  \int_{\mathbb{R}^d} v_\text{Hxc}(\b{r},t) \partial_t  \rho(\b{r},t) \d \b{r}.
\label{dEhxcdt}
\end{eqnarray}

\subsubsection{Geometric TDDFT}

Let us consider the non-interacting geometric TDDFT scheme in Eq.~\eqref{eq:TDDFT-TDGKScomm} in which the external potential is replaced by the universal potential functional $v[\rho,\Psi_0]$. The Hamiltonian is
\begin{eqnarray}
\hat{H}_\s^\G(t):= \hat{T} + \hat{V}(t) + \hat{\Sigma}_\Hxc(t),
\end{eqnarray}
with $\hat{V}(t) := \int_{\mathbb{R}^d} \d \b{r} \, v[\rho,\Psi_0](\b{r},t) \hat{\rho}(\b{r})$ and $\hat{\Sigma}_\Hxc(t):= \sum_{\sigma \in \{\uparrow,\downarrow\}} \int_{\mathbb{R}^d} \!\int_{\mathbb{R}^d} {\scriptstyle \Sigma}_\Hxc^\sigma[\rho,\Phi_0,\Psi_0](\b{r},\b{r}',t) \hat{\gamma}^\sigma(\b{r}',\b{r}) \d \b{r} \d \b{r}'$ where we have introduced the universal nonlocal geometric Hxc potential
\begin{eqnarray}
{\scriptstyle \Sigma}_\Hxc^\sigma[\rho,\Phi_0,\Psi_0](\b{r},\b{r}',t):={\scriptstyle \Sigma}_\s^\sigma[\rho,\Phi_0,v[\rho,\Psi_0]](\b{r},\b{r}',t),\;\;\;
\end{eqnarray} 
within a spin-collinear formalism.
The associated time-dependent total energy is
\begin{eqnarray}
E_\s^\G(t) := \bra{\Phi^\G(t)}\hat{H}_\s^\G(t)\ket{\Phi^\G(t)},
\label{eq:EsGt}
\end{eqnarray}
which is again defined up to an arbitrary additive time-dependent constant.
Taking the time-derivative of Eq.~\eqref{eq:EsGt}, and using the Ehrenfest equation of motion for the Hamiltonian $\hat{H}_\s^\G(t)$, we obtain
\begin{eqnarray}
\frac{\d}{\d t} T_\text{s}^\G(t)  =  -\int_{\mathbb{R}^d} v(\b{r},t) \partial_t \rho(\b{r},t) \d \b{r} \phantom{xxxxxxxxxxx}
\nonumber\\
-\sum_{\sigma \in \{\uparrow,\downarrow\}} \int_{\mathbb{R}^d} \int_{\mathbb{R}^d} {\scriptstyle \Sigma}_\Hxc^\sigma(\b{r},\b{r}',t) \partial_t \gamma_{\Phi^\G(t)}^\sigma(\b{r}',\b{r}) \d \b{r} \d \b{r}',\;\;
\label{dTsGdt}
\end{eqnarray}
where $T_\text{s}^\G(t):=\bra{\Phi^\G(t)} \hat{T} \ket{\Phi^\G(t)}$ is the time-dependent geometric non-interacting kinetic energy. Note that, even though $\bra{\Phi^\G(t)} \hat{\Sigma}_\Hxc(t) \ket{\Phi^\G(t)} = 0$, the nonlocal geometric Hxc potential has an effect on the energy evolution.

The non-interacting geometric TDDFT scheme gives us a decomposition of the energy $E^\V(t)$ in Eq.~\eqref{eq:EVtcontrib} in a similar way as in Eq.~\eqref{eq:EVtdecomp},
\begin{eqnarray}
E^\V(t) = T_\text{s}^\G(t) + V(t) + E_\text{Hxc}^\G(t),
\end{eqnarray}
where $E_\text{Hxc}^\G(t) := T^\V(t) + V_\text{ee}^\V(t) -  T_\text{s}^\G(t)$ is the time-dependent geometric Hxc energy, which is a functional of $\rho$, $\Phi_0$, and $\Psi_0$. Using Eqs.~\eqref{dTdtdVeedt} and~\eqref{dTsGdt}, we then obtain the relation between $E_\text{Hxc}^\G(t)$ and ${\scriptstyle \Sigma}_\Hxc(\b{r},\b{r}',t)$
\begin{eqnarray}
\frac{\d}{\d t} E_\text{Hxc}^\G(t)  = \phantom{xxxxxxxxxxxxxxxxxxxxxxxxxx}
\nonumber\\
 \sum_{\sigma \in \{\uparrow,\downarrow\}}\int_{\mathbb{R}^d} \int_{\mathbb{R}^d} {\scriptstyle \Sigma}_\Hxc^\sigma(\b{r},\b{r}',t) \partial_t \gamma_{\Phi^\G(t)}^\sigma(\b{r}',\b{r}) \d \b{r} \d \b{r}'.
\label{dEhxcGdt}
\end{eqnarray}

\subsection{Time-dependent total momentum}

\subsubsection{Variational TDDFT}

We consider now the time-dependent total momentum (see Ref.~\onlinecite{Ullrichs-11}). The time-dependent total momentum associated with the variational TDDFT dynamics in Eq.~\eqref{eq:TDSE-TDDFT} is
\begin{eqnarray}
\b{P}^\V(t) := \bra{\Psi^\V(t)} \hat{\b{P}} \ket{\Psi^\V(t)},
\end{eqnarray}
where $\hat{\b{P}} := -i \int_{\mathbb{R}^d} [\nabla_\b{r} \hat{\gamma}(\b{r},\b{r}')]_{\b{r}'=\b{r}} \d \b{r} =\int_{\mathbb{R}^d} \hat{\b{j}}(\b{r}) \d\b{r}$ is the total momentum operator and $\hat{\b{j}}(\b{r})$ is the current-density operator. The time-derivative of the total momentum is
\begin{eqnarray}
\frac{\d}{\d t} \b{P}^\V(t) &=& i \bra{\Psi^\V(t)} [\hat{H}^\V(t),\hat{\b{P}}] \ket{\Psi^\V(t)} 
\nonumber\\
&=& - \int_{\mathbb{R}^d} \rho(\b{r},t) \nabla v(\b{r},t) \d\b{r}.
\label{eq:dPVdt}
\end{eqnarray}
Similarly, for the non-interacting variational TDDFT dynamics in Eq.~\eqref{eq:TDDFT-TDKS}, the time-dependent total momentum is
\begin{eqnarray}
\b{P}_\s^\V(t) := \bra{\Phi^\V(t)} \hat{\b{P}} \ket{\Phi^\V(t)},
\end{eqnarray}
and its time-derivative is
\begin{eqnarray}
\frac{\d}{\d t} \b{P}_\s^\V(t) =  - \int_{\mathbb{R}^d} \rho(\b{r},t) \nabla v_\s(\b{r},t) \d\b{r}.
\label{eq:dPsVdt}
\end{eqnarray}

In variational TDDFT, the total momentum can in fact be entirely expressed in terms of the density,
\begin{eqnarray}
\b{P}^\V(t) &=& \int_{\mathbb{R}^d} \b{j}_{\Psi^\V(t)}(\b{r}) \d\b{r}
\nonumber\\
&=& - \int_{\mathbb{R}^d} \b{r} \; \nabla \cdot \b{j}_{\Psi^\V(t)}(\b{r}) \d\b{r}
\nonumber\\
&=& \int_{\mathbb{R}^d} \b{r} \; \partial_t \rho(\b{r},t) \d\b{r},
\end{eqnarray}
where we have used an integration by parts (assuming the boundary term vanishes) and the continuity equation in Eq.~\eqref{eq:conteq}.  Similarly, for non-interacting variational TDDFT, we have
\begin{eqnarray}
\b{P}_\s^\V(t) &=& \int_{\mathbb{R}^d} \b{j}_{\Phi^\V(t)}(\b{r}) \d\b{r}
\nonumber\\
&=& - \int_{\mathbb{R}^d} \b{r} \; \nabla \cdot \b{j}_{\Phi^\V(t)}(\b{r}) \d\b{r}
\nonumber\\
&=& \int_{\mathbb{R}^d} \b{r} \; \partial_t \rho(\b{r},t) \d\b{r},
\end{eqnarray}
where we have used now the continuity equation in Eq.~\eqref{eq:conteqTDKS}. This shows that the total momentum is the same in interacting and non-interacting variational TDDFT, i.e. $\b{P}^\V(t) = \b{P}_\s^\V(t)$. Taking the difference between Eqs.~\eqref{eq:dPVdt} and ~\eqref{eq:dPsVdt}, we arrive at the zero-force theorem for the Hxc potential~\cite{Vig-PRL-95}
\begin{eqnarray}
\int_{\mathbb{R}^d} \rho(\b{r},t) \nabla v_\Hxc(\b{r},t) \d\b{r} = \b{0},
\label{eq:zeroforce}
\end{eqnarray}
i.e., the Hxc potential does not generate a net force.

\subsubsection{Geometric TDDFT}

We consider again the non-interacting geometric TDDFT scheme in Eq.~\eqref{eq:TDDFT-TDGKScomm} in which the external potential is replaced by the universal potential functional $v[\rho,\Psi_0]$. The time-dependent total momentum is
\begin{eqnarray}
\b{P}_\s^\G(t) := \bra{\Phi^\G(t)} \hat{\b{P}} \ket{\Phi^\G(t)},
\end{eqnarray}
and its time-derivative is
\begin{multline}
\begin{aligned}
\frac{\d}{\d t} \b{P}_\s^\G(t)  
&= i \bra{\Phi^\G(t)} [\hat{H}_\s^\G(t),\hat{\b{P}}] \ket{\Phi^\G(t)} 
\\
&=  - \int_{\mathbb{R}^d} \rho(\b{r},t) \nabla v(\b{r},t) \d\b{r} 
\end{aligned}
\\
-\sum_{\sigma \in \{\uparrow,\downarrow\}} \int_{\mathbb{R}^d} \int_{\mathbb{R}^d} \gamma_{\Phi^\G(t)}^\sigma(\b{r}',\b{r})  (\nabla_{\b{r}} + \nabla_{\b{r}'}) {\scriptstyle \Sigma}_\Hxc^\sigma(\b{r},\b{r}',t) \d \b{r} \d \b{r}'.
\\
\label{eq:dPsGdt}
\end{multline}
Note that the total momentum $\b{P}_\s^\G(t)$ generally differs from the total momentum $\b{P}^\V(t)$ of variational TDDFT,
\begin{align}
\b{P}_\s^\G(t) &= \int_{\mathbb{R}^d} \b{j}_{\Phi^\G(t)}(\b{r}) \d\b{r}
\nonumber\\
&= - \int_{\mathbb{R}^d} \b{r} \; \nabla \cdot \b{j}_{\Phi^\G(t)}(\b{r}) \d\b{r}
\nonumber\\
&= \int_{\mathbb{R}^d} \b{r} \; \partial_t \rho(\b{r},t) \d\b{r}
\nonumber\\
& \quad -2\int_{\mathbb{R}^d}\int_{\mathbb{R}^d} \b{r} \; S_{\Phi^\G(t)}(\b{r},\b{r}') w_\Hxc(\b{r}',t) \d\b{r} \d \b{r}',
\end{align}
where we have used an integration by parts and the modified continuity equation in Eq.~\eqref{eq:modconteqTDGKS}. Taking the difference of Eqs.~\eqref{eq:dPVdt} and~\eqref{eq:dPsGdt}, we find the time derivative of the difference between the total momenta
\begin{align}
& \frac{\d}{\d t} \left( \b{P}^\V(t)  - \b{P}_\s^\G(t) \right)= 
\nn \\
& \sum_{\sigma \in \{\uparrow,\downarrow\}} \int_{\mathbb{R}^d} \int_{\mathbb{R}^d} \gamma_{\Phi^\G(t)}^\sigma(\b{r}',\b{r})  (\nabla_{\b{r}} + \nabla_{\b{r}'}) {\scriptstyle \Sigma}_\Hxc^\sigma(\b{r},\b{r}',t) \d \b{r} \d \b{r}',
\end{align}
which is the analogue of the zero-force theorem in Eq.~\eqref{eq:zeroforce}, but in this case the nonlocal geometric Hxc potential generally generates a non-vanishing net force.

\subsection{Self-interaction for one-electron densities}
\label{sec:selfinteraction}

Let us consider one-electron systems ($N=1$). For this particular case, within a fixed spin sector, the admissible initial wave function is uniquely determined (up to a global phase factor) by the density trajectory~\cite{MaiBur-01}. The interacting and KS potential have thus no independent initial-state dependence and must of course coincide
\begin{eqnarray}
v[\rho,\Psi_0] \eqv v_\s[\rho,\Phi_0] \eqv v[\rho] \;\;\; (N=1),\;
\end{eqnarray}
and thus the Hxc potential must vanish,
\begin{eqnarray}
v_\Hxc[\rho,\Phi_0,\Psi_0]\eqv 0  \;\;\; (N=1),
\end{eqnarray}
i.e., $v_\Hxc$ is self-interaction free. Similarly, the Hxc geometric potential must vanish for one-electron systems,
\begin{eqnarray}
w_\Hxc[\rho,\Phi_0,\Psi_0]=0  \;\;\; (N=1),
\end{eqnarray}
provided that we impose the additional initial compatibility condition not only on $\Psi_0$ but also on $\Phi_0$, i.e. $\nabla \cdot \b{j}_{\Phi_0}(\b{r}) = - \partial_t \rho(\b{r},t)|_{t=0}$, otherwise the mismatch on the value of $\partial_t \rho(\b{r},t)|_{t=0}$ will have to be compensated by a non-zero $w_\Hxc[\rho,\Phi_0,\Psi_0]$.

In the case of the TDKS scheme with geometric correction in Eq.~\eqref{eq:TDKS+G}, if we use an approximate Hxc potential which is not self-interaction free, i.e. $v_\Hxc^\text{app}[\rho,\Phi_0,\Psi_0] \not\eqv0$, then there exists a non-trivial geometric correction potential, $w^\cor[\rho,\Phi_0,\Psi_0] \not=0$, which gives the exact self-interaction-free density.

\subsection{Adiabatic limit}

Finally, we consider the adiabatic limit, which is an important exact condition in TDDFT (see, e.g., Ref.~\onlinecite{Ullrichs-11}). We give here a heuristic derivation of the adiabatic limit. Note, however, that giving a rigorous mathematical justification of these calculations is more involved than what is done in the previous Sections~\ref{sec:causality}--\ref{sec:selfinteraction}. In particular, every statement below involving $O(\eps)$ is only formal since the corresponding norm should be specified.

The adiabatic limit may be formulated as follows. Let us define the time-scaled density,
\begin{eqnarray}
\rho_\epsilon(\b{r},t) := \rho(\b{r},\epsilon t),
\label{}
\end{eqnarray}
for $0 \leq t \leq T/\epsilon$ where $\epsilon \in (0,1]$ is a scaling factor. As we decrease $\epsilon$, the density $\rho_\epsilon(\b{r},t)$ has slower time variations. In particular, in the adiabatic limit $\epsilon \to 0$, we have $\partial_t \rho_\epsilon(\b{r},t)  = O(\epsilon)$. We will determine the adiabatic limit of both variational and geometric TDDFT.

Let us start with variational TDDFT. We define the exact-adiabatic interacting Hamiltonian associated with the density $\rho(\epsilon t)$, depending parametrically on $s:=\epsilon t$,
\begin{eqnarray}
\hat{H}^{\ea}(\epsilon t) := \hat{T} + \hat{V}_\text{ee} + \hat{V}^{\ea}(\epsilon t),
\label{}
\end{eqnarray}
with the exact-adiabatic potential operator $\hat{V}^{\ea}(\epsilon t) := \int_{\mathbb{R}^d} \d\b{r} \, v^\ea[\rho_\epsilon](\b{r},t) \, \hat{\rho}(\b{r})$ where $v^\ea[\rho_\epsilon](\b{r},t) := v^\gs[\rho_\epsilon(t)](\b{r})=v^\gs[\rho(\epsilon t)](\b{r})$ and $v^\gs[\rho](\b{r})$ is the exact potential of static ground-state DFT which reproduces the ground-state density $\rho$ in the interacting system. We have thus assumed that the density $\rho(\epsilon t)$ is interacting ground-state $v$-representable. We further assume that, for any parameter $\epsilon t$, $\hat{H}^{\ea}(\epsilon t)$ has a non-degenerate ground-state wave function $\Psi_0^{\gs}(\epsilon t)$, which yields by construction the density $\rho(\epsilon t)$ and which is separated from the excited states by a finite gap, and that the Hamiltonian $\hat{H}^{\ea}(\epsilon t)$ depends sufficiently smoothly on $\epsilon t$. Next, we define the exact-adiabatic effective TDSE,
\begin{eqnarray}
\left\{
\begin{aligned}
i \partial_t \Psi^{\ea}_\epsilon(t) &= \hat{H}^{\ea}(\epsilon t) \Psi^{\ea}_\epsilon(t), \\
\Psi^{\ea}_\epsilon(0) &= \Psi_0^{\gs}[\rho_0],
\end{aligned}
\right.
\label{eq:TDSEea}
\end{eqnarray}
starting from the ground-state wave function $\Psi_0^{\gs}[\rho_0] := \Psi_0^{\gs}(0)$ associated with the initial density $\rho_0(\b{r}):=\rho(\b{r},0)$. As we decrease $\epsilon$, $\hat{H}^{\ea}(\epsilon t)$ has slower time variations. According to the adiabatic theorem of quantum mechanics (see Ref.~\onlinecite{Teu-BOOK-03}, Thm. 2.2), as $\epsilon \to 0$, we have, up to a time-dependent phase factor, $\Psi^{\ea}_\epsilon(t)\eqpsi \Psi_0^{\gs}(\epsilon t) + O(\epsilon)$. This implies that the density of $\Psi^{\ea}_\epsilon(t)$ is the density of the instantaneous ground-state wave function $\Psi_0^{\gs}(\epsilon t)$, i.e. $\rho_\epsilon$, within $O(\epsilon)$,
\begin{eqnarray}
\rho_{\Psi^{\ea}_\epsilon} = \rho_\epsilon+ O(\epsilon).
\label{eq:rhoPsiea}
\end{eqnarray}
Besides, since the wave function $\Psi^{\ea}_\epsilon$ satisfies the effective TDSE in Eq.~\eqref{eq:TDSEea} with potential $v^\ea[\rho_\epsilon](\b{r},t)$ and initial condition $\Psi_0^{\gs}[\rho_0]$ and has density $\rho_{\Psi^{\ea}_\epsilon}$, according to the Runge-Gross theorem, we have
\begin{eqnarray}
v[\rho_{\Psi^{\ea}_\epsilon},\Psi_0^{\gs}[\rho_0]](\b{r},t) \eqv v^\ea[\rho_\epsilon](\b{r},t).
\label{eq:vrhoPsiea}
\end{eqnarray}
Equations~\eqref{eq:rhoPsiea} and~\eqref{eq:vrhoPsiea} characterize the adiabatic limit of variational TDDFT: the exact potential evaluated at the density $\rho_{\Psi^{\ea}_\epsilon}$, which approaches $\rho_\epsilon$ as $\epsilon \to 0$, and at the initial wave function $\Psi_0^{\gs}[\rho_0]$ reduces to the exact-adiabatic potential evaluated at the density $\rho_\epsilon$. If we assume in addition that the TDDFT density-to-potential map $\rho \mapsto v[\rho,\Psi_0^{\gs}[\rho_0]]$ is  locally Lipschitz continuous, in appropriate norms, along the density path considered, we arrive at the following version of the adiabatic limit of variational TDDFT:
\begin{eqnarray}
v[\rho_\epsilon,\Psi_0^{\gs}[\rho_0]](\b{r},t) \eqv v^\ea[\rho_\epsilon](\b{r},t) + O(\epsilon),
\label{}
\end{eqnarray}
provided that the additional initial compatibility condition is fulfilled, i.e. $\partial_t \rho_\epsilon(\b{r},t)|_{t=0} = - \nabla \cdot \b{j}_{\Psi_0^{\gs}[\rho_0]}(\b{r}) = 0$.

Proceeding exactly in the same way for the non-interacting TDSE [Eq.~\eqref{eq:TDDFT-TDKS}], assuming that the densities are also non-interacting ground-state $v$-representable and making the same regularity assumptions, we obtain similarly the adiabatic limit for the KS potential $v_s$,
\begin{eqnarray}
v_\s[\rho_\epsilon,\Phi_0^{\gs}[\rho_0]](\b{r},t) \eqv v_\s^\ea[\rho_\epsilon](\b{r},t) + O(\epsilon),
\end{eqnarray}
with $v_\s^\ea[\rho](\b{r},t):= v_\s^\gs[\rho(t)](\b{r})$ where $v_\s^\gs[\rho](\b{r})$ is the exact KS potential of static ground-state DFT, and $\Phi_0^{\gs}[\rho_0]$ is the non-degenerate ground-state single-determinant KS wave function reproducing the density $\rho_0$. Using Eq.~\eqref{eq:vHxc}, we thus obtain the adiabatic limit of the Hxc potential,
\begin{eqnarray}
v_\Hxc[\rho_\epsilon,\Phi_0^{\gs}[\rho_0],\Psi_0^{\gs}[\rho_0]](\b{r},t) \eqv  v_\Hxc^\ea[\rho_\epsilon](\b{r},t)  + O(\epsilon),\;\;\;\;
\label{eq:vHxcadiabaticexpansion}
\end{eqnarray}
where $v_\Hxc^\ea[\rho]:=v_\s^\ea[\rho] - v^\ea[\rho]$ is the exact-adiabatic Hxc potential. Of course, at fixed $t$, in the limit $\epsilon \to 0$, the Hxc potential becomes the exact static ground-state Hxc potential associated with the initial density,
\begin{eqnarray}
\lim_{\epsilon \to 0} v_\Hxc[\rho_\epsilon,\Phi_0^{\gs}[\rho_0],\Psi_0^{\gs}[\rho_0]](\b{r},t) \eqv v_\Hxc^\gs[\rho_0](\b{r}).
\label{eq:vHxcepsilonto0}
\end{eqnarray}
Note also that Eq.~\eqref{eq:vHxcadiabaticexpansion} gives us a way to extract the exact-adiabatic approximation from the exact Hxc potential for any density trajectory $\rho$ satisfying the above assumptions, namely
\begin{eqnarray}
v_\Hxc^\ea[\rho](\b{r},t) \eqv \lim_{\epsilon \to 0} v_\Hxc[\rho_\epsilon,\Phi_0^{\gs}[\rho_0],\Psi_0^{\gs}[\rho_0]](\b{r},t/\epsilon).\;
\end{eqnarray}

Let us go now to geometric TDDFT. The equivalent of Eq.~\eqref{eq:vHxcepsilonto0} provides a definition to what we may call the ``ground-state'' limit of the geometric Hxc potential $w_\Hxc$,
\begin{eqnarray}
w^\text{gs}_\Hxc[\rho_0](\b{r},t) := \lim_{\epsilon \to 0} w_\Hxc[\rho_\epsilon,\Phi_0^\text{gs}[\rho_0],\Psi_0^\text{gs}[\rho_0]](\b{r},t),\;\;
\label {}
\end{eqnarray}
which is the geometric Hxc potential needed in the geometric TDKS scheme in Eq.~\eqref{eq:TDGKS} with $v^\ph=v^\gs[\rho_0]$ in order to reproduce the density $\rho_0(\b{r})$ at all times. Even though the density $\rho_0(\b{r})$ is time independent, the ``ground-state'' geometric Hxc potential $w^\text{gs}_\Hxc[\rho_0] (\b{r},t)$ has a priori a non-trivial time dependence. Indeed, while the initial KS ground-state wave function $\Phi_0^\text{gs}[\rho_0]$ is stationary under the ground-state KS Hamiltonian $\hat{T} + \int_{\R^d} v_\s^\gs[\rho_0](\br) \hat{\rho}(\br) \d\br$, it is generally not stationary under the non-interacting reference Hamiltonian $\hat{T} + \int_{\R^d} v^\gs[\rho_0](\br) \hat{\rho}(\br) \d\br$ so that a time-dependent geometric Hxc potential can be needed to maintain the constant density. By contrast, the non-adiabatic geometric potential correction in Eq.~\eqref{eq:wna} is expected to go to zero as $\epsilon \to 0$,
\begin{eqnarray}
w^\text{na}[\rho_\epsilon,\Phi_0^\text{gs}[\rho_0],\Psi_0^\text{gs}[\rho_0]] = 0 + O(\epsilon),
\label{}
\end{eqnarray}
which suggests that it is preferable to try to approximate $w^\text{na}$ in the TDKS scheme with geometric correction in Eq.~\eqref{eq:TDKS+G}, rather than $w_\Hxc$ in the geometric TDKS scheme in Eq.~\eqref{eq:TDGKS}. Plots of $w_\Hxc(t)$ and $w^\text{na}(t)$ near the adiabatic limit in the Hubbard dimer are provided in the left panel of Fig. 6 of Ref.~\onlinecite{MAQUI_finite-26}.

\section{Mathematical analysis of variational and geometric TDDFT}
\label{sec:math}

This section develops the mathematical foundations of variational and geometric TDDFT introduced in Section~\ref{sec:var_geom_TDDFT} and is of a different nature from the rest of the text since all the results are entirely mathematically rigorous. We will first study the properties of the two operators $S_\Psi$ and $K_\rho$, introduced in Eqs.~\eqref{eq:SPsirrp} and~\eqref{eq:K}, respectively. We recall that the invertibility of $S_\Psi$ was assumed in the geometric interpretation of the abstract formulation of TDDFT developed in Section~\ref{sec:geometric_interpretation}. Furthermore, our geometric variant of TDDFT requires the invertibility of $S_\Psi$ (Section~\ref{sec:geometricTDDFT}), while conventional  TDDFT based on the variational principle requires the invertibility of $K_\rho$ (Section~\ref{sec:standardTDDFT}). Then, in Section~\ref{sec:manifold}, we explain in which sense the set  $\cM_\rho$ of wave functions having a prescribed density $\rho$ is really a manifold, so that the tangent and normal spaces used in Section~\ref{sec:geometric_interpretation} are really tangent and normal spaces in the sense of differential geometry. This is more subtle than expected. A summary of the section is given in Section~\ref{sec:summary}.

In order to cover a variety of physical situations of interest (e.g., the one-dimensional (1D) soft-Coulomb model considered in Ref.~\onlinecite{MAQUI_TDDFT-26}), we work in the whole section in a space $\Omega$ that will be either the whole space $\Omega=\R^d$ or the unit torus $\Omega=\bT^d=\R^d/\Z^d$ in dimension $d\in\{1,2,3\}$. The unit torus can be identified with the cube $(-1/2,1/2)^d$ with periodic boundary conditions. The torus case avoids the difficulties associated with density decay at infinity, allowing us to obtain stronger results. For many of our results the set $\Omega$ could also be a general open subset of $\R^d$ but then boundary conditions need to be specified when we deal with operators involving space derivatives. We will not discuss the latter case for simplicity. Throughout this section, we denote by $\mathcal H:=\bigwedge^N  L^2(\Omega \times \{ \uparrow,\downarrow \};\mathbb{C})$ the $N$-electron Hilbert space with inner product $\braket{\,\cdot\,}{\,\cdot\,}$ and norm $\| \cdot \|_\cH$.

In the whole section, we assume that the density $\rho$ associated with the considered wave function $\Psi \in \cH$ satisfies
\begin{equation}
\rho = \rho_\Psi>0\qquad \text{almost everywhere on $\Omega$} 
\end{equation}
(i.e., everywhere except on a set of Lebesgue measure zero).
This is a physically reasonable assumption and it will be useful for the mathematical analysis. It can be weakened at some places, but we do not discuss this further. In fact, in many cases we will need to assume more on the density (typically continuity and strict positivity everywhere). For ground states, under suitable assumptions on the external and interaction potentials, the almost-everywhere strict positivity of the density follows from the unique continuation principle~\cite{Garrigue-20}. In fact, neutral molecular Coulombic electronic systems on $\Omega=\R^3$ always have a continuous ground-state density (see Theorem~\ref{thm:H2} below) and it is conjectured that it is strictly positive everywhere and not just almost everywhere~\cite{FouHofSor-07}. The validity of this conjecture seems unknown at present. For an odd number of 1D spinless fermions on the torus $\Omega=\bT$, everywhere strict positivity of the ground-state density has recently been established for a broad class of external and interaction potentials~\cite{Corso-26,Corso-26b}. For general interacting systems in $\Omega=\R^d$ with $d\in\{2,3\}$, the strict positivity of ground-state densities can fail, even with very smooth potentials. We provide in Appendix~\ref{app:counterex-positivity} the example of an interacting model system in $\R^3$ with smooth potentials for which the spin-polarized ground-state density vanishes along the whole $z$ axis. We believe that such examples are rare and that the density is ``generically'' strictly positive everywhere for ground states. For the time-dependent case, strict positivity almost everywhere is an additional reasonable assumption on the density trajectory but does not follow in general from strict positivity almost everywhere of the initial density.

\subsection{Properties of the operator $S_\Psi$}
\label{sec:S}
In this subsection, we properly define the operator $S_\Psi$ and study its invertibility properties. 

\subsubsection{Functional setting and basic properties of $S_\Psi$}

The quadratic form associated with the operator $S_\Psi$, formally introduced in Eq.~\eqref{eq:SPsirrp}, is
\begin{align}
&\pscal{w,S_\Psi w}\nn
:=\int_{\Omega}\int_{\Omega}S_\Psi(\br,\br')w(\br)w(\br')\,\dr\,\dr'\nn\\
& =\int_{\Omega}\rho_\Psi(\br)w(\br)^2\dr+\int_{\Omega}\int_{\Omega}\rho_{2,\Psi}(\br,\br')w(\br)w(\br')\,\dr\,\dr',
\label{eq:formula_S}
\end{align}
for functions $w:\Omega \to \mathbb{R}$ in a function space to be determined, and where we recall that $\rho_\Psi(\b{r})$ and $\rho_{2,\Psi}(\b{r},\b{r}')$ are the density and pair density of the wave function $\Psi  \in {\cal H}$. The first term in this formula suggests that the natural function space for $w$ is the density-weighted real $L^2$ Hilbert space
\begin{equation}
L^2_{\rho_\Psi}(\Omega):=\left\{f:\Omega\to\R\ \mid \int_{\Omega}f(\br)^2 \rho_\Psi(\br)\,\dr<\ii\right\}
\label{eq:L2rho}
\end{equation}
containing the real-valued functions that are square integrable with the weight $\rho_\Psi$. For simplicity of notation, we omit the target space of a function space whenever it is $\mathbb{R}$, i.e. we write $L^2(\Omega)$ instead of $L^2(\Omega;\R)$, since it is understood that densities and potentials are always assumed to be real-valued. 
The norm associated with the space $L^2_{\rho_\Psi}(\Omega)$ is $\|f\|_{L^2_{\rho_\Psi}}:=( \int_{\Omega} f(\b{r})^2 \rho_\Psi(\b{r})\d\b{r})^{1/2}$.

Let us show that the quadratic form of Eq.~\eqref{eq:formula_S} can indeed be defined on $L^2_{\rho_\Psi}(\Omega)$. 
From Eq.~\eqref{eq:SPsirrp}, we see that 
\begin{align}
&\pscal{w,S_\Psi w}
= \left\| \int_{\Omega} \dr \; w(\br) \hat{\rho}(\br) \Psi \right\|_{{\cal H}}^2
= \left\| \sum_{j=1}^N w(\br_j) \Psi \right\|_{{\cal H}}^2,
\label{eq:wSwnorm}
\end{align}
and using the Cauchy-Schwarz inequality
\begin{equation}
\left(\sum_{j=1}^N w(\br_j)\right)^2\leq N\sum_{j=1}^N w(\br_j)^2,
\label{eq:Cauchy-Schwarz_W}
\end{equation}
we find
\begin{equation}
\pscal{w,S_\Psi w}\leq N\int_{\Omega}\rho_\Psi(\br)w(\br)^2\,\dr = N \| w \|_{L^2_{\rho_\Psi}}^2.
\label{eq:S_bounded}
\end{equation}
This inequality shows that the non-negative quadratic form $\pscal{w,S_\Psi w}$ is bounded on $L^2_{\rho_\Psi}(\Omega)$, and therefore that we can define $S_\Psi$ as a bounded operator from the space $L^2_{\rho_\Psi}(\Omega)$ to its dual space (with respect to the $L^2$ inner product), which can be identified with the space $L^2_{1/\rho_\Psi}(\Omega)$.
The expression of $S_\Psi$ is
\begin{equation}
(S_\Psi w)(\br):=\rho_\Psi(\br)w(\br)+\int_{\Omega}\rho_{2,\Psi}(\br,\br') w(\br')\,\dr'.
\label{eq:formula_S2}
\end{equation}

It is sometimes more convenient to represent the same quadratic form $\pscal{w,S_\Psi w}$ by an operator $S_\Psi'$ acting from $L^2_{\rho_\Psi}(\Omega)$ to itself instead of its dual. This is achieved by defining $S_\Psi' := \rho_{\Psi}^{-1} S_\Psi$, or, more explicitly,
\begin{equation}
(S'_\Psi w)(\br):=w(\br)+\frac1{\rho_\Psi(\br)}\int_{\Omega}\rho_{2,\Psi}(\br,\br') w(\br')\,\dr'.
\label{eq:formula_S2'}
\end{equation}
Here and below, since $\rho_\Psi$ is assumed strictly positive almost everywhere, division by $\rho_\Psi$ is also understood almost everywhere. We have $\pscal{w,S_\Psi w} = \pscal{w,S_\Psi' w}_{L^2_{\rho_\Psi}}$ where $\pscal{f,g}_{L^2_{\rho_\Psi}}:=\int_{\Omega} f(\b{r}) g(\b{r}) \rho_\Psi(\b{r}) \d\b{r}$ designates the real inner product of $L^2_{\rho_\Psi}(\Omega)$, and $S_\Psi'$ is indeed a bounded self-adjoint operator on $L^2_{\rho_\Psi}(\Omega)$. 

Finally, it is sometimes useful to remove the weight completely and instead work on the usual real Hilbert space $L^2(\Omega)$. This can be achieved by defining the operator $S_\Psi'' := \rho_{\Psi}^{1/2} S_\Psi' \rho_{\Psi}^{-1/2}$, or, more explicitly,
\begin{equation}
(S''_\Psi \tilde{w})(\br):=\tilde{w}(\br)+\int_{\Omega}\frac{\rho_{2,\Psi}(\br,\br')}{\sqrt{\rho_\Psi(\br)\rho_\Psi(\br')}} \tilde{w}(\br')\,\dr'.
\label{eq:formula_S2''}
\end{equation}
The latter is unitarily equivalent to $S'_\Psi$, and we have $\pscal{w,S_\Psi' w}_{L^2_{\rho_\Psi}}=\pscal{\tilde{w},S_\Psi'' \tilde{w}}_{L^2}$ 
with $\tilde{w} = \sqrt{\rho_\Psi}w \in L^2(\Omega)$. Therefore, $S''_\Psi$ is a bounded  self-adjoint operator on $L^2(\Omega)$. Depending on the property we study, it is convenient to use either $S_\Psi$, $S'_\Psi$ or $S_\Psi''$. These three operators represent the same quadratic form. In particular, the invertibility of $S_\Psi$ is equivalent to the invertibility of $S'_\Psi$ or $S_\Psi''$ on the corresponding spaces. 

The three operators have the same spectrum
\begin{equation}
\sigma(S_\Psi) = \sigma (S_\Psi')=\sigma(S''_\Psi),
\end{equation}
where the spectrum of $S_\Psi$ has to be understood as the generalized spectrum with weight $\rho_\Psi$, i.e. the corresponding (generalized) eigenvalue equation is $S_\Psi w = \lambda \rho_\Psi w$. For studying spectral properties it is more convenient to work with self-adjoint operators on a Hilbert space, hence use either $S'_\Psi$ on $L^2_{\rho_\Psi}(\Omega)$ or $S''_\Psi$ on $L^2(\Omega)$. Note that $N$ is an eigenvalue of $S_\Psi$ or $S_\Psi'$ with associated eigenfunction $w=1\in L^2_{\rho_\Psi}(\Omega)$ because $\int_{\Omega}\rho_{2,\Psi}(\br,\br')\,\dr'=(N-1)\rho_\Psi(\br)$. For $S''_\Psi$ the corresponding eigenfunction is $\tilde{w} = \sqrt{\rho_\Psi}\in L^2(\Omega)$. Under the condition considered below in Theorem~\ref{thm:unique-continuation} that the spin-summed $N$-electron probability density is strictly positive almost everywhere, this eigenvalue $N$ is non-degenerate for $N\geq2$ since equality in Eq.~\eqref{eq:Cauchy-Schwarz_W} implies that $w$ is constant almost everywhere. It also follows from Eq.~\eqref{eq:wSwnorm} that $S_\Psi'$ is a non-negative self-adjoint operator, which implies it has a non-negative spectrum. So we have
\begin{equation}
\{N\} \subset 
\sigma (S_\Psi')\subset [0,N].
\end{equation}

For the geometric interpretation of TDDFT, we are not interested in the largest eigenvalue of $S_\Psi'$, but rather in its invertibility, that is, the bottom of its spectrum. Invertibility has to be understood in the strong sense, that is, the inverse must exist and be a bounded operator. The existence of the inverse requires the triviality of the kernel in the sense $\ker(S_\Psi)=\ker(S'_\Psi)=\ker(S_\Psi'')=\{0\}$, i.e. $0$ is not an eigenvalue. In the self-adjoint case, the boundedness of the inverse holds if and only if the bottom of the spectrum is separated from $0$ by a strictly positive gap $g_S$,
\begin{equation}
\sigma(S_\Psi')\subset [g_S,N],\qquad g_S>0,
\end{equation}
which is the same as requiring that the quadratic form $\pscal{w,S_\Psi w}$ satisfies the \emph{gap inequality}
\begin{equation}
\pscal{w,S_\Psi w}\geq g_S\int_{\Omega}\rho_\Psi(\br)w(\br)^2\,\dr,
\label{eq:gap_S}
\end{equation}
for all $w \in L^2_{\rho_\Psi}(\Omega)$.
This is a reversed inequality compared to Eq.~\eqref{eq:S_bounded}. A strictly positive gap excludes the existence of potentials $w$ with finite density-weighted norm whose action on the wave function $\Psi$ becomes arbitrarily small. It is therefore essential for stable extraction of the potential $w$ by inversion of $S_\Psi$. In the next two subsections we discuss the triviality of the kernel and the gap inequality.

\subsubsection{Triviality of the kernel of $S_\Psi$}
\label{sec:kernelSPsi}

The existence of the inverse of $S_\Psi$, $S_\Psi'$ or $S_\Psi''$ on the appropriate spaces requires that $\ker(S_\Psi)=\{0\}$. From Eq.~\eqref{eq:wSwnorm}, this is equivalent to
\begin{equation}
\sum_{j=1}^N w(\br_j)\Psi=0\quad\Longrightarrow\quad w=0
\label{eq:unique-V-rep}
\end{equation}
where the two equalities are understood almost everywhere and $w$ is assumed to belong to $L^2_{\rho_\Psi}(\Omega)$. The property~\eqref{eq:unique-V-rep} is often called the \textit{unique $v$-representability property}~\cite{PenLeu-21}. Indeed, if $\Psi$ happens to be an $N$-electron eigenstate for some external potential $v$, then Eq.~\eqref{eq:unique-V-rep} implies the uniqueness of $v$ modulo constants, after subtracting the eigenvalue equations for two potentials. Here we are not requiring that $\Psi$ is an eigenstate but can nevertheless require the property~\eqref{eq:unique-V-rep}. This property was also recently studied in Ref.~\onlinecite{Corso-26_ppt}.

To explain the meaning of the property~\eqref{eq:unique-V-rep}, we now give an example where it fails. Consider an $N$-electron wave function $\Psi$ such that exactly $K$ electrons are confined to a non-trivial subdomain $D\subset\Omega$ and $N-K$ electrons are in the complement $\Omega\setminus D$, for some $0<K<N$. Then it is clear that
\begin{equation}
\sum_{j=1}^N w(\br_j)\Psi=0\quad\text{for}\quad w(\br)=\1_D(\br)-\frac{K}{N},
\label{eq:counter_v_rep}
\end{equation}
where $\1_D(\br)$ designates the indicator function of $D$, 
because $\sum_{j=1}^N\1_D(\br_j)$ gives the number of electrons in $D$ which is always $K$. In this example, the wave function is supported only on configurations having exactly $K$ electrons in $D$, and therefore vanishes on large subsets of $\Omega^{N}$, for instance on $D^N\subset \Omega^{N}$. This is the situation we want to avoid.

The following theorem gives a simple sufficient condition for the unique $v$-representability property to hold, and thus for the triviality of the kernel of $S_\Psi$.

\begin{theorem}[Sufficient condition for unique $v$-representability in terms of $\Psi$]\label{thm:unique-continuation}
Let $\Omega$ be either $\R^d$ or the torus $\bT^d$ with $d\in\{1,2,3\}$. If the spin-summed $N$-electron probability density is strictly positive,
\begin{equation}
\sum_{\sigma_1,...,\sigma_N \in \{\uparrow,\downarrow\}}|\Psi(\b{r}_1,\sigma_1,...,\b{r}_N,\sigma_N)|^2>0,
\label{eq:P_N}
\end{equation}
almost everywhere on $\Omega^N$, then the unique $v$-representability property~\eqref{eq:unique-V-rep} holds.
\end{theorem}

The sum over all spins in~\eqref{eq:P_N} is to be able to cover fixed-spin states that have many zero spin components by definition. It is sufficient that $\Psi \not=0$ almost everywhere for at least one set of spin variables to obtain the unique $v$-representability property~\eqref{eq:unique-V-rep}.

\begin{proof}
Assume $\sum_{j=1}^N w(\br_j)\Psi=0$. Take then the modulus square and sum over all spins. Using that $w$ is spin-independent, the assumption~\eqref{eq:P_N} then implies that $\sum_{j=1}^N w(\br_j)=0$ almost everywhere on $\Omega^N$. We then integrate this relation over $\rho_\Psi(\br_2)\cdots\rho_\Psi(\br_N)$. This yields
\begin{equation}
w(\br_1)=-\frac{N-1}{N}\int_{\Omega}\rho_\Psi(\br)w(\br)\,\dr
\label{eq:W-N-1W}
\end{equation}
almost everywhere. Note that the integral on the right side is well defined since $\sqrt{\rho_\Psi}$ and $\sqrt{\rho_\Psi}w$ are both square-integrable, hence their product is integrable. On the other hand, integrating $\sum_{j=1}^N w(\br_j)\Psi=0$ over $\Psi^*$ we find $\int_{\Omega}\rho_\Psi(\br)w(\br)\,\dr=0$, and thus conclude that $w=0$ almost everywhere.
\end{proof}

As a consequence of unique continuation, under suitable assumptions on the external and interaction potentials, the strict positivity almost everywhere of the spin-summed $N$-electron probability density is known to hold for exact eigenstates~\cite{Garrigue-20}. In the physically relevant case $\Omega=\R^3$, it is also valid for \emph{any} wave function $\Psi$ that is a finite linear combination of Slater determinants made with orbitals that are real-analytic on $\R^3\setminus\{R_1,...,R_M\}$ for finitely many points $R_1,...,R_M\in\R^3$ (typically the nuclear positions). This is because $\Psi$ is then real-analytic on $(\R^3\setminus\{R_1,...,R_M\})^N$, and a nonzero real-analytic function can only vanish on a set of measure zero. This covers almost all the wave functions encountered in practical quantum-chemistry calculations, for instance in a finite Gaussian basis set, whether or not they are eigenstates.

Next we give a criterion that only involves a physical regularity assumption on $\Psi$ ensuring that $\rho_\Psi$ is a continuous function. The latter is then assumed to be strictly positive everywhere. To state the theorem, we need to define the $N$-electron Sobolev spaces
\begin{equation}
\cH^1 :=\left\{\Psi\in\cH\ \vert\ \sum_{j=1}^N\int_{(\Omega\times \{ \uparrow,\downarrow \})^N}|\nabla_{\br_j}\Psi|^2<\ii\right\}
\label{eq:def_H1_Psi}
\end{equation}
and
\begin{equation}
\cH^2 :=\left\{\Psi\in\cH\ \vert\ \sum_{j=1}^N\int_{(\Omega\times \{ \uparrow,\downarrow \})^N}|\nabla_{\br_j}^2\Psi|^2<\ii\right\},
\label{eq:def_H2,Psi}
\end{equation}
which are subspaces of the $N$-electron Hilbert space ${\cal H}$. In the case of the torus $\Omega=\bT^d$, we can simply think of the cube $(-1/2,1/2)^d$ but then $\Psi$ must satisfy suitable periodic boundary conditions. The spaces $\cH^1$ and $\cH^2$ are very natural physically. The space $\cH^1$ contains the wave functions that have a finite kinetic energy, whereas $\cH^2$ contains those that have a finite variance of the kinetic energy as well. We have the strict inclusions $\cH^2\subset\cH^1\subset\cH$. It is very reasonable to assume that $\Psi\in\cH^2$ because the latter is the domain of the $N$-electron Hamiltonian under suitable assumptions on the external and interaction potentials, see Ref.~\onlinecite{Lewin-spectral} (Thm.~6.2). The associated one-particle Sobolev spaces read
\begin{equation}
H^1(\Omega):=\left\{f\in L^2(\Omega)\ \vert\ \int_{\Omega}|\nabla f(\br)|^2\,\rd\br<\ii\right\},
\end{equation}
and
\begin{equation}
H^2(\Omega):=\left\{f\in L^2(\Omega)\ \vert\ \int_{\Omega}|\nabla^2 f(\br)|^2\,\rd\br<\ii\right\}.
\end{equation}
Again, when $\Omega=\bT^d$ we can think of the cube $(-1/2,1/2)^d$ but then periodic boundary conditions are further assumed on $f$.
An important property is that $H^1(\Omega)$ is an algebra for $d=1$ and that $H^2(\Omega)$ is an algebra for $d \in\{1,2,3\}$. 
This means that $fg\in H^k(\Omega)$ whenever $f,g\in H^k(\Omega)$ with a bound in the form
\begin{equation}
    \norm{fg}_{H^k}\leq C\norm{f}_{H^k}\norm{g}_{H^k}
    \label{eq:Hk_algebra}
\end{equation}
with some constant $C>0$, for $k=1$ when $d=1$ or $k=2$ when $d \in \{1,2,3\}$, and where we have introduced the norms
$\norm{f}_{H^1}:=(\norm{f}_{L^2}^2+\norm{\nabla f}_{L^2}^2)^{1/2}$ and $\norm{f}_{H^2} := (\norm{f}_{L^2}^2 + \norm{\nabla^2 f}_{L^2}^2)^{1/2}$. 
In addition, functions in $H^k(\Omega)$ are also continuous, again for the same $k$ and $d$ (as a consequence of the Sobolev embedding theorems~\cite{LieLos-01,Lewin-spectral}). This gives a natural role to $H^1(\Omega)$ in 1D models, and to $H^2(\Omega)$ in two-dimensional (2D) and three-dimensional (3D) models. 

We can now state the following result which is inspired by Ref.~\onlinecite{Corso-26_ppt}.

\begin{theorem}[Sufficient condition of unique $v$-representability from strict positivity of $\rho_\Psi$]\label{thm:H2}
Let $\Omega$ be either $\R^d$ or the torus $\bT^d$ with $d\in\{1,2,3\}$. Assume that $\Psi\in \cH^k$ with $k=1$ if $d=1$ and $k=2$ if $d \in \{2,3\}$. Then $\rho_\Psi\in H^k(\Omega)$ and it is thus a continuous function. If $\rho_\Psi>0$ everywhere, then the unique $v$-representability property~\eqref{eq:unique-V-rep} holds. 
\end{theorem}

Compared with Theorem~\ref{thm:unique-continuation}, the advantage of Theorem~\ref{thm:H2} is that it provides a criterion involving only $\rho_\Psi$, together with a natural regularity assumption on $\Psi$. 

\begin{proof}
We write the proof for dimension $d \in\{2,3\}$ and $k=2$. The argument is similar for the 1D case. The fact that $\Psi\in\cH^2$ means $\sum_{j=1}^N\nabla_{\br_j}^2\Psi\in\cH$. This implies that the (spin-dependent) one-electron density matrix $\gamma_\Psi$ satisfies $\tr(\nabla^2\gamma_\Psi\nabla^2)<\ii$ where $\tr$ designates the trace in $L^2(\Omega\times\{\uparrow,\downarrow\};\C)$. Of course we also have $\tr(\gamma_\Psi)=N$ with $\gamma_\Psi\geq0$ in the sense of operators. Consider a spectral decomposition of the form
\begin{equation}
\gamma_\Psi = \sum_j n_j |\phi_j\rangle\langle\phi_j|, \quad \langle \phi_i,\phi_j\rangle_{L^2}=\delta_{i,j}, \quad n_j > 0,
\end{equation}
where the sum over $j$ can be finite or infinite.
The $\phi_j$'s form an orthonormal family of natural spin-orbitals with non-zero natural occupation numbers $n_j>0$, and we have 
\begin{equation}
\tr(\gamma_\Psi)=\sum_{\sigma\in \{\uparrow,\downarrow\}} \sum_{j} n_j=N, \\ 
\end{equation}
and
\begin{equation}
\sum_{\sigma\in \{\uparrow,\downarrow\}} \sum_{j} n_j\int_{\Omega}|\nabla_\br^2\phi_j(\br,\sigma)|^2\,\rd\br=\tr(\nabla^2\gamma_\Psi\nabla^2)<\ii.
\end{equation}
This proves that $\phi_j\in H^2(\Omega\times\{\uparrow,\downarrow\};\C)$ for all $j$. Hence all the  natural spin-orbitals $\phi_j$ are continuous functions. Recall that the density is
\begin{equation}
\rho_\Psi(\br)=\sum_{\sigma\in \{\uparrow,\downarrow\}} \sum_j n_j|\phi_j(\br,\sigma)|^2.
\end{equation}
Using that $H^2(\Omega)$ is an algebra, we conclude from Eq.~\eqref{eq:Hk_algebra} that $\rho_\Psi\in H^2(\Omega)$ with the bound
\begin{equation}
\norm{\rho_\Psi}_{H^2}\leq C\sum_j n_j\norm{\phi_j}^2_{H^2} \le  C(N+\tr(\nabla^2\gamma_\Psi\nabla^2))<\infty.
\end{equation}
In particular, $\rho_\Psi$ is a continuous function. 

Next we prove the unique $v$-representability property~\eqref{eq:unique-V-rep} when $\rho_\Psi>0$ everywhere. We consider a $w\in L^2_{\rho_\Psi}(\Omega)$ for which the left-hand side of Eq.~\eqref{eq:unique-V-rep} holds. This implies that the $N$-electron density matrix $|\Psi\rangle\langle\Psi|$ commutes with $\hat{W} = \sum_{j=1}^Nw(\br_j)$ (in the sense of Ref.~\onlinecite{Lewin-spectral}, Thm.~4.41). Taking the partial trace of this commutator over $N-1$ electrons, we obtain that the operator $\gamma_\Psi$ commutes with the operator of multiplication by $w$. Since the two operators commute, the eigenspaces of $\gamma_\Psi$ are invariant under the action of any bounded function of $w$. The eigenspaces corresponding to strictly positive natural occupation numbers are all finite-dimensional since $\sum_j n_j = N$. This implies that we can choose the natural spin-orbitals $\phi_j$'s so that $w\phi_j=a_j\phi_j$ for some real numbers $a_j\in\R$. In other words, $\phi_j$ is spatially supported on the level set $\{\br\in\Omega\ \vert\ w(\br)=a_j\}$ of $w$ (up to a set of zero Lebesgue measure). Many $a_j$'s can be identical here. For any given value $a$ among the $a_j$'s, let us then consider the open set
\begin{equation}
\Omega_a:=\bigg\{\br\in\Omega\ \vert\  \sum_{\sigma \in \{\uparrow,\downarrow\}}\sum_{a_j=a}|\phi_j(\br,\sigma)|^2>0\bigg\},
\end{equation}
that is the set of spatial points where at least one of the natural spin-orbitals supported in the level set $\{\br\in\Omega\ \vert\ w(\b{r})= a\}$ (up to a set of zero Lebesgue measure) is non zero. By definition $\Omega_a\subset \{\br\in\Omega\ \vert\ w(\b{r})= a\}$ up to a set of zero measure and thus $\Omega_a\cap \Omega_b=\emptyset$ for $a\neq b$ because those sets are open. 
Since $\rho_\Psi(\br)=\sum_\sigma\sum_j n_j|\phi_j(\br,\sigma)|^2$ is assumed to be strictly positive, for any $\br\in\Omega$ one term must be non-zero and then $\br\in \Omega_{a_j}$ with $j$ the corresponding index. We have thus proved that the family of disjoint open sets $\Omega_a$ covers the whole of $\Omega$. Since $\Omega$ is a connected space, this implies that there can be only one such set! So we have the equality $\Omega=\Omega_a$ for some $a$, which implies that $\{\br\in\Omega\ \vert\ w(\b{r})= a\}=\Omega$ up to a set of measure zero. Hence $w$ is almost-everywhere equal to the constant $a$. Finally, we get $a=0$ after integrating the left-hand side of Eq.~\eqref{eq:unique-V-rep} against $\Psi^*$.
\end{proof}

In Ref.~\onlinecite{Corso-26_ppt}, Carvalho Corso studied the case of states $\Psi\in\cH^1$ in 2D and 3D. This is slightly more involved because $\rho_\Psi$ is then not necessarily continuous. It is then necessary to assume that $\rho_\Psi>0$ \emph{quasi}-everywhere, instead of everywhere, which means that the nodal set $\{\br\in\Omega\ \vert\ \rho_\Psi(\br) = 0\}$ of the density has zero \emph{capacity} (in $\Omega=\R^3$, this means that it cannot support a unit charge with finite Coulomb energy). Furthermore, in Ref.~\onlinecite{Corso-26_ppt} $w$ is assumed to be a form-bounded distribution and not just a function in $L^2_{\rho_\Psi}(\Omega)$.

\subsubsection{Boundedness of the inverse $S_\Psi^{-1}$}
\label{sec:boundedinverseS}

As we have explained, the triviality of the kernel of $S_\Psi$ is not the only interesting property. When we invert the operator we also want the inverse to be a bounded operator, which requires the existence of a gap as in Eq.~\eqref{eq:gap_S}.

For finite-rank wave functions, the following theorem shows that a trivial kernel automatically implies the existence of a gap.

\begin{theorem}[Existence of a gap for $S_\Psi$ with finite-rank wave functions]\label{thm:compact}
Let $\Omega$ be either $\R^d$ or the torus $\bT^d$ with $d\in\{1,2,3\}$. Assume that $\Psi$ has finite rank in the sense that it is a finite linear combination of Slater determinants, that is, it has a finite number $K$ of natural spin-orbitals. Then $S''_\Psi-1$ has rank at most $K^2$ and thus $S''_\Psi$ has at most $K^2$ eigenvalues different from $1$ counting multiplicities. Hence, if $\ker(S''_\Psi)=\{0\}$ in $L^2(\Omega)$, there must exist a $g_S>0$ so that the gap inequality~\eqref{eq:gap_S} holds.
\end{theorem}

The result covers most wave functions encountered in practical quantum-chemistry calculations, which are often finite rank. Since $S''_\Psi-1$ is a finite-rank self-adjoint operator, the spectrum of $S''_\Psi$ consists of $1$ and only finitely many eigenvalues different from $1$, none can accumulate at $0$. Hence if $0$ is not an eigenvalue, it must be at a finite distance from the spectrum and a gap exists. The value of the gap is, of course, the bottom of the spectrum, i.e. $g_S=\min\sigma(S_\Psi'')>0$.

\begin{proof}
Let $\phi_1,...,\phi_K\in L^2(\Omega\times\{\uparrow,\downarrow\};\C)$ be the natural spin-orbitals of $\Psi$ and $n_1,...,n_K>0$ the corresponding occupation numbers. For simplicity, we assume the latter are ordered, $n_1\geq n_2\geq \cdots\geq n_K>0$. Hence $n_K=\min_j n_j$. By Löwdin's theorem\cite{Lowdin-55a}, we can expand $\Psi$ over this basis in the form\cite{Lewin-04a}
\begin{multline}
\Psi(\br_1,\sigma_1,...,\br_N,\sigma_N)\\
=\sum_{1\leq j_1,...,j_N\leq K}c_{j_1,...,j_N}\phi_{j_1}(\br_1,\sigma_1)\cdots\phi_{j_N}(\br_N,\sigma_N)
\end{multline}
where the coefficients $c_{j_1,...,j_N}$ are antisymmetric with respect to permutations of their indices, so that the resulting wave function is fermionic. Then
\begin{equation}
n_j=N\sum_{j_2,...,j_N}|c_{j,j_2,...,j_N}|^2.
\end{equation}
The pair density can be expressed as
\begin{multline}
\rho_{2,\Psi}(\br,\br')=N(N-1)\sum_{\sigma,\sigma'\in\{\uparrow,\downarrow\}}\sum_{\substack{j_1,j_2,k_1,k_2,\\
j_3,...,j_N}}c_{j_1,j_2,j_3,...,j_N}\\
\times c_{k_1,k_2,j_3,...,j_N}^*\phi_{j_1}(\br,\sigma)\phi_{j_2}(\br',\sigma')\phi_{k_1}^*(\br,\sigma)\phi_{k_2}^*(\br',\sigma').
\end{multline}
This proves that 
\begin{equation}
\frac{\rho_{2,\Psi}(\br,\br')}{\sqrt{\rho_\Psi(\br)\rho_\Psi(\br')}}=\sum_{j_1,j_2,k_1,k_2}R_{j_1,j_2,k_1,k_2}F_{j_1,k_1}(\br)F_{j_2,k_2}(\br')
\label{eq:finite_rank_rho2}
\end{equation}
with
\begin{equation}
F_{j,k}(\br):=\sum_{\sigma \in\{\uparrow,\downarrow\}}\frac{\phi_j(\br,\sigma)\phi_k^*(\br,\sigma)}{\sqrt{\rho_\Psi(\br)}}
\end{equation}
and 
\begin{equation}
R_{j_1,j_2,k_1,k_2}:=N(N-1)\!\sum_{j_3,...,j_N}c_{j_1,j_2,j_3,...,j_N} c_{k_1,k_2,j_3,...,j_N}^*.
\end{equation}
Recall from Eq.~\eqref{eq:formula_S2''} that $S''_\Psi-1$ is by definition the operator whose integral kernel is the left-hand side of Eq.~\eqref{eq:finite_rank_rho2}. The formula~\eqref{eq:finite_rank_rho2} thus shows that it is finite rank, with its range included in the space spanned by the functions $F_{j,k}$. There are exactly $K^2$ such functions, hence the statement in the theorem follows. We however still have to justify that the $F_{j,k}$'s are square integrable on $\Omega$. Using the Cauchy-Schwarz inequality 
$\left|\sum_n x_ny_n\right|^2\leq \left(\sum_n|x_n|^2\right)\left(\sum_n|y_n|^2\right)$
for the spin sum, we obtain
\begin{align}
|F_{j,k}(\br)|^2&\leq \dfrac{(\sum_{\sigma\in\{\uparrow,\downarrow\}}|\phi_j(\br,\sigma)|^2)(\sum_{\sigma\in\{\uparrow,\downarrow\}}|\phi_k(\br,\sigma)|^2)}{\rho_\Psi(\br)}
\nonumber\\
&\leq \frac1{n_K}\sum_{\sigma\in\{\uparrow,\downarrow\}}|\phi_j(\br,\sigma)|^2
\end{align}
where we have used that $\sum_{\sigma\in\{\uparrow,\downarrow\}}|\phi_k(\br,\sigma)|^2\leq \rho_\Psi(\br)/n_k\leq \rho_\Psi(\br)/n_K$.
This proves that $F_{j,k}$ is square-integrable, with
\begin{equation}
\int_{\Omega}|F_{j,k}(\br)|^2\,\rd\br\leq \frac1{n_K}
\end{equation}
and concludes the proof of the theorem.
\end{proof}

Let us now discuss the specific case of the physical space $\Omega = \R^3$.
For Coulombic atoms and molecules with $N \geq 2$, exact ground-state wave functions always have infinite rank~\cite{Friesecke-03b,HeaSob-23}, that is, they cannot be expressed as a finite linear combination of Slater determinants. Therefore, Theorem~\ref{thm:compact} cannot be applied to these exact Coulombic ground states. In that case we \emph{conjecture} that  $S''_\Psi-1$ is nevertheless a compact operator. Compact operators can be viewed, roughly speaking, as infinite-dimensional analogues of finite matrices: their spectral structure is essentially discrete away from zero. Therefore, if $S''_\Psi-1$ is a compact operator, it implies that the eigenvalues of $S_\Psi''$ can only accumulate at $1$, and hence there is again a gap when $\ker(S''_\Psi)=\{0\}$. In fact, it seems reasonable to expect that for exact $N$-electron molecular ground-state wave functions $\Psi$ we have 
\begin{equation}
\rho_{2,\Psi}(\br,\br')\leq C\rho_\Psi(\br)\rho_\Psi(\br')
\end{equation}
for some constant $C$. This would imply that 
\begin{equation}\label{eq:HS}
\iint_{\R^3\times\R^3}\frac{\rho_{2,\Psi}(\br,\br')^2}{\rho_\Psi(\br)\rho_\Psi(\br')}\,\rd\br\,\rd\br'\leq C^2 N^2.
\end{equation}
The finiteness of the double integral on the left would mean that  $S''_\Psi-1$ is a Hilbert-Schmidt operator, hence compact  by Ref.~\onlinecite{Lewin-spectral} (Prop.~5.20). Note, however, that $S''_\Psi-1$ is not always compact for an arbitrary wave function $\Psi$. An example of $\Psi$ for which $S_\Psi''-1$ is not compact is provided in Appendix~\ref{app:noncompactness}. 

In the case of the torus $\Omega=\bT^d$, we can in fact prove that $S''_\Psi-1$ is compact under the assumptions of Theorem~\ref{thm:H2}, leading to the desired existence of a gap.

\begin{theorem}[Existence of a gap for $S_\Psi$ on the torus]\label{thm:gap_S_torus}
Assume here that $\Omega=\bT^d$ with $d\in\{1,2,3\}$. Let $\Psi\in \cH^k$ with $k=1$ if $d=1$ and $k=2$ if $d \in \{2,3\}$ and assume that $\rho_\Psi>0$ everywhere. Then $S''_\Psi-1$ is a compact operator on $L^2(\Omega)$ and therefore $S_\Psi$ has a gap as in Eq.~\eqref{eq:gap_S}.
\end{theorem}

\begin{proof}
First, since $\rho_\Psi>0$ everywhere, Theorem~\ref{thm:H2} tells us that $\ker(S''_\Psi)=\{0\}$. The two-electron density matrix $\gamma_{2,\Psi}$ satisfies 
\begin{equation}
\tr\left(((-\nabla_{\br_1}^2)^{k}+(-\nabla_{\br_2}^2)^{k})\gamma_{2,\Psi}\right)<\ii
\end{equation}
with $k$ as in the theorem and where $\tr$ designates here the trace in $L^2((\bT^d\times\{\uparrow,\downarrow\})^2;\C)$. From the periodic Sobolev inequality for bounded operators in dimension $2d$, this implies that the pair density $\rho_{2,\Psi}\in L^p(\bT^d\times\bT^d)$ for all $p<d/(d-k)$.
Our assumption  $k>d/2$ ensures that $d/(d-k)>2$ hence $\rho_{2,\Psi}\in L^2(\bT^d\times\bT^d)$. In particular, $\rho_{2,\Psi}(\br,\br')/\sqrt{\rho(\br)\rho(\br')}$ is also square-integrable, since $\rho$ is continuous and strictly positive on the compact torus, hence bounded away from zero. This proves that this is the kernel of a Hilbert-Schmidt operator. In other words $S''_\Psi-1$ is compact. This proves the existence of a gap $g_S>0$. 
\end{proof}

\begin{figure}[t]
\includegraphics[width=0.9\columnwidth]{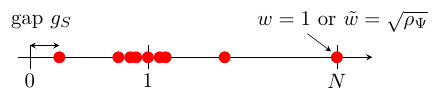}
\caption{Expected typical form of the spectrum of the operator $S_\Psi$, $S_\Psi'$ or $S_\Psi''$. The largest eigenvalue is the electron number $N$, with eigenfunction $w=1$ for $S_\Psi$ or $S_\Psi'$, and $\tilde{w} = \sqrt{\rho_\Psi}$ for $S_\Psi''$. The invertibility of $S_\Psi$ with a bounded inverse requires the spectrum to start at some $g_S>0$. A sufficient condition for $0$ not to be an eigenvalue is that the spin-summed $N$-electron probability density associated with $\Psi$ is strictly positive almost everywhere (Theorem~\ref{thm:unique-continuation}) or $\rho_\Psi>0$ everywhere for $\Psi$ sufficiently regular (Theorem~\ref{thm:H2}). If $\Psi$ is a finite linear combination of Slater determinants, $S''_\Psi-1$ has finite rank and hence the spectrum of $S''_\Psi$ has only finitely many eigenvalues different from $1$ (Theorem~\ref{thm:compact}). More generally, for $\Psi$ sufficiently regular, $S''_\Psi-1$ is compact on the torus (Theorem~\ref{thm:gap_S_torus}) and we expect this to also hold in most cases of physical interest, in which case the spectrum of $S''_\Psi$  has eigenvalues that can only accumulate at $1$. In both cases, this implies the existence of a gap $g_S>0$ if $0$ is not an eigenvalue.
\label{fig:spectrum-S}}
\end{figure}

To conclude this section, we show in Figure~\ref{fig:spectrum-S} the expected typical form of the spectrum of $S_\Psi$, $S_\Psi'$ or $S''_\Psi$.

\subsection{Properties of the operator $K_\rho$}
\label{sec:K}
Next, we turn to the analysis of the operator $K_\rho$.

\subsubsection{Functional setting and basic properties of $K_\rho$}

The operator $K_\rho$ is a somewhat easier object than $S_\Psi$ because it only depends on the density $\rho$ and does not see the detailed correlation structure of $\Psi$. But the precise definition of $K_\rho$ is somewhat more involved  and requires stronger conditions on the density $\rho$. This is a well studied question in mathematics, since those operators appear naturally as generators of Markov semi-groups~\cite{BouHir-91,Eberle-99,FukOshTak-11}.

The quadratic form associated with the operator $K_\rho$ introduced in Eq.~\eqref{eq:K} is, formally, after one integration by parts,
\begin{equation}
\pscal{v,K_\rho v}:=\int_{\Omega}|\nabla v(\br)|^2\rho(\br)\,\dr,
\label{eq:K_rho_quadratic_form}
\end{equation}
for functions $v:\Omega \to \R$ in a function space to be determined. It seems again natural to work in the Hilbert space $L^2_\rho(\Omega)$ with weight $\rho$, under the assumption that $\rho>0$ almost everywhere. Next we discuss under which conditions one can define the operator $K_\rho$ solely based on the quadratic form~\eqref{eq:K_rho_quadratic_form}. Unlike the quadratic form~\eqref{eq:formula_S} associated with $S_\Psi$ which is bounded on the whole space $L^2_{\rho_\Psi}(\Omega)$, the quadratic form~\eqref{eq:K_rho_quadratic_form} involves derivatives and is generally unbounded on $L^2_\rho(\Omega)$. We therefore need to choose a suitable form domain on which the quadratic form~\eqref{eq:K_rho_quadratic_form} is finite to define the operator $K_\rho$. To additionally obtain a self-adjoint operator $K_\rho'$ acting within $L^2_\rho(\Omega)$, the Riesz-Friedrichs theorem (see Ref.~\onlinecite{Lewin-spectral}, Thm.~3.17) requires this form domain to be dense and the quadratic form to be closed in the (real) Hilbert space $L^2_\rho(\Omega)$.

The first difficulty concerns our interpretation of $\nabla v$. We interpret this gradient in the sense of distributions~\cite{LieLos-01} and then require that each component of $\nabla v$ belongs to the space $L^2_\rho(\Omega)$. To avoid difficulties associated with possible singular behavior of $\nabla v$ near the nodal set of $\rho$, it is convenient to not just assume that $\rho>0$ almost everywhere but that
\begin{equation}
\text{$\rho$ is continuous and $\rho>0$ everywhere on $\Omega$.}
\label{eq:rho_continuous}
\end{equation}
As we have seen in Theorem~\ref{thm:H2}, the continuity of $\rho_\Psi$ holds automatically when $\Psi\in \cH^1$ in 1D or $\Psi\in \cH^2$ in 2D and 3D, hence is physically very reasonable. Under the assumption~\eqref{eq:rho_continuous}, every $v \in  L^2_\rho(\Omega)$ is locally integrable and hence defines a distribution, so its distributional gradient $\nabla v$ is well defined. Alternatively, the continuity assumption on $\rho$ can be weakened by requiring $1/\rho$ to be locally integrable, which again ensures that every $v \in  L^2_\rho(\Omega)$ is locally integrable, but we shall not pursue this direction for simplicity.

Assumption~\eqref{eq:rho_continuous} permits us to define the quadratic form~\eqref{eq:K_rho_quadratic_form} on the density-weighted first-order Sobolev space
\begin{equation}
H^1_\rho(\Omega):=\left\{ v\in L^2_\rho(\Omega)\ \mid\ \int_{\Omega} |\nabla v(\br)|^2 \rho(\br) \d\br <\infty\right\},
\label{eq:H1rho}
\end{equation}
where again for $\Omega=\bT^d$ we can think of the cube $(-1/2,1/2)^d$ with periodic boundary conditions on $v$.
It is known that $H^1_\rho(\Omega)$ is a Hilbert space equipped with the norm $\|v\|_{H^1_\rho}:=(\|v\|_{L^2_\rho}^2 + \|\nabla v\|_{L^2_\rho}^2)^{1/2}$, and the quadratic form~\eqref{eq:K_rho_quadratic_form} is closed on $H^1_\rho(\Omega)$.

On the torus $\Omega=\bT^d$, the assumption~\eqref{eq:rho_continuous} implies that $\rho$ is bounded above and bounded away from zero, i.e. $0<c\leq \rho \leq C < \infty$ for some constants $c$ and $C$. As a consequence, we have in fact $L^2_\rho(\bT^d)=L^2_{1/\rho}(\bT^d)=L^2(\bT^d)$ and $H^1_\rho(\bT^d)=H^1(\bT^d)$, with equivalent norms. In particular,  for every $v\in H^1_\rho(\bT^d)$, $\nabla v$ is a square-integrable function on the torus.

On the whole space $\Omega=\R^d$, these spaces generally do not coincide because $\rho$ cannot be uniformly bounded away from zero, because it is integrable. However, the assumption~\eqref{eq:rho_continuous} ensures that $\rho$ is bounded above and bounded away from zero on every compact subset. In particular,
\begin{equation}
\int_{\R^d}|\nabla v(\br)|^2\rho(\br)\,\dr\geq c_R\int_{B(0,R)}|\nabla v(\br)|^2\,\dr
\label{eq:loc_coercive}
\end{equation}
where $c_R$ is the minimum of $\rho$ over the closed ball $\overline{B(0,R)}$, which is attained and therefore strictly positive since $\rho$ is continuous and strictly positive. For every $v\in H^1_\rho(\R^d)$, the bound~\eqref{eq:loc_coercive} implies that $\nabla v$ is a locally square-integrable function.

In both cases, since the quadratic form~\eqref{eq:K_rho_quadratic_form} is bounded on $H^1_\rho(\Omega)$, we can define $K_\rho$ as an operator from the space $H^1_\rho(\Omega)$ to its dual space $H^{-1}_\rho(\Omega)$, with the expression
\begin{equation}
(K_{\rho} v) (\b{r}) :=  - \nabla \cdot (\rho(\b{r}) \, \nabla v(\b{r}) ).
\end{equation}

Like for $S_\Psi$, it is more convenient to represent the same quadratic form $\pscal{v,K_\rho v}$ by an operator $K_\rho'$ acting within $L^2_\rho(\Omega)$. Formally, this is achieved by defining $K_\rho' := \rho^{-1} K_\rho$, or, more explicitly,
\begin{equation}
(K_\rho' v)(\br):=-\frac{1}{\rho(\b{r})} \nabla \cdot (\rho(\b{r}) \, \nabla v(\b{r}) ).
\label{eq:def_K}
\end{equation}
The following theorem gives the conditions under which $K_\rho'$ is well defined as the Friedrichs extension of the differential operator in Eq.~\eqref{eq:def_K}, initially defined on $C^\ii_c(\Omega)$, the space of infinitely differentiable functions with compact support.

\begin{theorem}[Definition of $K_\rho'$]\label{thm:K}
Let $\Omega$ be either $\R^d$ or the torus $\bT^d$ with $d\in\{1,2,3\}$. Assume that the density $\rho\geq0$ with $\int_{\Omega}\rho(\br)\dr=N$
\begin{itemize}
\item is continuous and strictly positive everywhere;
\item satisfies $\sqrt{\rho}\in H^1(\Omega)$.
\end{itemize}
Then the operator $K_\rho':v\mapsto -\rho^{-1}\nabla\cdot(\rho\nabla v)$ defined first on $C^\ii_c(\Omega)$ admits a unique self-adjoint extension in $L^2_\rho(\Omega)$ whose domain is included in the space $H^1_\rho(\Omega)$.
\end{theorem}

\begin{proof}
We write the proof for $\Omega=\R^d$. Note that the operator in Eq.~\eqref{eq:def_K} is well defined and symmetric on $C^\ii_c(\R^d)$. Indeed, for every $v\in C_c^\infty(\R^d)$, we have $\sqrt{\rho}\,K_\rho'v=-\sqrt{\rho}\,\nabla^2 v-2\nabla(\sqrt{\rho})\cdot\nabla v \in L^2(\R^d)$ since $\sqrt{\rho}\in H^1(\R^d)$. Thus, $K_\rho'v\in L_\rho^2(\R^d)$, and symmetry follows by integration by parts. 
The uniqueness of the extension follows immediately from Corollary~3.18 in Ref.~\onlinecite{Lewin-spectral}, using that $C^\ii_c(\R^d)$ is dense in $H^1_{\rho}(\R^d)$ for the associated norm. The latter uses that, on the compact closure of any bounded ball $B$, the continuous and strictly positive function $\rho$ attains a strictly positive minimum and a finite maximum. Thus, there exist constants $0<c_B\leq C_B<\infty$ such that $c_B\leq\rho\leq C_B$ on $B$.
On the torus the proof is easier. One should only be careful that functions in $C^\ii_c(\bT^d)$ really live on the geometric torus and not only on $(-1/2,1/2)^d$ otherwise the boundary is not seen.
\end{proof}

The condition $\sqrt\rho\in H^1(\Omega)$ means physically that the density has a finite von Weizsäcker kinetic energy 
\begin{equation}
\frac12\int_{\Omega}|\nabla\sqrt{\rho(\br)}|^2\,\dr<\ii.
\end{equation}
This is valid whenever $\rho=\rho_\Psi$ with a wave function $\Psi$ having itself a finite kinetic energy, that is, belonging to the Sobolev space $\cH^1$ defined in Eq.~\eqref{eq:def_H1_Psi}. The proof is due to Hoffmann-Ostenhof~\cite{Hof-77} and it is different from the one we gave in the proof of Theorem~\ref{thm:H2}, because $H^1(\Omega)$ is not an algebra when $d\geq2$. 

It is known that the density of a molecular $N$-electron ground state is Lipschitz-continuous and uniformly bounded from above on $\R^3$~\cite{Kato-57,Hof-77}. As mentioned above, strict positivity everywhere is conjectured for molecular Coulombic ground states~\cite{FouHofSor-07}, but remains unproved in general. It is known for non-interacting ground states and, under suitable assumptions, for interacting spinless fermions in one dimension~\cite{Corso-26,Corso-26b}. For more general interactions, strict positivity can fail, as shown in Appendix~\ref{app:counterex-positivity}. The operator $K_\rho$ was studied on bounded domains with Dirichlet boundary conditions in Refs.~\onlinecite{RugPenBau-JPA-09,PenRug-11,Penz-PhD}.

At this point, under the assumptions of Theorem~\ref{thm:K}, we have thus defined the operator $K_\rho'$ as an unbounded self-adjoint operator representing the same quadratic form within $L^2_{\rho}(\Omega)$, i.e. $\pscal{v,K_\rho v} = \pscal{v,K_\rho' v}_{L^2_{\rho}}$. From the definition of the Friedrichs extension, the domain of the unbounded operator $K_\rho'$ is
\begin{align}
\cD(K_\rho')&=\left\{v\in H^1_\rho(\Omega)\ \mid\ \frac{\nabla\cdot(\rho\nabla v)}{\rho}\in L^2_\rho(\Omega)\right\}
\end{align}
where the numerator is understood as a distribution.
We can also work within the usual real Hilbert space $L^2(\Omega)$ by defining the unitarily equivalent operator $K_\rho'' := \rho^{1/2} K_\rho' \rho^{-1/2}$, or, more explicitly,
\begin{equation}
(K''_\rho \tilde{v})(\b{r}) :=-\nabla^2  \tilde{v}(\b{r}) +\frac{\nabla^2\sqrt{\rho(\b{r})}}{\sqrt{\rho(\b{r})}}  \tilde{v}(\b{r})
\label{eq:formulaK''}
\end{equation}
where the Bohm potential $v_\text{B} :={\nabla^2\sqrt\rho}/{\sqrt\rho}$ is \emph{a priori} only a distribution.
We have then  $\langle v,K_\rho' v\rangle_{L^2_{\rho}}=\langle \tilde{v},K_\rho'' \tilde{v}\rangle_{L^2}$ with $\tilde{v} = \sqrt{\rho}v \in L^2(\Omega)$. This transformation thus converts the weighted differential operator $K_\rho'$ into a Schrödinger operator $K_\rho''$. The self-adjointness of Schrödinger operators with a distributional potential like in Eq.~\eqref{eq:formulaK''} is usually not an easy problem, but here everything works through the quadratic form interpretation in $L^2_\rho(\Omega)$. Thus, $K''_\rho$ is an unbounded self-adjoint operator with domain included in $L^2(\Omega)$. The three operators $K_\rho$, $K_\rho'$, and $K_\rho''$ represent the same quadratic form in different function spaces.  The three operators have the same spectrum
\begin{equation}
    \sigma(K_\rho) = \sigma(K_\rho') = \sigma(K_\rho''),
\end{equation}
where again the spectrum of $K_\rho$ has to be understood as the generalized spectrum with weight $\rho$, i.e. the corresponding eigenvalue equation is $K_\rho v = \lambda \rho v$.

Since $K_\rho'$ is a non-negative self-adjoint operator, it has a non-negative spectrum
\begin{equation}
\sigma(K_\rho') \subset [0,+\infty).
\end{equation}
Moreover, $0$ is an eigenvalue of $K_\rho$ or $K_\rho'$ with the constant eigenfunction $v = 1 \in L^2_\rho(\Omega)$. This corresponds to the invariance of the density under adding a constant to the external potential. For $K_\rho''$, the corresponding eigenfunction is $\tilde{v} = \sqrt\rho \in L^2(\Omega)$. Thus, the kernel of $K_\rho$ or $K_\rho'$ contains the constant functions. The question now is whether the kernel of $K_\rho$ or $K_\rho'$ only contains the constant functions, i.e. whether the kernel is trivial in the sense $\ker(K_\rho) = \ker(K_\rho') = \text{span}\{1\}$. If this is the case, then the restriction of $K_\rho'$ to the orthogonal complement of the constant functions in $L_\rho^2(\Omega)$ is injective and therefore admits an inverse on its range. This inverse is defined on the whole orthogonal complement and is bounded if and only if there is a strictly positive gap $g_K$ above $0$,
\begin{equation}
\sigma(K_\rho')\setminus \{0\} \subset [g_K,+\infty), \quad g_K >0.
\end{equation}
In the next subsections, we study these questions.

\subsubsection{Triviality of the kernel of $K_\rho$}
It is easy to verify that under the assumptions of Theorem~\ref{thm:K}, the kernel of $K_\rho$ only contains the constant functions. Hence $0$ is a non-degenerate eigenvalue of $K_\rho$.

\begin{theorem}[Kernel of $K_\rho$]\label{thm:kernelK}
Let $\Omega$ be either $\R^d$ or the torus $\bT^d$ with $d\in\{1,2,3\}$. Assume that $\rho$ satisfies the assumptions of Theorem~\ref{thm:K}. Then any $v\in H^1_\rho(\Omega)$ so that $\int_{\Omega}|\nabla v(\br)|^2\rho(\br)\,\dr=0$ is necessarily constant, i.e. $\ker(K_\rho) = \operatorname{span}\{1\}$.
\end{theorem}

\begin{proof}
On the unit torus $\Omega= \bT^d$, we can use that $\rho\geq c>0$ under our assumptions, hence we obtain 
\begin{equation}
\int_{\bT^d}|\nabla v(\br)|^2\,\dr=\sum_{\bk \in 2\pi\Z^d} |\bk|^2|\widehat{v}(\bk)|^2=0,
\end{equation}
where $\widehat{v}(\bk)$ are the Fourier coefficients of $v$. This implies that all these coefficients are $0$ except from the zero mode, hence $v$ is constant. 

On $\Omega=\R^d$, we deduce from Eq.~\eqref{eq:loc_coercive} that $\nabla v=0$ over any ball $B(0,R)$, hence also in the sense of distributions over $\R^d$. This is enough to deduce that $v$ is constant by Ref.~\onlinecite{LieLos-01} (Thm.~6.11).
\end{proof}

\subsubsection{Boundedness of the inverse $K_\rho^{-1}$}

Parallel to our study of $S_\Psi$, we now ask whether the operator $K_\rho$ possesses a gap $g_K >0$ above its trivial eigenvalue $0$, which is equivalent to
\begin{eqnarray}
\pscal{v,K_\rho v}=\int_{\Omega}|\nabla v(\br)|^2\rho(\br)\,\dr
\geq
g_K\int_{\Omega}|v(\br)|^2\rho(\br)\,\dr,\nn\\
\quad\text{for all }v\in H^1_\rho(\Omega) \; \text{with } \int_{\Omega}v(\br)\rho(\br)\,\dr=0, \;\;\;\;
\label{eq:gap_K}
\end{eqnarray}
and is often called a \emph{Poincaré inequality} in the literature. 

On the torus $\Omega=\bT^d$, a gap always exists since $K_\rho$ is comparable to the periodic Laplacian.

\begin{theorem}[Existence of a gap for $K_\rho$ on the torus]\label{thm:gap_K_torus}
Let $\Omega=\bT^d$ be the torus and assume that $\rho$ is continuous and strictly positive. Then $K_\rho$ has a gap in the sense of Eq.~\eqref{eq:gap_K}.
\end{theorem}

\begin{proof}
Let $v\in H^1(\bT^d)$ be such that $\int_{\bT^d}v(\br)\rho(\br)\,\rd\br=0$ and denote by $\bar v:=\int_{\bT^d}v(\br)\,\rd\br$ its average on the unit torus $\bT^d$. Then, using that the periodic Laplacian has a gap $4\pi^2$, we have
\begin{align}
\pscal{v,K_\rho v}=&\int_{\bT^d}|\nabla (v-\bar v)(\br)|^2\rho(\br)\,\dr\nn\\
\geq &\left(\min_{\bT^d}\rho\right) \int_{\bT^d}|\nabla (v-\bar v)(\br)|^2\,\dr\nn\\
\geq &4\pi^2\left(\min_{\bT^d}\rho\right)\int_{\bT^d}|v(\br)-\bar v|^2\,\dr\nn\\
\geq &4\pi^2\left(\frac{\min_{\bT^d}\rho}{\max_{\bT^d}\rho}\right)\int_{\bT^d}|v(\br)-\bar v|^2\rho(\br)\,\dr.
\label{eq:compare_rho_cnst_K}
\end{align}
Using that $\int_{\bT^d}v(\br)\rho(\br)\,\rd\br=0$ we find
\begin{equation}
\int_{\bT^d}|v(\br)-\bar v|^2\rho(\br)\,\dr=\int_{\bT^d}v(\br)^2\rho(\br)\,\dr+\bar v^2N.
\label{eq:variance_v}
\end{equation}
Neglecting the positive last term, we thus obtain Eq.~\eqref{eq:gap_K} with 
$g_K\geq 4\pi^2(\min_{\bT^d}\rho)/(\max_{\bT^d}\rho)$.
\end{proof}

On the whole space $\Omega=\R^d$, the situation is much more complicated and the decay of $\rho$ at infinity plays a crucial role. This is a very well studied question in mathematics since a positive gap ensures the large-time convergence of Markov semi-groups~\cite{BakGenLed-14}.  One famous result is that there is a gap whenever $\rho$ is comparable to a log-concave function. 

\begin{theorem}[Existence of a gap for $K_\rho$ for densities comparable to log-concave functions]\label{thm:Poincaré}
Let $\Omega=\R^d$ with $d\in\{1,2,3\}$. Assume that $\rho$ is continuous with 
\begin{equation}
\frac{e^{-u(\br)}}{C}\leq \rho(\br)\leq C e^{-u(\br)}
\label{eq:log-concave}
\end{equation}
for some constant $C>0$ and a convex function $u:\R^d\to\R$. Then $K_\rho$ has a gap in the sense of Eq.~\eqref{eq:gap_K}.
\end{theorem}

As typical examples we mention the harmonic case $u(\br)=a|\br|^2$ where $\rho$ is comparable to a Gaussian function, or the case $u(\br)=a|\br|$ where $\rho$ is comparable to the density of the ground state of a hydrogen-like atom. In these two physical examples the existence of a gap follows.

We will just give the general ideas behind Theorem~\ref{thm:Poincaré}. A famous inequality of  Brascamp--Lieb~\cite{BraLie-76b,CarCorLie-13} states that for $\mu(\br)=Z^{-1}e^{-u(\br)}$ with $Z=\int_{\R^d} e^{-u(\br)}\,\rd \br<\ii$ and $u\in C^2(\R^d)$ with a positive-definite Hessian matrix $D^2u(\br)$ at every point, we have
\begin{multline}
\int_{\R^d}\left|v(\br)-\bar{v} \right|^2\mu(\br)\,\dr\\
\leq \int_{\R^d}\nabla v(\br)^T(D^2u(\br))^{-1}\nabla v(\br)\,\mu(\br)\,\rd\br,    
\label{eq:Brascamp-Lieb}
\end{multline}
where $\bar{v} := \int_{\R^d} \mu(\br) v(\br) \d\br$ and $(D^2u(\br))^{-1}$ is the inverse of the Hessian matrix. If $u$ is uniformly strictly convex in the sense that $D^2u(\br)\geq a^2>0$ as a matrix for all $\br$, then the inequality~\eqref{eq:Brascamp-Lieb} implies that we have a gap $g_K \geq a^2>0$ as in Eq.~\eqref{eq:gap_K} for $\rho(\br)=N\mu(\br)$. This includes Gaussian densities for instance, but not the hydrogen-like exponential density corresponding to $u(\br)=a|\br|$, whose Hessian is not strictly positive as a matrix (it has a zero eigenvalue for every $\br\neq0$). For $\rho(\br)=Ne^{-u(\br)}/Z$ with $u$ a general convex function, the existence of a gap as stated in Theorem~\ref{thm:Poincaré} is much more involved and can be read in Refs.~\onlinecite{Bobkov-99,BakBarCatGui-08}. Finally, when $\rho$ is only comparable to $\mu(\br)=Z^{-1}e^{-u(\br)}$ as in Eq.~\eqref{eq:log-concave}, we can reduce it to the case of equality treated in Refs.~\onlinecite{Bobkov-99,BakBarCatGui-08} using the same argument that we used in Eqs.~\eqref{eq:compare_rho_cnst_K} and~\eqref{eq:variance_v}. 
This concludes our description of the general ideas behind Theorem~~\ref{thm:Poincaré}.

Applying Theorem~\ref{thm:Poincaré} to ground states of molecular systems would require precise estimates of the decay of $\rho$ at infinity. Unfortunately, we cannot expect $\rho$ to be comparable everywhere to an exponential $e^{-a |\br|}$ function~\cite{GoGalBae-16}.

Although we cannot prove that ground states satisfy the condition~\eqref{eq:log-concave}, we can show the existence of a gap by another means, using an inequality of Hoffmann-Ostenhof~\cite{Hof-77}.

\begin{theorem}[Existence of a gap for $K_\rho$ for ground-state densities]\label{thm:gapKgroundstate}
Let $\Omega=\R^d$ with $d\in\{1,2,3\}$. 
Let $\rho=\rho_\Psi$ where $\Psi$ is an $N$-electron ground-state wave function in a given spin sector, with external potential $v$ and interaction $v_{\rm \ee}\geq0$. We assume that $v,v_{\ee}\in L^2(\R^d)+L^\ii(\R^d)$ and tend to 0 at infinity. We also assume that $\rho$ is strictly positive everywhere. Finally, we denote by $\Sigma$ the bottom of the essential spectrum of the $N$-electron Hamiltonian $\hat{H}$ and assume that the ground-state energy $E_0<\Sigma$ (both in the fixed spin sector).
Then the essential spectrum, which in particular contains the continuous spectrum, of $K_\rho''$ satisfies
\begin{equation}
\sigma_{\rm ess}(K_\rho'')\subset [2(\Sigma-E_0),+\ii).
\end{equation}
Discrete eigenvalues may occur below this threshold $2(\Sigma-E_0)>0$. Nevertheless, because $0$ is a non-degenerate eigenvalue of $K_\rho''$ lying below the essential spectrum, it is isolated and a strictly positive gap above $0$ exists.
\end{theorem}

\begin{figure}[t]
\includegraphics[width=0.9\columnwidth]{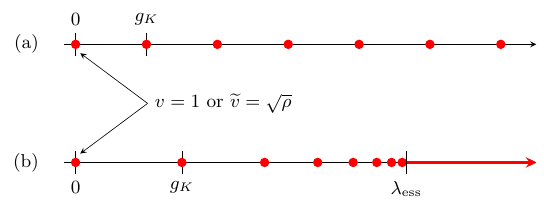}
\caption{
Schematic form of the spectrum of the operator $K_\rho$, $K_\rho'$ or $K_\rho''$ for (a) a Gaussian density on $\R^d$ and (b) an exponential density on $\R^3$. The spectrum is discrete for a Gaussian density while it also contains an essential (continuous) part for an exponential density. Under the assumptions of Theorem~\ref{thm:kernelK}, $0$ is a non-degenerate eigenvalue, with eigenfunction $v=1$ for $K_\rho$ or $K_\rho'$ and $\tilde v=\sqrt{\rho}$ for $K_\rho''$. The invertibility of $K_\rho$ modulo constants with a bounded inverse requires the nonzero spectrum to start at some $g_K>0$. Theorem~\ref{thm:Poincaré} establishes the existence of such a gap for Gaussian or exponential densities, or more generally densities that are comparable to a log-concave function. For exponential densities, or more generally ground-state densities, satisfying the assumptions of Theorem~\ref{thm:gapKgroundstate}, the onset of the essential spectrum $\lambda_{\rm ess}$ satisfies $\lambda_{\rm ess}\geq 2(\Sigma-E_0)$ where $E_0$ and $\Sigma$ are the ground-state energy and the bottom of the essential spectrum of the $N$-electron Hamiltonian.
\label{fig:spectrum-K}}
\end{figure}

The proof of Theorem~\ref{thm:gapKgroundstate} is provided in Appendix~\ref{app:proof_thm:gapKgroundstate}. The schematic form of the spectrum of $K_\rho$, $K_\rho'$ or $K''_\rho$ for a Gaussian density on $\R^d$ and an exponential density on $\R^3$ is illustrated in Figure~\ref{fig:spectrum-K}. For a Gaussian density $\rho$, the operator $K''_\rho$ in Eq.~\eqref{eq:formulaK''} is a harmonic-oscillator Hamiltonian shifted by its ground-state energy, and thus the spectrum is fully discrete. For an exponential density $\rho$, the operator $K''_\rho$ in Eq.~\eqref{eq:formulaK''} is a hydrogen-like atom Hamiltonian shifted by its ground-state energy, and thus the spectrum also contains a continuous spectrum. For invertibility, we focus on the lower part of the spectrum, which is purely discrete. On the torus, the spectrum is entirely discrete.

We close this section with a link between the gaps of $S_\Psi$ that we studied in the previous section and the one of $K_\rho$. 

\begin{theorem}[Link between the three gaps]\label{thm:threegaps}
Let $\Omega=\R^d$ with $d\in\{1,2,3\}$. Assume that $\rho=\rho_\Psi$ where $\Psi$ is a \emph{non-degenerate} $N$-electron ground-state wave function in a given spin sector, with external potential $v$ and interaction $v_{\ee}\geq0$. We assume that $v,v_{\ee}\in L^2(\R^d)+L^\ii(\R^d)$ and tend to 0 at infinity. We also assume that $\rho$ is strictly positive everywhere. Then
\begin{equation}
g_K\geq 2g_H g_S
\end{equation}
where $g_H$ is the gap in the spectrum of the $N$-electron Hamiltonian $\hat{H}$ above the ground-state energy $E_0$ of $\Psi$ in the given spin sector, and $g_S=\min\sigma(S_\Psi)$.
\end{theorem}

Hence, if $g_H >0$, and since $\ker(S_\Psi)=\{0\}$ under our assumptions by Theorem~\ref{thm:H2}, the compactness of $S''_\Psi-1$ conjectured in Section~\ref{sec:boundedinverseS} would automatically imply that both $K_\rho$ and $S_\Psi$ have a gap.
The proof of Theorem~\ref{thm:threegaps} is provided in Appendix~\ref{app:proof_thm:threegaps}.

\subsection{Geometry of the set of density-constrained wave functions}
\label{sec:manifold}

In Section~\ref{sec:geometric_interpretation} we have introduced the tangent and normal spaces at a wave function $\Psi$ for the set $\cM_\rho$ of wave functions having density $\rho=\rho_\Psi$. Our purpose in this section is to explain in which mathematical sense $\cM_\rho$ is a (smooth) manifold. In fact, we will explain that, although the linear tangent-normal decomposition can be established in the $L^2$ space, a manifold structure requires stronger regularity assumptions.

\subsubsection{``Tangent'' and ``normal'' spaces}

Let us first discuss the proper definition of the two spaces $\cT_\Psi\cM_\rho$ and $\cN_\Psi\cM_\rho$ that we have introduced in Section~\ref{sec:geometric_interpretation}, without investigating yet the geometric structure of $\cM_\rho$.
For a one-electron density $\rho \in L^1(\Omega)$ such that $\rho \ge 0$ and  $\int_\Omega \rho(\b{r}) \d\b{r} = N$, we consider the set of states $\Psi\in  \mathcal H$ yielding this density,
\begin{equation}
{\cal M}_\rho := \big\{ \Psi \in   \mathcal H \; | \; \bra{\Psi}\hat{\rho}(\b{r})\ket{\Psi} \!=\!\rho(\b{r}) \mbox{ for a.e. } \b{r} \in \Omega\big\},
\end{equation}
where $\hat{\rho}(\b{r})$ is the density operator and ``a.e."~stands for ``almost every'' in the Lebesgue sense. Under a wave-function variation $\Psi \to \Psi + \eps \delta \Psi$ where $\delta \Psi \in {\cal H}$ and $\eps \in \mathbb{R}$, the density becomes
\begin{multline}
\bra{\Psi+\eps \delta \Psi}\hat{\rho}(\b{r})\ket{\Psi+\eps \delta \Psi} = \bra{\Psi}\hat{\rho}(\b{r})\ket{\Psi} 
\\
+ 2 \eps \Re\bra{\delta \Psi}\hat{\rho}(\b{r})\ket{\Psi}
+\eps^2 \bra{\delta \Psi}\hat{\rho}(\b{r})\ket{\delta \Psi}.
\end{multline}
We then define the corresponding ``tangent space'' at $\Psi \in {\cal M}_\rho$ as the vector space of wave-function variations $\delta \Psi \in {\cal H}$ preserving the density constraint to first order,
\begin{align}  
{\cal T}_{\Psi} {\cal M}_{\rho} 
& 
 :=\Big\{  \delta \Psi \in {\cal H} \; |\;\!  \Re\bra{\delta \Psi}\hat{\rho}(\b{r})\ket{\Psi} \!=\!0 \mbox{ for a.e. } \b{r} \in \Omega\Big\}.
\label{eq:TMOmega}
\end{align}
At this stage, ``tangent'' only means that the density constraint is preserved to first order, and not that these variations are tangent vectors to a smooth manifold. The corresponding ``normal space'' is
\begin{equation}
 \mathcal N_{\Psi}\mathcal M_{\rho}:= \mathcal T_{\Psi}\mathcal M_{\rho}^{\perp_\R},
\end{equation}
 where $A^{\perp_\R}$ denotes the orthogonal complement of a set $A \subset \mathcal H$ for the real inner product ${\rm Re}\braket{\,\cdot\,}{\,\cdot\,}$. From the Cauchy–Schwarz inequality
\begin{equation}
\int_{\Omega} |\Re\bra{\delta \Psi}\hat{\rho}(\b{r})\ket{\Psi}| \,\d\b{r} \leq N \|\Psi \|_{\cal H}  \|\delta \Psi \|_{\cal H},
\label{eq:inequalityL1}
\end{equation}
we can deduce that the tangent space is closed, which implies that the Hilbert space ${\cal H}$ can be decomposed into the (real) direct sum
\begin{equation}
{\cal H} = {\cal T}_{\Psi} {\cal M}_{\rho} \oplus_\R \mathcal N_{\Psi}\mathcal M_{\rho}.
\label{eq:Hdirectsum}
\end{equation}
The two spaces are orthogonal for the real Hilbert structure, not necessarily with respect to the complex inner product.

We now justify the explicit expression of the normal space found in Section~\ref{sec:geometric_interpretation}. The inequality~\eqref{eq:inequalityL1} means that the first-order density variation $\b{r} \mapsto  \Re\bra{\delta \Psi}\hat{\rho}(\b{r})\ket{\Psi}$ is a function in $L^1(\Omega)$. Therefore, by duality, the first-order density variation vanishes if and only if its integral over all functions $w$ in the dual space $L^\infty(\Omega)$ (the space of real-valued functions bounded almost everywhere) vanishes,
\begin{multline}
\Re\bra{\delta \Psi}\hat{\rho}(\b{r})\ket{\Psi} = 0 \mbox{ for a.e. } \b{r} \in \Omega\\
\Leftrightarrow \forall w \in L^\infty(\Omega), \;  \int_\Omega w(\br) \Re\bra{\delta \Psi}\hat{\rho}(\b{r})\ket{\Psi} \d\b{r} = 0.
\end{multline}
Therefore, for a wave-function variation $\delta\Psi \in \cal H$, we have the equivalence
 \begin{align*}
 & \delta \Psi \in {\cal T}_{\Psi} {\cal M}_{\rho} \\
 & \quad \Leftrightarrow  \forall w \in L^\infty(\Omega), \;  \int_\Omega w(\br) \Re\bra{\delta \Psi}\hat{\rho}(\b{r})\ket{\Psi} \, \d\b{r} = 0 \\
 & \quad \Leftrightarrow  \forall w \in L^\infty(\Omega), \;  \Re  \pscal{\delta \Psi\bigg|\int_\Omega \d\b{r} \, w(\br) \hat{\rho}(\b{r}) \Psi} = 0, 
 \end{align*}
leading to the characterization of the tangent space
\begin{equation}
{\cal T}_{\Psi} {\cal M}_{\rho} = \left\{ \int_\Omega \d\b{r} \, w(\br) \hat{\rho}(\b{r}) \Psi, \, w \in L^\infty(\Omega) \right\}^{\perp_\R},
\end{equation}
 and thus to the characterization of the normal space
\begin{align}
\mathcal N_{\Psi}\mathcal M_{\rho} &= \left\{ \int_\Omega \d\b{r} \, w(\br) \hat{\rho}(\b{r}) \Psi, \, w \in L^\infty(\Omega) \right\}^{\perp_\R\perp_\R}
\nonumber\\
& = \overline{ \left\{ \int_\Omega \d\b{r} \, w(\br) \hat{\rho}(\b{r}) \Psi, \, w \in L^\infty(\Omega) \right\}},
\label{eq:NPsiMrhoclosure}
\end{align}
where the overline denotes the closure in ${\cal H}$. This means that any vector $\delta \Psi_\perp \in \mathcal N_{\Psi}\mathcal M_{\rho}$ can be written as $\delta \Psi_\perp = \lim_{n \to \infty} \int_\Omega \d\b{r} \, w_n(\br) \hat{\rho}(\b{r}) \Psi$ (limit in the sense of the ${\cal H}$-norm) where $(w_n)$ is a sequence in $L^\infty(\Omega)$.

If we assume that the density $\rho$ is strictly positive almost everywhere, we can characterize the normal space with the density-weighted $L^2$ space [Eq.~\eqref{eq:L2rho}]
\begin{equation}
L^2_{\rho}(\Omega):=\left\{f:\Omega\to\R\ \mid \int_{\Omega}f(\br)^2 \rho(\br)\,\dr<\ii\right\}.
\label{eq:L2Omega}
\end{equation}
Indeed, the space $L^\infty(\Omega)$ is dense in $L^2_{\rho}(\Omega)$ and, due to the inequality~\eqref{eq:S_bounded},
\begin{equation}
\left\| \int_\Omega \d \b{r} \, w(\br) \hat{\rho}(\b{r}) \Psi \right\|_{\cal H}^2 = 
\pscal{w,S_\Psi w}\leq N \| w \|_{L^2_{\rho}}^2,
\label{eq:inequality1}
\end{equation}
the map $w \mapsto \int_\Omega \d \b{r} \, w(\br) \hat{\rho}(\b{r}) \Psi$ is continuous from $L^2_{\rho}(\Omega)$ to ${\cal H}$. This implies that the normal space can also be written as
\begin{align}
\mathcal N_{\Psi}\mathcal M_{\rho} &= \overline{ \left\{ \int_\Omega \d\b{r} \, w(\br) \hat{\rho}(\b{r}) \Psi, \, w \in L^2_\rho(\Omega) \right\}}.
\label{eq:NPsiMrhoclosure2}
\end{align}
We can thus take $w$ in the larger space $L^2_\rho(\Omega)$ instead of $L^\ii(\Omega)$. 
If we further assume that the operator $S_\Psi$ is boundedly invertible, i.e. we have the gap inequality~\eqref{eq:gap_S}
\begin{equation}
g_S \| w \|_{L^2_{\rho}}^2 \leq \left\| \int_\Omega \d \b{r} \, w(\br) \hat{\rho}(\b{r}) \Psi \right\|_{\cal H}^2 = 
\pscal{w,S_\Psi w},
\end{equation}
with $g_S >0$, then together with Eq.~\eqref{eq:inequality1} this implies that the range of the map $w \mapsto \int_\Omega \d \b{r} \, w(\br) \hat{\rho}(\b{r}) \Psi$ is closed, and thus the closure in Eq.~\eqref{eq:NPsiMrhoclosure2} can be removed and we arrive at the simple characterization of the normal space
\begin{align}
\mathcal N_{\Psi}\mathcal M_{\rho} &= \left\{ \int_\Omega \d\b{r} \, w(\br) \hat{\rho}(\b{r}) \Psi, \, w \in L^2_\rho(\Omega) \right\}
\label{eq:NPsiMrhofinal}
\end{align}
that we had in Section~\ref{sec:geometric_interpretation} but now with the space for $w$ specified. 

Note that $\Psi\in\cN_\Psi\cM_\rho$ which can be seen by choosing the constant potential $w=1/N$ in Eq.~\eqref{eq:NPsiMrhofinal}, but $i\Psi\in\cT_\Psi\cM_\rho$ due to the real part in Eq.~\eqref{eq:TMOmega}. This emphasizes the real-linearity of these two spaces. The fact that $i\Psi$ belongs to the tangent space corresponds to the invariance of the density under multiplication of $\Psi$ by a global \emph{complex} phase factor $e^{i \eps}$ (i.e., $U(1)$ gauge symmetry). By contrast, multiplying $\Psi$ by a real number $1+\eps$ changes its normalization and therefore takes it outside the set $\cM_\rho$.

The direct-sum decomposition~\eqref{eq:Hdirectsum} and the expression of $\cN_{\Psi}\cM_{\rho}$ in Eq.~\eqref{eq:NPsiMrhofinal} mean that any vector $\delta \Psi\in\cH$ can be written as
\begin{equation}
\delta\Psi=\underbrace{\delta\Psi_\parallel}_{\in\cT_\Psi\cM_\rho}+\underbrace{\int_\Omega \d\b{r} \, w_\perp(\br) \hat{\rho}(\b{r}) \Psi}_{=\delta\Psi_\perp\in\cN_\Psi\cM_\rho}
\label{eq:decomp_ortho_deltaPsi}
\end{equation}
with a unique $w_\perp \in L^2_\rho(\Omega)$. Since the tangent component $\delta\Psi_\parallel$ does not change the density to first order, $\Re\bra{\delta\Psi_\parallel}\hat{\rho}(\br)\ket{\Psi}=0$, the first-order density variation is entirely determined by the normal component $\delta\Psi_\perp$, i.e. $\Re\bra{\delta\Psi}\hat{\rho}(\br)\ket{\Psi}=\Re\bra{\delta\Psi_\perp}\hat{\rho}(\br)\ket{\Psi}$. Inserting the expression of $\delta\Psi_\perp$ in terms of $w_\perp$ leads to the linear equation for $w_\perp$
\begin{equation}
S_\Psi \, w_\perp = \rho_{\Psi,\delta \Psi},
\label{eq:Sw_deltarho}
\end{equation}
where we have introduced the real mixed density $\rho_{\Psi_1,\Psi_2}(\br):=\Re\bra{\Psi_2}\hat\rho(\br)\ket{\Psi_1}$. Note that, in Section~\ref{sec:geometric_interpretation}, essentially the same equation appeared in the form of Eq.~\eqref{eq:Swperp}. Hence, $w_\perp$ is given by
\begin{equation}
w_\perp= S_\Psi^{-1} \, \rho_{\Psi,\delta \Psi}.
\label{eq:w_for_any_deltaPsi}
\end{equation}
Note that by Cauchy--Schwarz, we have
\begin{equation}
    |\rho_{\Psi,\delta \Psi}(\br)|^2\leq \rho_{\delta\Psi}(\br)\rho_\Psi(\br) \quad \text{for a.e.}\;  \br \in \Omega
    \label{eq:CS_w}
\end{equation}
and thus
\begin{equation}
    \int_\Omega \frac{|\rho_{\Psi,\delta \Psi}(\br)|^2}{\rho_\Psi(\br)} \d\br \leq \int_\Omega \rho_{\delta\Psi}(\br) \d\br = N \| \delta \Psi \|_{\cH}^2,
 \label{}
\end{equation}
showing that $\rho_{\Psi,\delta \Psi} \in L^2_{1/\rho_\Psi}$. Under the assumed bounded invertibility of $S_\Psi$, this implies that $\rho_{\Psi,\delta \Psi}$ is in the range of $S_\Psi$, as required for solving Eq.~\eqref{eq:Sw_deltarho}. The invertibility of $S_\Psi$ with the gap $g_S>0$ is crucial to ensure both the expression~\eqref{eq:NPsiMrhofinal} for $\cN_\Psi\cM_\rho$ and the existence and uniqueness of $w_\perp$ in Eq.~\eqref{eq:w_for_any_deltaPsi}.

\subsubsection{Manifold structure}

\begin{figure}[t]
\includegraphics[scale=1,angle=0]{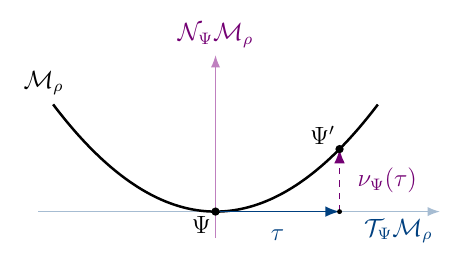}
        \caption{Schematic local parametrization of a smooth density-constrained manifold $\cM_\rho$ near $\Psi$.
Each sufficiently small tangent displacement $\tau\in\cT_\Psi\cM_\rho$ determines a unique small normal correction
$\nu_\Psi(\tau)\in\cN_\Psi\cM_\rho$ such that $\Psi'=\Psi+\tau+\nu_\Psi(\tau)\in\cM_\rho$.
The correction depends smoothly on $\tau$ and satisfies $\nu_\Psi(0)=0$ and $D\nu_\Psi(0)=0$.
This parametrization can fail if the normal correction $\nu_\Psi(\tau)$ does not exist, is not unique, or does not depend smoothly on the tangent displacement.
Failure of such a smooth local parametrization in every neighborhood of $\Psi$ rules out a smooth submanifold structure
with tangent space $\cT_\Psi\cM_\rho$ at $\Psi$.}
\label{fig:manifold_local_graph}
\end{figure}

The direct-sum decomposition in Eq.~\eqref{eq:decomp_ortho_deltaPsi} and the explicit expression of the normal space in Eq.~\eqref{eq:NPsiMrhofinal} 
provide the \emph{linear} geometric structure used in our interpretation of TDDFT. A stronger question is whether the set $\cM_\rho$ itself is locally a smooth manifold, giving a precise mathematical meaning to
the local geometric picture in Figs.~\ref{fig:manifold} and~\ref{fig:geometric_principle}. Answering this question requires control of the \emph{nonlinear} density constraint, beyond its first-order variation.
Standard textbook treatments of manifolds can be found in Refs.~\onlinecite{Lee-12,AbrMarRat-88}. We give here a concrete discussion specific to our setting.

The set $\cM_\rho$ being a $C^k$ manifold with tangent space $\cT_\Psi\cM_\rho$ in the neighborhood of a given $\Psi\in\cM_\rho$ means that it is given there by a deformation of class $C^k$ of the tangent space $\cT_\Psi\cM_\rho$, with an invertible map called a \emph{local parametrization} (or the inverse of a local chart). To make this precise, let us consider another state $\Psi' \in \cH$. From Eq.~\eqref{eq:Hdirectsum}, it can be uniquely written as
\begin{equation}
\Psi'=\Psi+\tau+\nu,
 \qquad
 \tau\in\cT_\Psi\cM_\rho,
 \quad
 \nu\in\cN_\Psi\cM_\rho,
 \label{eq:decomp_T_N_Psi}
\end{equation}
with $\tau$ and $\nu$ the (real) projections of $\Psi'-\Psi$ onto $\cT_\Psi\cM_\rho$ and its orthogonal $\cN_\Psi\cM_\rho$. The set $\cM_\rho$ has a manifold structure in the neighborhood of $\Psi\in\cM_\rho$ if, whenever $\Psi' \in\cM_\rho$ is close enough to $\Psi$, $\nu$ is uniquely determined by $\tau$ via a smooth function $\tau \mapsto \nu_\Psi(\tau)$. 
More precisely, we thus need to find neighborhoods $U$ and $V$ of zero in the tangent and normal spaces, respectively, and a $C^k$ map $\nu_\Psi:U\to V$ such that every $\Psi'\in\cM_\rho$ close enough to $\Psi\in\cM_\rho$ can be uniquely written in the form
\begin{equation}
\Psi'=\Psi+\tau+\nu_\Psi(\tau),
\label{eq:local_chart}
\end{equation}
and, conversely, for every $\tau\in U$, the state $\Psi+\tau+\nu_\Psi(\tau)$ belongs to $\cM_\rho$. The local parametrization is the map $\tau\mapsto \Psi+\tau+\nu_\Psi(\tau)$. The regularity of $\nu_\Psi$ as a function of $\tau$ is the regularity of the manifold. This function must satisfy
\begin{equation}
 \nu_\Psi(0)=0,\qquad D\nu_\Psi(0)=0,
    \label{eq:condition_local_chart}
\end{equation}
where $D$ is the (infinite-dimensional) differential, because tangent variations already preserve the density to first order: the normal correction has no first-order term. Figure~\ref{fig:manifold_local_graph} illustrates the idea of a local parametrization of $\cM_\rho$.

The function $\nu_\Psi(\tau)$ is implicitly defined by the nonlinear density-constraint equation
\begin{equation}
\label{eq:equal-densities}
    \rho_{\Psi+\tau+\nu}=\rho.
\end{equation}
Using again the notation $\rho_{\Psi_1,\Psi_2}(\br):=\Re\bra{\Psi_2}\hat\rho(\br)\ket{\Psi_1}$, and the fact that the tangent component does not contribute to the first-order density change $\rho_{\Psi,\tau+\nu}=\rho_{\Psi,\nu}$, the density-constraint equation~\eqref{eq:equal-densities} can be rewritten as
\begin{equation}
2\rho_{\Psi,\nu} +\rho_{\tau+\nu}=0.
\end{equation}
The first-order density change $2\rho_{\Psi,\nu}$ must be compensated by the quadratic density change $\rho_{\tau+\nu} = \rho_\tau + \rho_\nu + 2\rho_{\tau,\nu}$. Another view of it is that the density error $\rho_\tau$ produced by the tangent displacement must be cancelled out by the normal displacement.

Assuming that $S_\Psi$ has a gap, we know from the discussion of the previous subsection that $\cN_\Psi\cM_\rho$ is given by Eq.~\eqref{eq:NPsiMrhofinal} and are thus looking for a $w\in L^2_\rho(\Omega)$ solving the nonlinear equation
\begin{equation}
2S_\Psi w =-\rho_{\tau+\nu},\qquad \nu=\int_\Omega \d\br \, w(\br)\hat\rho(\br)\,\Psi.
\label{eq:manifold_eq_to_be_solved}
\end{equation}
This corresponds to Eq.~\eqref{eq:Sw_deltarho} with $\delta\Psi = \tau + \nu$ since by~\eqref{eq:equal-densities} we have $-\rho_{\tau+\nu} = 2\rho_{\Psi,\tau+\nu} = 2\rho_{\Psi,\delta \Psi}$, but now in Eq.~\eqref{eq:manifold_eq_to_be_solved} we view the right-hand side as depending on the unknown $w$.
The standard way to solve the nonlinear equation~\eqref{eq:manifold_eq_to_be_solved} is by an iterative procedure called a fixed-point iteration in the mathematical literature. Namely we define recursively the function $w_n$ by
\begin{equation}
2S_\Psi w_{n+1} =-\rho_{\tau+\nu_n}
\label{eq:iteration_w_n}
\end{equation}
with, of course, $\nu_n:=\int_\Omega \d\br \, w_n(\br)\hat\rho(\br)\,\Psi$ and $w_0=0$. We then hope that the sequence converges to the desired solution, under the assumption that $\tau$ is small enough. This is essentially the content of the Banach fixed point or implicit function theorem. To make this program work we need first to ensure that the left-hand side of Eq.~\eqref{eq:iteration_w_n} defines an invertible linear map. This is where the invertibility of $S_\Psi$ from $L^2_\rho(\Omega)$ to its dual $L^2_{1/\rho}(\Omega)$ comes in. But the problem is that the right-hand side of Eq.~\eqref{eq:manifold_eq_to_be_solved} is not necessarily in the right space $L^2_{1/\rho}(\Omega)$ where $S_\Psi$ takes its values. Unlike the mixed density in Eq.~\eqref{eq:CS_w}, the quadratic density $\rho_{\tau+\nu}$ in Eq.~\eqref{eq:manifold_eq_to_be_solved} does not admit the same Cauchy--Schwarz estimate with a factor $\rho_\Psi$.
For instance, for $n=0$, the density $\rho_\tau$ is in $L^2_{1/\rho}(\Omega)$ if and only if
\begin{equation}
\int_{\Omega}\frac{\rho_\tau(\br)^2}{\rho(\br)}\,\rd\br<\ii.
\end{equation}
Even if we work on a compact set and assume that $\rho>0$, this is not true in general because we only assume $\tau\in \cT_\Psi \cM_\rho \subset \cH$, which tells us at best that $\int_\Omega\rho_\tau(\br)\,\rd\br<\ii$ without the square.
The variations in the tangent plane are too general and can provide wild densities that are hard to compensate with a normal component representable by a function $w\in L^2_\rho(\Omega)$, that is, with a behavior related to the given density $\rho$. 

These observations obstruct the proposed construction of a local parametrization of $\cM_\rho$ if we work with $L^2$ spaces.
The $L^2$ topology is too weak to control the density variations in a good manner. In Appendix~\ref{sec:not_a_manifold} we give an explicit counterexample on the torus showing that the fixed-density set $\cM_\rho$ can indeed fail to be a smooth submanifold of $\cH$.
Appendix~\ref{app:noncompactness} also shows that  changes of $\Psi$ that are arbitrarily small in the $L^2$ norm, even at fixed density, can destroy the spectral gap of $S_\Psi$.
This is another manifestation of the fact that $L^2$ is not really a good space for our problem.

\subsubsection{Making $\cM_\rho$ a manifold on the torus}

We finally provide an appropriate functional setting for which $\cM_\rho$ is a manifold in the mathematical sense. We have two difficulties to face. The first is that we need to insert more regularity of the wave functions in order to better control the densities. We cannot just work in $L^2$. The second is that the space $L^2_\rho(\R^d)$ depends on the behavior of $\rho$ at infinity. To avoid this second difficulty and only address the first, we will work on the torus $\Omega=\bT^d$ and assume that $\rho$ is continuous with $\rho>0$ everywhere. Then $L^2_\rho(\bT^d)=L^2_{1/\rho}(\bT^d)=L^2(\bT^d)$ with equivalent norms.

Recall the $N$-electron Sobolev spaces $\cH^1$ and $\cH^2$ introduced in Eqs.~\eqref{eq:def_H1_Psi} and~\eqref{eq:def_H2,Psi}, respectively. By Theorem~\ref{thm:H2} we have $\rho_\Psi\in H^k(\bT^d)$ for any $\Psi\in\cH^k$, with $k=1$ for $d=1$, and $k=2$ for $d\in\{2,3\}$. We use the notation
\begin{equation}
    \cM_\rho^k:=\cM_\rho\cap \cH^k
    \label{eq:manifold_Sobolev}
\end{equation}
for the restriction of our fixed-density set $\cM_\rho$ to the wave functions in $\cH^k$. 

\begin{theorem}[Manifold structure in Sobolev spaces on the torus]\label{thm:manifold}
Let $\Omega=\bT^d$ be the torus in dimension $d\in\{1,2,3\}$. 
Let $k=1$ for $d=1$ and $k=2$ for $d\in\{2,3\}$. Let $\rho\in H^k(\bT^d)$ be any real continuous function, with $\rho>0$ everywhere and $\int_{\bT^d}\rho(\br)\,\rd\br=N$. 
Then the set $\cM^k_\rho$ is 
a $C^\ii$ real submanifold of the (real) Hilbert space $\cH^k$. At any $\Psi\in\cM_\rho^k$, the tangent space is given by 
\begin{align}
\cT_\Psi\cM_\rho^k &=\cT_\Psi\cM_\rho\cap \cH^k \\
&=\left\{\tau\in\cH^k\ \mid\ \Re\pscal{\tau|\hat{\rho}(\br)|\Psi} = 0, \; \forall \b{r}\in\bT^d\right\}.\nonumber
\end{align}
We have the (real) direct sum
\begin{equation}
\cH^k=\cT_\Psi\cM^k_\rho\oplus_\R\cN_\Psi\cM^k_\rho,
\label{eq:direct_sum}
\end{equation}
where the normal space is given by
\begin{align}
\cN_\Psi\cM^k_\rho&=\cN_\Psi\cM_\rho \cap \cH^k
\nonumber\\
&=\left\{\int_{\bT^d} \operatorname{d}\!\br \, w(\br) \hat{\rho}(\br) \Psi, \; w\in H^k(\bT^d)\right\}.
\end{align}
The spaces $\cT_\Psi\cM^k_\rho$ and $\cN_\Psi\cM^k_\rho$ are orthogonal with respect to the real inner product of $\cH$ but not necessarily with respect to the real Sobolev
inner product of $\cH^k$.
\end{theorem}

Let us clarify the content of this result.
For $\Psi\in\cM_\rho^k$, recall from Eq.~\eqref{eq:decomp_T_N_Psi} that any perturbation $\eta\in\cH^k$ can be written in the form $\eta=\tau+\nu$ with $\tau\in\cT_\Psi\cM_\rho$ and $\nu\in\cN_\Psi\cM_\rho$. By Theorem~\ref{thm:gap_S_torus} we know that $S_\Psi$ is invertible, which implies from Eqs.~\eqref{eq:NPsiMrhofinal} and~\eqref{eq:Sw_deltarho} that $\nu=\int_{\bT^d}\d\br\, w(\br)\widehat{\rho}(\br) \;\Psi$ with $w=S_\Psi^{-1}\rho_{\Psi,\eta}\in L^2(\bT^d)$. The statement~\eqref{eq:direct_sum} means that the two vectors $\tau$ and $\nu$ are in fact in $\cH^k$ and that $w\in H^k(\bT^d)$. 

The proof of Theorem~\ref{thm:manifold} is provided in Appendix~\ref{app:proof_thm:manifold}. The key step is to show that $S_\Psi^{-1}$ is not only bounded on $L^2(\bT^d)$, it is also bounded as an operator from $H^k(\bT^d)$ to itself. Since $\rho_{\Psi,\eta}\in H^k(\bT^d)$ because $\Psi,\eta\in\cH^k$, this is how we get that  $w\in H^k(\bT^d)$. Then we infer $\nu\in\cH^k$ and finally $\tau=\eta-\nu \in\cH^k$ as well. This general structure is essential and allows us to apply the implicit function theorem to solve the nonlinear equation~\eqref{eq:manifold_eq_to_be_solved} for $w \in H^k(\bT^d)$ as a function of the tangent displacement $\tau$.

Thus, in this stronger topology, every sufficiently small tangent displacement admits a unique small normal correction that restores the density exactly, and this correction depends smoothly on the displacement.

\subsection{Summary of the mathematical analysis and consequences for TDDFT}
\label{sec:summary}

Our mathematical results about the operators $S_\Psi$ and $K_\rho$, and about the fixed-density set $\cM_\rho$, clarify when the different constructions underlying TDDFT are expected to be well posed, at a given time, with the instantaneous wave function $\Psi$ and density $\rho=\rho_\Psi$.

The general geometric interpretation of TDDFT in Section~\ref{sec:geometric_interpretation} first relies on the decomposition of the wave-function velocity $\partial_t \Psi$ into tangent and normal components [Eq.~\eqref{eq:dtPsidecomp}], which is always possible [Eq.~\eqref{eq:Hdirectsum}]. Whether we can write the normal component of $\partial_t \Psi$ in terms of a local potential $w_\perp$ depends on the invertibility of the operator $S_\Psi$ [Eq.~\eqref{eq:Swperp}]. The natural functional setting for studying $S_\Psi$ requires $\rho>0$ almost everywhere. Uniqueness of the potential $w_\perp$ requires that $0$ is not an eigenvalue of $S_\Psi$, i.e. $\ker(S_\Psi)=\{0\}$. This is expected quite generally for physically relevant wave functions $\Psi$ and is guaranteed, for instance, when $\Psi\neq0$ almost everywhere (Theorem~\ref{thm:unique-continuation}) or $\Psi$ is sufficiently regular with $\rho>0$ everywhere (Theorem~\ref{thm:H2}). For the potential $w_\perp$ to exist and depend continuously on the prescribed density velocity $\partial_t \rho$, we need in addition that $S_\Psi$ has a positive spectral gap $g_S$ above $0$, which ensures that the inverse of $S_\Psi$ is a bounded operator. This prevents small variations of $\partial_t \rho$ from producing arbitrarily large variations of $w_\perp$. Provided that $0$ is not an eigenvalue of $S_\Psi$, we can prove the existence of such a gap for finite-rank wave functions $\Psi$, e.g. finite linear combinations of Slater determinants (Theorem~\ref{thm:compact}), and expect a gap also in many infinite-rank cases of physical interest. On the torus, i.e. a box with periodic boundary conditions, such a gap always exists for $\Psi$ sufficiently regular and $\rho>0$ everywhere (Theorem~\ref{thm:gap_S_torus}). Note, however, that arbitrarily small changes of $\Psi$ in the $L^2$ norm can change widely the spectrum of $S_\Psi$ and destroy the spectral gap, even at fixed density (Appendix~\ref{app:noncompactness}). Thus, continuity of the wave-function evolution in $L^2$ alone does not guarantee that an initially positive gap persists.

Geometric TDDFT in Section~\ref{sec:geometricTDDFT} uses the same operator $S_\Psi$ to determine the density-enforcing local geometric potential $w$, through inversion of $S_\Psi$ in the modified continuity equation [Eq.~\eqref{eq:modconteq}]. Therefore, under the same conditions for which $S_\Psi$ is boundedly invertible, $w$ is expected to be uniquely and stably determined by the left-hand side of the modified continuity equation, namely $\partial_t \rho + \nabla \cdot \b{j}_{\Psi}$. In this sense, the mathematical conditions required for this instantaneous determination of $w$ are the same as those required for the local-potential representation of the normal component of $\partial_t \Psi$ in Section~\ref{sec:geometric_interpretation}.

Variational TDDFT in Section~\ref{sec:standardTDDFT} relies on the elliptic operator $K_\rho$ which appears in the force-balance equation [Eq.~\eqref{eq:divlocforce}], the density-enforcing local potential $v$ being obtained by inverting $K_\rho$. The natural functional setting for studying $K_\rho$ requires stronger assumptions on the density than for $S_\Psi$, for instance $\rho$ be continuous and $\rho >0$ everywhere. Uniqueness of the potential $v$, up to the usual freedom of adding a purely time-dependent constant, requires that constant functions be the only eigenfunctions with eigenvalue $0$ of $K_\rho$, i.e. $\ker(K_\rho)=\operatorname{span}\{1\}$. Under the assumptions considered here, this is indeed the case (Theorem~\ref{thm:kernelK}). For a stable determination of the potential $v$, one needs in addition that $K_\rho$ has a positive spectral gap $g_K$ above $0$, which ensures that the inverse of $K_\rho$ modulo constants is a bounded operator. This prevents small variations of the source term $q_\Psi- \partial_t^2 \rho$ in the force-balance equation from producing arbitrarily large variations of the potential $v$. On the torus, we can prove the existence of such a gap (Theorem~\ref{thm:gap_K_torus}). On the whole space, a gap is not guaranteed for an arbitrary density, but we prove that there is a gap for a broad class of log-concave densities (Theorem~\ref{thm:Poincaré}) and for physically relevant ground-state densities under suitable assumptions (Theorem~\ref{thm:gapKgroundstate}). This latter result is particularly relevant to the common TDDFT situation in which the time evolution starts from a ground state: the required gap is then present at the initial time and may be expected to persist, at least for sufficiently short times under a regular evolution, although this persistence is not proved here.

Overall, the basic functional assumptions entering geometric TDDFT appear less restrictive than those entering variational TDDFT. In particular, the functional setting of $S_\Psi$ only requires $\rho>0$ almost everywhere, whereas the present construction of $K_\rho$ requires the stronger condition that $\rho$ be continuous and $\rho>0$ everywhere, in addition to further regularity assumptions. This does not, however, imply a simple hierarchy between the corresponding spectral gap conditions. In fact, we have also seen that for non-degenerate ground states the two gaps are directly related through $g_K\geq 2g_Hg_S$ where $g_H$ is the gap of the many-electron Hamiltonian (Theorem~\ref{thm:threegaps}). Consequently, if $S_\Psi$ has a bounded inverse and if $g_H>0$, then $K_\rho$ also has a bounded inverse modulo constants. Thus, at the initial time of an evolution starting from such a ground state, and provided all the technical assumptions are satisfied, the spectral condition ensuring stable inversion in geometric TDDFT is sufficient to ensure the corresponding spectral condition in variational TDDFT. Away from this ground-state setting, however, no such general implication between the two gap conditions has been established.

Finally, we have discussed the rigorous geometric properties of the set $\cM_\rho$ of wave functions having a prescribed density $\rho$. This set is generally not a smooth manifold in the sense of differential geometry when distances between wave functions are measured in the $L^2$ norm. Indeed, this norm allows large changes of a wave function on sufficiently small spatial regions to count as small variations, even when these changes preserve the density exactly (Appendix~\ref{sec:not_a_manifold}). This does not prevent the decomposition of the wave-function velocity into tangent and normal components [Eq.~\eqref{eq:dtPsidecomp}], nor the construction of variational and geometric TDDFT, but it obstructs the smooth geometric picture suggested by Figures~\ref{fig:manifold} and~\ref{fig:geometric_principle}. On the torus, we recover a smooth manifold structure by imposing sufficient spatial regularity on the considered wave functions and $\rho >0$ everywhere (Theorem~\ref{thm:manifold}). We have not established the corresponding manifold structure on the whole space, where the behavior of the density at large distances introduces additional difficulties.

\section{Conclusion}
\label{sec:conclusion}

In this work, we have extended our previous developments in TDDFT~\cite{MAQUI_TDDFT-26,MAQUI_finite-26} by formulating in detail the general geometric interpretation of TDDFT, as well as variational and geometric TDDFT, in the infinite-dimensional setting relevant to continuous electronic systems. This provides a unified framework in which these two formulations of TDDFT, as well as their corresponding TDKS schemes, can be compared on the same footing.

Within this framework, we have derived a number of exact conditions for geometric TDDFT and provided a thorough mathematical analysis of the operators $K_\rho$ and $S_\Psi$ underlying variational and geometric TDDFT, respectively. Our present mathematical results establish conditions ensuring uniqueness and stable determination of the density-enforcing potentials $v$ and $w$ of variational and geometric TDDFT, respectively, at a given instantaneous wave function $\Psi$. They do not, by themselves, establish that these instantaneous constructions can be integrated into a wave-function trajectory reproducing the prescribed density over a nonzero time interval. This requires additional control on the nonlinear evolution of the time-dependent wave function and the density-enforcing potentials, which we do not provide here. 

On the torus, short-time well-posedness has been recently established under suitable assumptions for variational TDDFT in  Refs.~\onlinecite{LauLewOld-V-26_ppt,LauLewOld-V-short-26_ppt} and for the geometric TDKS scheme in Ref.~\onlinecite{LauLewOld-W-26_ppt}. On the Hubbard dimer, long-time existence of $v$ and $w$ can be established~\cite{CanDue-26_ppt}. In addition, for two-electron spin-singlet systems, long-time existence of $v$ and $w$ in TDKS schemes can be proved on the torus using the hydrodynamic reformulation of the problem~\cite{MAQUI_TDDFT-26} and the theory of viscosity solutions of Hamilton--Jacobi equations~\cite{CraLio-83}. In both these long-time existence studies, there may exist singular times after which uniqueness of $v$ and $w$ is lost. Extending these short- and long-time well-posedness results to continuous many-electron systems in the whole space remains a major challenge.

The main next step is to develop practical approximations for the Hxc geometric potential $w_\Hxc$ and/or for the geometric potential correction $w^{\rm cor}$. The exact conditions derived here provide natural constraints for constructing and assessing such approximations. In particular, $w^{\rm cor}$ offers a way of supplementing existing approximations to the usual Hxc potential $v_\Hxc$ in order to account for missing nonadiabatic effects. Testing such approximations on realistic electronic dynamics will be essential for assessing the practical potential of geometric TDDFT.

\begin{acknowledgments}
We thank Neepa Maitra, Eberhard K.U. Gross, and Kieron Burke for insightful discussions.
This work has benefited from French State support managed by ANR under the France 2030 program through the MaQuI CNRS Risky and High-Impact Research programme (RI)$^2$ (grant agreement ANR-24-RRII-0001).
\end{acknowledgments}

\section*{AI Statement}
Large Language Models (LLMs) were used in the final stage of preparing the manuscript for proofreading. Moreover, the counterexample in Appendix~\ref{app:counterex-positivity} was suggested by ChatGPT-6 Astra. We verified it and rewrote the presentation (except Fig.~\ref{fig:symmetry}).


\appendix
\section{The potentials $v$ and $w$ as Lagrange multipliers}
\label{app:Lagrange}

In this appendix, we interpret the potentials $v$ and $w$ of variational and geometric TDDFT as Lagrange multipliers and we express them in terms of functional derivatives. We do not claim here full mathematical rigor. 

\subsection{The potential $v$ from the global action density functional}

For variational TDDFT, we follow the work of Vignale~\cite{Vignale-08}. First, we define the universal action wave-function functional,
\begin{eqnarray}
{\cal A}_T[\tilde{\Psi}] := \int_0^T \!\!\! \d t\, \bra{\tilde{\Psi}(t)} i \partial_t - \hat{T} - \hat{V}_\ee \ket{\tilde{\Psi}(t)},
\label{}
\end{eqnarray}
for an arbitrary final time $T$. For a given density trajectory $\rho$ and initial wave function $\Psi_0$, the time-dependent wave function $\Psi^\V[\rho,\Psi_0]$ of variational TDDFT is a stationary point of the action under the constraint of fixed endpoints and fixed density, which can be written as the implicit stationarity condition [Eq.~\eqref{eq:statPsirho}]
\begin{eqnarray}
\Psi^\V[\rho,\Psi_0] \in \!\!\! \argstat{\substack{\tilde{\Psi} \rightsquigarrow \rho \\ \tilde{\Psi}(0)=\Psi_0,\; \tilde{\Psi}(T)=\Psi^\V_T[\rho,\Psi_0]}} {\cal A}_T[\tilde{\Psi}],
\label{eq:PsiVargstatapp}
\end{eqnarray}
where $\Psi^\V_T[\rho,\Psi_0]:= \Psi^\V[\rho,\Psi_0](T)$ is the wave function at final time $T$ and $\tilde{\Psi} \rightsquigarrow \rho$ means that the variations of $\tilde{\Psi}$ are taken under the constraint $\rho_{\tilde{\Psi}(t)}(\b{r}) = \rho(\b{r},t)$. To impose this density constraint, we introduce the Lagrangian of variational TDDFT,
\begin{eqnarray}
{\cal L}_T^\V[\tilde{\Psi},\tilde{v};\rho]  :=  {\cal A}_T[\tilde{\Psi}] \phantom{xxxxxxxxxxxxxxxxxxxxxxx}
\nonumber\\
- \int_0^T \int_{\mathbb{R}^d}\! \tilde{v}(\b{r},t) \left( \rho_{\tilde{\Psi}(t)}(\b{r}) - \rho(\b{r},t) \right) \d \b{r} \, \d t, \;
\label{eq:LPsiv}
\end{eqnarray}
where $\tilde{v}(\b{r},t)$ is the Lagrange multiplier associated with the density constraint. The functional differential of ${\cal L}_T^\V[\tilde{\Psi},\tilde{v};\rho]$ with respect to $\tilde{\Psi}$ with unconstrained endpoints is~\cite{Vignale-08}
\begin{eqnarray}
\delta_{\tilde{\Psi}} {\cal L}_T^\V[\tilde{\Psi},\tilde{v};\rho](\delta\tilde{\Psi})  = \phantom{xxxxxxxxxxxxxxxxxx}
\nonumber\\
 2 \int_0^T \!\!\!\! \d t\, \Re\bra{\delta\tilde{\Psi}(t)} i \partial_t - \hat{T} - \hat{V}_\ee - \hat{\tilde{V}}(t) \ket{\tilde{\Psi}(t)} 
\nonumber\\
+ i\left[ \braket{\tilde{\Psi}(t)}{\delta\tilde{\Psi}(t)} \right]_{t=0}^T,
\label{eq:dLdPsi}
\end{eqnarray}
where $\hat{\tilde{V}}(t) := \int_{\mathbb{R}^d} \d\b{r} \, \tilde{v}(\b{r},t) \, \hat{\rho}(\b{r})$. The last term in Eq.~\eqref{eq:dLdPsi} comes from integration by parts over time. For variations with fixed endpoints, in Eq.~\eqref{eq:PsiVargstatapp}, this last boundary term vanishes. The resulting stationarity condition can equivalently be written for unrestricted variations by subtracting this boundary term, giving 
\begin{eqnarray}
\left\{ 
\begin{aligned}
&\delta_{\tilde{\Psi}} {\cal L}_T^\V[\tilde{\Psi},\tilde{v};\rho] (\delta \tilde{\Psi}) - i\left[ \braket{\tilde{\Psi}(t)}{\delta\tilde{\Psi}(t)} \right]_{t=0}^T = 0,\\[0.3cm]
&\delta_{\tilde{v}} {\cal L}_T^\V[\tilde{\Psi},\tilde{v};\rho] (\delta \tilde{v}) = 0,
\end{aligned}
\right.
\label{eq:Lstat}
\end{eqnarray}
which leads to the effective TDSE in Eq.~\eqref{eq:TDSE-TDDFT}. The solution gives the wave function $\Psi^\V[\rho,\Psi_0]$ and potential $v[\rho,\Psi_0]$ of variational TDDFT. Note that, although the stationary condition in Eq.~\eqref{eq:PsiVargstatapp} involves a final time $T$, by the Runge-Gross theorem $\Psi^\V[\rho,\Psi_0](t)$ and $v[\rho,\Psi_0](t)$ are causal functionals of the density, depending only on its history up to time $t$, and are therefore independent of the arbitrary final time $T>t$ (see Section~\ref{sec:causality}).

We now introduce the universal action density functional,
\begin{eqnarray}
A_T[\rho,\Psi_0] &:=& {\cal A}_T[\Psi^\V[\rho,\Psi_0]] 
\nonumber\\
&=& {\cal L}_T^\V[\Psi^\V[\rho,\Psi_0],v[\rho,\Psi_0];\rho], \;
\label{}
\end{eqnarray}
where the last equality comes from the fact that $\Psi^\V[\rho,\Psi_0]$ satisfies the density constraint and therefore the last term in Eq.~\eqref{eq:LPsiv} vanishes. We choose a phase convention for $\Psi^\V[\rho,\Psi_0]$ and assume differentiability with respect to admissible density variations.
Considering variations of $\rho$ at fixed initial wave function $\Psi_0$,  preserving the electron number and the initial density and current-divergence compatibility conditions, and using the stationary conditions in Eq.~\eqref{eq:Lstat}, we obtain the functional differential of $A_T[\rho,\Psi_0]$ with respect to $\rho$,
\begin{align}
\delta_{\rho} A_T [\rho,\Psi_0](\delta \rho) &=  \delta_{\tilde{\Psi}}  {\cal L}_T^\V [\tilde{\Psi},v[\rho,\Psi_0];\rho] (\delta_\rho \Psi^\V[\rho,\Psi_0]) 
\nonumber\\
& \quad + \delta_{\tilde{v}}  {\cal L}_T^\V[\Psi^\V[\rho,\Psi_0],\tilde{v};\rho] (\delta_\rho v[\rho,\Psi_0])
\nonumber\\
& \quad + \delta_{\rho} {\cal L}_T^\V[\Psi^\V[\rho,\Psi_0],v[\rho,\Psi_0];\rho](\delta \rho)
\nonumber\\
&= i\braket{\Psi^\V_T[\rho,\Psi_0]}{\delta_\rho \Psi^\V_T[\rho,\Psi_0]}
\nonumber\\
&\quad  + \int_0^T \!\! \int_{\mathbb{R}^d} v[\rho,\Psi_0](\b{r},t) \delta \rho(\b{r},t) \d \b{r} \, \d t,
\end{align}
where we have used the fact that the wave function $\Psi^\V[\rho,\Psi_0]$ is fixed at $t=0$ by the initial condition, i.e. $\delta_\rho \Psi^\V[\rho,\Psi_0](0) = 0$, but is not fixed at $t=T$, i.e. $\delta_\rho \Psi^\V_T[\rho,\Psi_0]$ need not vanish. We thus obtain the potential $v[\rho,\Psi_0](\b{r},t)$ as~\cite{Vignale-08}
\begin{eqnarray}
v[\rho,\Psi_0](\b{r},t) \eqv \frac{\delta A_T[\rho,\Psi_0]}{\delta \rho(\b{r},t)} -i\braket{\Psi^\V_T[\rho,\Psi_0]}{\frac{\delta \Psi^\V_T[\rho,\Psi_0]}{\delta \rho(\b{r},t)}},
\nonumber\\
\label{eq:vVignale}
\end{eqnarray}
with equality modulo time-dependent constants. Even though the two terms on the right-hand side of Eq.~\eqref{eq:vVignale} depend separately on the arbitrary final time $T$, their $T$-dependences cancel out each other, as shown in Ref.~\onlinecite{Vignale-08}, and we obtain a properly causal potential $v[\rho,\Psi_0](\b{r},t)$ which does not depend on $T$. Omission of the last time-boundary term leads to the so-called causality paradox of variational TDDFT~\cite{GroDobPet-96,vanLeeuwen-98,Lee-IJMPB-01}. This subtlety regarding causality stems from the global action formulation of variational TDDFT. By contrast, geometric TDDFT is based on a local-in-time condition, which makes causality more apparent, as shown in the next section.

\subsection{The geometric potential $w$ from the instantaneous residual density-velocity functional}

For geometric TDDFT, we first define a residual wave-function-velocity functional associated with the Hamiltonian $\hat{H}(t):=\hat{T}+\hat{V}_\ee+\hat{{\cal V}}(t)$ at fixed time $t$ and for a fixed wave function $\Psi$,
\begin{eqnarray}
{\cal R}[\tilde{\Xi};\Psi] &:=& \left\| i \, \tilde{\Xi} -\hat{H}(t) \Psi \right\|_{{\cal H}}^2
\nonumber\\
&=& \braket{\tilde{\Xi} + i \hat{H}(t) \Psi\;}{\;\tilde{\Xi} + i \hat{H}(t) \Psi},
\label{}
\end{eqnarray}
where the trial wave-function velocity $\tilde{\Xi}$ will be varied. According to the geometric principle in Eq.~\eqref{eq:mindtPsi}, for a given density velocity $\xi$ and wave function $\Psi$, the optimal wave-function velocity $\Xi_{*}[\xi;\Psi]$ minimizes ${\cal R}[\tilde{\Xi};\Psi]$ under the constraint of fixed density velocity,
\begin{eqnarray}
\Xi_{*}[\xi;\Psi] := \argmin{\substack{\tilde{\Xi} \rightsquigarrow  \xi}} \;\; {\cal R}[\tilde{\Xi};\Psi],
\label{eq:XitxiPsi}
\end{eqnarray}
where $\tilde{\Xi} \rightsquigarrow  \xi$ means that the variations of $\tilde{\Xi}$ are taken under the constraint $ 2 \Re \bra{\tilde{\Xi}} \hat{\rho}(\b{r}) \ket{\Psi} =  \xi(\b{r})$. For a system with fixed electron number, the density velocity $\xi$ must satisfy $\int_{\R^d} \xi(\b{r}) \d\b{r} =0$. To impose the density-velocity constraint in Eq.~\eqref{eq:XitxiPsi}, we introduce the instantaneous Lagrangian of geometric TDDFT,
\begin{eqnarray}
{\cal L}^\G[\tilde{\Xi},\tilde{w};\Psi,\xi] := {\cal R}[\tilde{\Xi};\Psi] \phantom{xxxxxxxxxxxxxxx}
\nonumber\\
- \int_{\mathbb{R}^d} \tilde{w}(\b{r}) \left( 2 \Re \bra{\tilde{\Xi}} \hat{\rho}(\b{r}) \ket{\Psi} - \xi(\b{r}) \right) \d \b{r},
\label{eq:LG}
\end{eqnarray}
where $\tilde{w}(\b{r})$ is the Lagrange multiplier associated with density-velocity constraint.
The functional differential of ${\cal L}^\G[\tilde{\Xi},\tilde{w};\Psi,\xi]$ with respect to $\tilde{\Xi}$ is
\begin{eqnarray}
\delta_{\tilde{\Xi}} {\cal L}^\G[\tilde{\Xi},\tilde{w};\Psi,\xi](\delta \tilde{\Xi}) = \phantom{xxxxxxxxxxx}
\nonumber\\
2 \Re \braket{\delta \tilde{\Xi}\;}{\;\tilde{\Xi} + i (\hat{H}(t)+i\hat{\tilde{W}}) \Psi},
\label{}
\end{eqnarray}
where $\hat{\tilde{W}} := \int_{\mathbb{R}^d} \d\b{r} \, \tilde{w}(\b{r}) \, \hat{\rho}(\b{r})$. The stationary equations of geometric TDDFT are 
\begin{eqnarray}
\left\{ 
\begin{aligned}
&\delta_{\tilde{\Xi}} {\cal L}^\G[\tilde{\Xi},\tilde{w};\Psi,\xi](\delta \tilde{\Xi}) = 0,\\[0.3cm]
&\delta_{\tilde{w}} {\cal L}^\G[\tilde{\Xi},\tilde{w};\Psi,\xi](\delta \tilde{w}) = 0,
\end{aligned}
\right.
\label{eq:LGstat}
\end{eqnarray}
which leads to the effective TDSE in Eq.~\eqref{eq:TDSE-TDDFTgeo}. The solution gives the instantaneous wave-function velocity $\Xi_{*}[\xi;\Psi]$ and potential $w_{*}[\xi;\Psi]$, which can be written as 
\begin{eqnarray}
\left\{ 
\begin{aligned}
& \Xi_{*}[\xi;\Psi] =  -i (\hat{H}(t)+i\hat{W}_{*}[\xi;\Psi]) \Psi\\
& w_{*}[\xi;\Psi] = \frac{1}{2} S_\Psi^{-1} \left( \xi + \nabla \cdot \b{j}_\Psi\right)
\end{aligned}
\right.,
\label{eq:Xi_and_w_sol}
\end{eqnarray}
where $\hat{W}_{*}[\xi;\Psi] := \int_{\mathbb{R}^d} \d\b{r} \, w_{*}[\xi;\Psi](\b{r}) \, \hat{\rho}(\b{r})$, $S_\Psi^{-1}$ is the inverse of the operator $S_\Psi$ defined in Eq.~\eqref{eq:SPsirrp} and studied in Section~\ref{sec:S}, and $\nabla \cdot \b{j}_\Psi$ is the divergence of the current density of $\Psi$. If we evaluate these solutions at the instantaneous density velocity $\partial_t \rho(t)$ and at the instantaneous geometric wave function $\Psi^\G[\rho,\Psi_0,\nu](t)$, we obtain the instantaneous wave-function velocity and the instantaneous potential of geometric TDDFT,
\begin{align}
\partial_t \Psi^\G[\rho,\Psi_0,\nu](t) 
& = \Xi_{*}[\partial_t \rho(t);\Psi^\G[\rho,\Psi_0,\nu](t)],\;\;\;\;
\\
w[\rho,\Psi_0,\nu](\b{r},t) 
& = w_{*}[\partial_t \rho(t);\Psi^\G[\rho,\Psi_0,\nu](t)](\b{r}).\;\;\;
\end{align}

We now introduce the residual density-velocity functional,
\begin{eqnarray}
R[\xi;\Psi] &:=& {\cal R}[\Xi_{*}[\xi;\Psi];\Psi] 
\nonumber\\
&=& {\cal L}^\G[\Xi_{*}[\xi;\Psi],w_{*}[\xi;\Psi];\Psi,\xi],
\label{eq:RxiPsi}
\end{eqnarray}
where the last equality comes from the fact that $\Xi_{*}[\xi;\Psi]$ satisfies the density-velocity constraint and therefore the last term in Eq.~\eqref{eq:LG} vanishes. Assuming differentiability with respect to $\xi$, we consider variations of $\xi$ at fixed $\Psi$ that preserve $\int_{\R^d} \xi(\b{r}) \d\b{r}=0$, and using the stationary conditions in Eq.~\eqref{eq:LGstat}, we obtain the functional differential of $R[\xi;\Psi]$ with respect to $\xi$,
\begin{eqnarray}
\delta_{\xi} R[\xi;\Psi](\delta \xi) &=& \delta_{\tilde{\Xi}} {\cal L}^\G[\tilde{\Xi},w_{*}[\xi;\Psi];\Psi,\xi](\delta_{\xi} \Xi_{*}[\xi;\Psi])
\nonumber\\
&& + \delta_{\tilde{w}} {\cal L}^\G[\Xi_{*}[\xi;\Psi],\tilde{w};\Psi,\xi](\delta_{\xi} w_*[\xi;\Psi])
\nonumber\\
&& + \delta_{\xi}  {\cal L}^\G[\Xi_{*}[\xi;\Psi],w_*[\xi;\Psi];\Psi,\xi] (\delta \xi)
\nonumber\\
&=& \int_{\mathbb{R}^d} w_{*}[\xi;\Psi](\b{r}) \delta \xi(\b{r}) \d \b{r}.
\label{}
\end{eqnarray}
We have thus found the potential $w_{*}[\xi;\Psi](\b{r})$ as a functional derivative,
\begin{eqnarray}
w_{*}[\xi;\Psi](\b{r}) = \frac{\delta R[\xi;\Psi]}{\delta \xi(\b{r})},
\label{}
\end{eqnarray}
where the arbitrary time-dependent constant in the functional derivative is fixed by the condition $\int_{\mathbb{R}^d} w_{*}(\b{r}) \rho_\Psi(\b{r}) \d\b{r}=0$.
We finally evaluate this quantity at the instantaneous density velocity $\partial_t \rho(t)$ and at the instantaneous geometric wave function $\Psi^\G[\rho,\Psi_0,\nu](t)$ to obtain the geometric potential as
\begin{eqnarray}
w[\rho,\Psi_0,\nu](\b{r},t) = \left. \frac{\delta R[\xi;\Psi]}{\delta \xi(\b{r})} \right|_{\substack{\xi=\partial_t \rho(t)\\\Psi=\Psi^\G[\rho,\Psi_0,\nu](t)}}.
\label{}
\end{eqnarray}
The functional derivative is taken at fixed time $t$ and does not depend on an arbitrary final time, which makes causality more apparent compared to Eq.~\eqref{eq:vVignale}. It also shows that the instantaneous geometric potential $w(t)$ can be thought of as a variable conjugate to the instantaneous density velocity $\partial_t \rho(t)$. 

Finally, we note that inserting the expressions~\eqref{eq:Xi_and_w_sol} in Eq.~\eqref{eq:RxiPsi} gives an explicit expression for the residual density-velocity functional
\begin{eqnarray}
R[\xi;\Psi] &=& \left\| \hat{W}_{*}[\xi;\Psi] \Psi \right\|_{{\cal H}}^2
\nonumber\\
&=& \pscal{w_{*}[\xi;\Psi], S_\Psi w_{*}[\xi;\Psi]}
\nonumber\\
&=& \frac{1}{4} \pscal{\left(\xi + \nabla \cdot \b{j}_\Psi\right), S_\Psi^{-1} \left( \xi + \nabla \cdot \b{j}_\Psi \right)},
\label{eq:RxiPsiSm1}
\end{eqnarray}
where we have used the quadratic form associated with $S_\Psi$ defined in Eq.~\eqref{eq:formula_S}. Thus, the residual density-velocity functional $R[\xi;\Psi]$ measures, in the metric defined by $S_\Psi^{-1}$, the squared magnitude of the source/sink term $\xi + \nabla \cdot \b{j}_\Psi = 2  S_\Psi w_{*}[\xi;\Psi]$ in the modified continuity equation. The expression of the residual functional in Eq.~\eqref{eq:RxiPsiSm1} is in fact just an instantaneous quadratic variational formulation of the modified continuity equation~\eqref{eq:modconteq}. In fact, an analogous instantaneous characterization of the potential $v$ of variational TDDFT follows from the force-balance equation~\eqref{eq:divlocforce}, as we discuss in the next section.

\subsection{The potential $v$ from the instantaneous residual density-acceleration functional}

For variational TDDFT, an instantaneous constrained minimization can be formulated in terms of the current-density velocity $\partial_t \b{j}_{\Psi^\V(t)}$ which enters the local-force equation~\cite{RunGro-84},
\begin{equation}
\partial_t \b{j}_{\Psi^\V(t)}(\br) = -\rho(\br,t) \nabla v(\br,t) + \b{f}_{\Psi^\V(t)}(\br),
\label{eq:localforceeq}
\end{equation}
where $\b{f}_\Psi(\b{r}):=i\bra{\Psi}[\hat{T}+\hat{V}_\ee,\hat{\b{j}}(\b{r})] \ket{\Psi}$ is the internal-force density. Taking the negative divergence of Eq.~\eqref{eq:localforceeq} gives the force-balance equation~\eqref{eq:divlocforce} with $q_\Psi(\br)=-\nabla\cdot\b{f}_\Psi(\br)$. To construct the instantaneous minimization, we introduce the residual current-density-velocity functional, with weight $1/\rho_\Psi$, for a fixed wave function $\Psi$,
\begin{equation}
\cQ[\tilde{\bm{\eta}};\Psi] := \frac{1}{2}\int_{\R^d} \frac{|\tilde{\bm{\eta}}(\b{r})-\b{f}_\Psi(\b{r})|^2}{\rho_\Psi(\b{r})}\d\b{r},
\end{equation}
where $\tilde{\bm{\eta}}$ is a trial current-density velocity. For a prescribed density acceleration $a$, satisfying $\int_{\R^d}a(\b{r})\d\b{r}=0$, the optimal current-density velocity $\bm{\eta}_{*}[a;\Psi]$ minimizes $\cQ[\tilde{\bm{\eta}};\Psi]$ under the constraint of fixed density acceleration,
\begin{equation}
\bm{\eta}_{*}[a;\Psi] := \argmin{\substack{\tilde{\bm{\eta}} \rightsquigarrow a}} \; \cQ[\tilde{\bm{\eta}};\Psi],
\label{eq:etaargmin}
\end{equation}
where $\tilde{\bm{\eta}} \rightsquigarrow a$ means that the variations of $\tilde{\bm{\eta}}$ are taken under the constraint $-\nabla\cdot\tilde{\bm{\eta}}=a$. This constraint follows by differentiating the continuity equation with respect to time. We then introduce the instantaneous acceleration Lagrangian of variational TDDFT,
\begin{align}
{\cal L}^\V_\text{acc}[\tilde{\bm{\eta}},\tilde v;\Psi,a]:= \cQ[\tilde{\bm{\eta}};\Psi] \phantom{xxxxxxxxxxxxxx}
\nonumber\\
+\int_{\R^d} \tilde v(\b{r}) \left(-\nabla\cdot\tilde{\bm{\eta}}(\b{r})-a(\b{r})\right)\d\b{r},
\end{align}
where $\tilde v$ is the Lagrange multiplier associated with the density-acceleration constraint. The functional differential of ${\cal L}^\V_\text{acc}[\tilde{\bm{\eta}},\tilde v;\Psi,a]$ with respect to $\tilde{\bm{\eta}}$ is, after integration by parts,
\begin{eqnarray}
\delta_{\tilde{\bm{\eta}}} {\cal L}^\V_\text{acc}[\tilde{\bm{\eta}},\tilde v;\Psi,a](\delta \tilde{\bm{\eta}}) = \phantom{xxxxxxxxxxxxxx}
\nonumber\\ 
\int_{\R^d} \left( \frac{\tilde{\bm{\eta}}(\b{r})-\b{f}_\Psi(\b{r})}{\rho_\Psi(\b{r})} + \nabla \tilde{v}(\br) \right) \cdot \delta \tilde{\bm{\eta}}(\br) \d\br,
\label{}
\end{eqnarray}
and the stationary equations are
\begin{eqnarray}
\left\{ 
\begin{aligned}
&\delta_{\tilde{\bm{\eta}}} {\cal L}^\V_\text{acc}[\tilde{\bm{\eta}},\tilde v;\Psi,a](\delta \tilde{\bm{\eta}}) = 0\\
&\delta_{\tilde v}  {\cal L}^\V_\text{acc}[\tilde{\bm{\eta}},\tilde v;\Psi,a](\delta \tilde v) = 0.
\end{aligned}
\right.
\label{eq:LVaccstat}
\end{eqnarray}
The solution gives the instantaneous current-density velocity $\bm{\eta}_{*}[a;\Psi]$ and potential $v_{*}[a;\Psi]$, which can be written as
\begin{eqnarray}
\left\{ 
\begin{aligned}
& \bm{\eta}_{*}[a;\Psi] = - \rho_\Psi \nabla v_{*}[a;\Psi] + \b{f}_\Psi\\
& v_{*}[a;\Psi]\equiv K_{\rho_\Psi}^{-1} \left( q_\Psi - a\right) 
\end{aligned}
\right.,
\label{eq:eta_and_v_sol}
\end{eqnarray}
where $K_{\rho_\Psi}^{-1}$ is the inverse (modulo constants) of the operator $K_{\rho_\Psi}$ defined in Eq.~\eqref{eq:K} and studied in Section~\ref{sec:K}. The first line of Eq.~\eqref{eq:eta_and_v_sol} is precisely the local-force equation~\eqref{eq:localforceeq} and the second line is the force-balance equation~\eqref{eq:divlocforce}, which confirms that the minimization principle in Eq.~\eqref{eq:etaargmin}, together with the effective TDSE in Eq.~\eqref{eq:TDSE-TDDFT}, is indeed equivalent to variational TDDFT. If we evaluate these solutions at the instantaneous density acceleration $\partial_t^2 \rho(t)$ and at the instantaneous variational wave function $\Psi^\V[\rho,\Psi_0](t)$, we obtain the instantaneous current-density velocity and the instantaneous potential of variational TDDFT,
\begin{align}
\partial_t \b{j}_{\Psi^\V[\rho,\Psi_0](t)}(\br)
& = \bm{\eta}_{*}[\partial_t^2 \rho(t);\Psi^\V[\rho,\Psi_0](t)](\br),\;\;\;\;
\\
v[\rho,\Psi_0](\b{r},t) 
& \equiv v_{*}[\partial_t^2 \rho(t);\Psi^\V[\rho,\Psi_0](t)](\br).\;\;\;
\end{align}

We now introduce the residual density-acceleration functional,
\begin{eqnarray}
Q[a;\Psi] &:=& \cQ[\bm{\eta}_{*}[a;\Psi];\Psi] 
\nonumber\\
&=& {\cal L}^\V_\text{acc}[\bm{\eta}_{*}[a;\Psi],v_{*}[a;\Psi];\Psi,a].
\label{eq:QaPsi}
\end{eqnarray}
As before, taking variations of $a$ at fixed $\Psi$ preserving $\int_{\R^d} a(\b{r}) \d\b{r}=0$, and using the stationary conditions in Eq.~\eqref{eq:LVaccstat}, gives
\begin{eqnarray}
\delta_{a} Q[a;\Psi](\delta a) &=& -\int_{\mathbb{R}^d} v_{*}[a;\Psi](\b{r}) \delta a(\b{r}) \d \b{r},
\label{}
\end{eqnarray}
leading to the potential $v_{*}[a;\Psi](\b{r})$ as the functional derivative
\begin{eqnarray}
v_{*}[a;\Psi](\b{r}) \equiv -\frac{\delta Q[a;\Psi]}{\delta a(\b{r})}.
\label{}
\end{eqnarray}
We finally evaluate this quantity at the instantaneous density acceleration $\partial_t^2 \rho(t)$ and at the instantaneous variational wave function $\Psi^\V[\rho,\Psi_0](t)$ to obtain the potential as
\begin{eqnarray}
v[\rho,\Psi_0](\b{r},t) \equiv -\left. \frac{\delta Q[a;\Psi]}{\delta a(\b{r})} \right|_{\substack{a=\partial_t^2 \rho(t)\\\Psi=\Psi^\V[\rho,\Psi_0](t)}}.
\label{}
\end{eqnarray}
This shows that the opposite of the instantaneous potential $-v(t)$ can also be thought of as a variable conjugate to the instantaneous density acceleration $\partial_t^2 \rho(t)$. Compared to Eq.~\eqref{eq:vVignale}, this expression of $v$ does not involve any final-time boundary term. Viewing variational TDDFT as an instantaneous problem with the density acceleration taken as the main variable is very much in line with the work of Ref.~\onlinecite{TarUll-21}. Moreover, the present discussion is similar to the work of Ref.~\onlinecite{Sch-JCP-15} in which the force $-\nabla v$ is obtained as a functional derivative with respect to the current-density velocity.

Finally, we note that inserting the expressions~\eqref{eq:eta_and_v_sol} in Eq.~\eqref{eq:QaPsi} gives an explicit expression for the residual density-acceleration functional
\begin{eqnarray}
Q[a;\Psi] &=& \frac{1}{2} \int_{\R^d} \rho_\Psi(\br) |\nabla v_{*}[a;\Psi] (\br)|^2 \d\br
\nonumber\\
&=& \frac{1}{2} \pscal{v_{*}[a;\Psi], K_{\rho_\Psi} v_{*}[a;\Psi]}
\nonumber\\
&=& \frac{1}{2} \pscal{\left(q_\Psi - a \right), K_{\rho_\Psi}^{-1} \left( q_\Psi - a  \right)},
\label{eq:QaPsiKm1}
\end{eqnarray}
where we have used the quadratic form associated with $K_{\rho_\Psi}$ defined in Eq.~\eqref{eq:K_rho_quadratic_form}. Thus, the residual density-acceleration functional $Q[a;\Psi]$ measures, in the metric defined by $K_{\rho_\Psi}^{-1}$, the squared magnitude of the potential contribution $q_\Psi - a = K_{\rho_\Psi} v_{*}[a;\Psi]$ to the force-balance equation. This construction is an instantaneous variational reformulation of the force-balance equation~\eqref{eq:divlocforce}.

\section{Counterexample to the strict positivity of the spin-polarized ground-state density and implications for $v$-representability}
\label{app:counterex-positivity}

\begin{figure}[t]
    \centering
     \includegraphics[scale=1,angle=0]{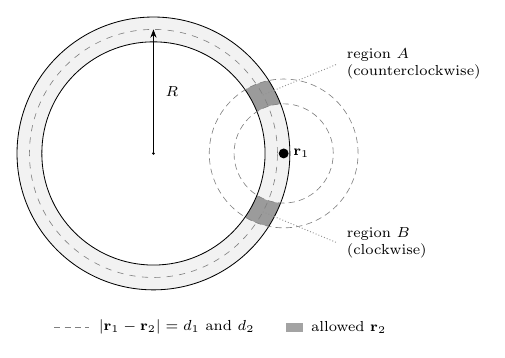}
    \caption{Schematic for the allowed region for $\br_2$ for the displayed value of $\br_1$ for the wave function $\Psi_\infty$.}
    \label{fig:symmetry}
\end{figure}

In this appendix, we provide a counterexample to the everywhere strict positivity of spin-polarized ground-state densities. 
This is based on a symmetry mechanism that was suggested by ChatGPT-6 Astra. We verified it and rewrote the presentation (except for Fig.~\ref{fig:symmetry}).

We consider $N=2$ spin-triplet electrons on $\R^3$, which have a spatial antisymmetric wave function $\Psi\in L^2(\R^3\times\R^3;\C)$. 
The corresponding density is $\rho_\Psi(\br):=2\int_{\R^3}|\Psi(\br,\br_2)|^2\,\rd\br_2$. Let us introduce the particle swapping operator $S(\br_1,\br_2) := (\br_2,\br_1)$ which acts on wave functions as $\hat{S}\Psi(\br_1,\br_2):=\Psi(\br_2,\br_1)$. The fermionic symmetry means that we consider wave functions with $\hat{S}\Psi=-\Psi$. We take the Hamiltonian
\begin{align}
\hat{H}:=-\frac12\nabla^2_{\br_1}-\frac12\nabla^2_{\br_2}+v(\br_1)+v(\br_2)+v_\text{ee}(\br_1-\br_2),
\label{eq:Happ}
\end{align}
with $v_\text{ee}$ radial and $v$ axially symmetric around the $z$-axis. We will construct an interaction $v_\ee$ and an external potential $v$ that are smooth and have compact support such that the associated ground-state density $\rho$ vanishes along the whole $z$-axis for symmetry reasons. 

\begin{theorem}[An interacting ground-state density vanishing on a whole line]\label{thm:counterexample}
There exist $v$ and $v_\ee$ that are smooth and have compact support such that the corresponding Hamiltonian~\eqref{eq:Happ} has a unique ground state, up to a phase, in the spin-triplet sector, whose density vanishes along the $z$-axis.
\end{theorem}

The mechanism that makes the density vanish is a certain symmetry of the wave function.
For describing this, we identify the group $O(2)$ of $2\times 2$ orthogonal matrices with determinant $\pm 1$ with the rotations and reflections of the
$(x,y)$-plane, acting trivially on $z$.
The reflection $\b{J}$ with respect to the $(x,z)$-plane acts on $\br=(x,y,z)$ as
\begin{equation}
\b{J}(x,y,z)=(x,-y,z).
\end{equation}
Then the group $O(2)$ is generated by $\b{J}$ and the rotations in the $(x,y)$-plane.
We let any element $\b{G}$ of this group act on wave functions through the unitary operator $\hat{U}_\b{G}$ defined by
\begin{equation}
  (\hat{U}_\b{G}\Psi)(\br_1,\br_2):=\Psi(\b{G}^{-1} \br_1, \b{G}^{-1} \br_2).
\end{equation}
The Hamiltonian is invariant under this action, since we have assumed that $v$ is axially symmetric,
\begin{equation}
\hat{U}_\b{G}^\dagger \hat{H} \hat{U}_\b{G}= \hat{H}.
\end{equation}
The ground state of the Hamiltonian inherits this symmetry. The key idea is to construct potentials such that the ground state has eigenvalue $-1$ under the action of reflections and from this conclude that whenever one particle is on the $z$-axis, the wave function vanishes.
We do so by confining the particles to a solid torus around the $z$-axis with major radius $R$ and minor radius $r$ (doughnut-shaped region) denoted as
\begin{equation}
\bT_{R,r} := \{ \br=(x,y,z) \mid (\sqrt{x^2 + y^2}-R)^2 + z^2 \leq r^2 \}.
\end{equation}
	
The steps of the proof are the following:
\begin{itemize}
	\item take $v$ to effectively trap the system to $\bT_{R,r}$, $v_\ee$ to confine the relative coordinate $|\br_1 - \br_2|$ to $[d_1,d_2]$, as illustrated in Fig.~\ref{fig:symmetry};
	\item for infinitely strong $v$ and $v_\ee$, the configuration space splits into two pieces: $\br_2$ can be either clockwise or counterclockwise with respect to $\br_1$;
	\item exchanging the particles swaps these two parts, hence this yields the symmetry $ \hat{U}_\b{G}\Psi=(\det \b{G})\Psi$ for $\b{G}\in O(2)$;
	\item ground states with finite-strength $v$ and $v_\ee$ inherit the symmetry of the infinitely strong ones, yielding $\rho((0,0,z))=0$ for all $z$.
\end{itemize}

\begin{proof}[Proof of Theorem \ref{thm:counterexample}]
Assume that $0<r<R/2$ and $2r<d_1 <d_2 <2(R-r)$.
Take any smooth and axially symmetric $v:\R^3\to [0,1]$ for which
\begin{equation}
v(\br)=0 \iff \br\in \bT_{R,r},
\end{equation}
and $v(\br)=1$ outside a ball of radius $2R$.
Also, take any smooth radial $v_\text{ee}:\R^3\to [0,1]$ satisfying
\begin{equation}
v_\text{ee}(\br)=0 \iff 2r<d_1\leq \abs{\br} \leq d_2 < 2(R-r),
\end{equation}
and $v_\text{ee}(\br)=1$ outside a ball of radius $3R$.
These potentials approximate the hard-wall confinement illustrated in Fig.~\ref{fig:symmetry} using finite potential barriers.
	
For $\lambda>0$, denote by $\hat{H}_\lambda$ the Hamiltonian with external potential $\lambda v$ and interaction $\lambda v_\text{ee}$. Note that $\hat{H}_\infty$ can be identified with the Laplacian operator $-(1/2)[\nabla^2_{\br_1}+\nabla^2_{\br_2}]$ with Dirichlet boundary conditions on the allowed configuration region
\begin{equation}
\Omega := \{(\br_1,\br_2)\in \bT^2_{R,r} \mid d_1\leq \abs{\br_1-\br_2} \leq d_2\}
\end{equation}
and formally $\hat{H}_\lambda\to \hat{H}_\ii$ as $\lambda \to \infty$.
Observe on Fig.~\ref{fig:symmetry} that $\Omega$ is made of two parts, $\Omega =\Omega_+ \cup \Omega_-$, where
\begin{equation}
\begin{split}
 \Omega_{+/-}:= \{&(\br_1,\br_2)\in \bT^2_{R,r} \mid d_1\leq \abs{\br_1-\br_2} \leq d_2,\\
             &\br_2 \textrm{ clockwise/counterclockwise from } \br_1\}.
\end{split}
\end{equation}
Here it is crucial that the particles are at a distance from one another that is bigger than the width of the annulus but much less than the circumference of the torus in order to make sense of the words clockwise/counterclockwise.
Swapping the two particles exchanges the two parts of $\Omega$:
\begin{equation}
S \Omega_+ = \Omega_-,\qquad S \Omega_- = \Omega_+.
\end{equation}
The group action of $O(2)$ has the following property: acting with $\b{G}$ on both particles preserves $\Omega_{+/-}$ if it is just a rotation, i.e. if $ \det\b{G}=1$, and swaps the two if it contains a reflection, i.e. if $\det\b{G}=-1$. This holds since rotations keep the relative position of the particles fixed whereas reflections swap which one is clockwise from the other.

Now we note that $\hat{H}_\ii$ has a gapped and non-degenerate ground state $\Psi_\ii$.
Indeed, due to the fermionic symmetry, $\hat{H}_\ii$ is unitarily equivalent to the Laplacian without symmetry on $\Omega_+$: any (nonsymmetric) function on $\Omega_+$ corresponds uniquely to a fermionic function on $\Omega$ as the values on $\Omega_-$ are determined by the symmetry.
The Dirichlet Laplacian on a sufficiently regular connected domain such as $\Omega_+$ is known to have a unique, positive, gapped ground state which translates to the ground state of $\hat{H}_\ii$. We choose $\Psi_\infty$ positive on $\Omega_+$ and fermionic antisymmetry then makes it negative on $\Omega_-$.
For rotations $\b{G}$, $\hat{U}_\b{G} \Psi_\ii$ corresponds to a positive ground state on $\Omega_+$ because of the symmetry (which also preserves $\Omega_+$), so it has to equal $\Psi_\ii$ by uniqueness.
Applying a reflection though gives eigenvalue $-1$ because it exchanges $\Omega_+$ and $\Omega_-$. In short,
\begin{equation}
	\hat{U}_\b{G}\Psi_\ii=(\det \b{G})\Psi_\ii,\qquad \b{G}\in O(2).
\end{equation}

Now we consider a large but finite $\lambda$. The first two eigenvalues in the antisymmetric sector, counting with multiplicity, converge to those of the limiting Dirichlet problem as $\lambda\to\infty$. Thus, for sufficiently large $\lambda$, $\hat{H}_\lambda$ has also a gapped and non-degenerate ground state $\Psi_\lambda$ in this sector. After a suitable choice of phase, we have the convergence $\Psi_\lambda\to\Psi_\infty$ in $L^2$. Since the reflection $\hat{U}_{\b{J}}$ commutes with $\hat{H}_\lambda$ and has only the eigenvalues $\pm1$, and $\hat{U}_{\b{J}}\Psi_\infty=-\Psi_\infty$, this convergence implies that $\hat{U}_{\b{J}}\Psi_\lambda=-\Psi_\lambda$ for sufficiently large $\lambda$.
More generally, non-degeneracy and the nonzero overlap with $\Psi_\infty$ imply that $\Psi_\lambda$ has the same symmetry properties as $\Psi_\infty$ for sufficiently large $\lambda$:
\begin{equation}
	\hat{U}_\b{G}\Psi_\lambda =(\det \b{G}) \Psi_\lambda
\end{equation}
for any $\b{G}\in O(2)$. Now note that both the external potential and the interaction are smooth functions. They are equal to $\lambda$ outside some bounded set; hence, in order to obtain compactly supported functions we subtract the irrelevant constant $\lambda$ from both $v$ and $v_\ee$.

Using this symmetry, we now show that the density vanishes on the $z$-axis. 
First note that $\Psi_\lambda$ is in fact a continuous function by usual regularity theory and so is its density $\rho$; this allows to give meaning to the values of $\rho$ on sets of measure zero such as an axis. We show that in fact the wave function $\Psi_\lambda$ vanishes if $\br_1=(0,0,z)$: For any $\br_2$ take $\b{G}$ to be the reflection with respect to the plane through it and the $z$-axis. This reflection has determinant $-1$ and does not change $\br_1$ and $\br_2$, hence
\begin{align}
\Psi_\lambda(\br_1, \br_2) &= \Psi_\lambda(\b{G}^{-1}\br_1, \b{G}^{-1} \br_2) 
\nonumber\\
&= (\det \b{G}) \Psi_\lambda(\br_1, \br_2)
\nonumber\\
&= - \Psi_\lambda(\br_1, \br_2).
\end{align}
Thus, $\Psi_\lambda(\br_1, \br_2)=0$ whenever $\br_1$ lies on the $z$-axis, and consequently $\rho((0,0,z))=0$ for any $z$.
\end{proof}

This example illustrates that the question whether the ground-state density is strictly positive everywhere is quite subtle. 
The potentials here are very regular and nevertheless this property fails, at least in the spin-triplet sector. Contrast that with Coulomb systems for which strict positivity is expected to hold~\cite{FouHofSor-07}. A crucial difference here is that $v_\text{ee}$ is not repulsive, namely it tends to trap particles at a certain distance.

Our spin-triplet two-electron ground-state density vanishing along a line also has implications for Kohn--Sham $v$-representability. In order to enforce this nodal set for the density of a non-interacting system, the Kohn-Sham potential $v_{\s}$ will have to blow up quite strongly along the entire $z$ axis (intuitively like $1/(x^2+y^2)$), even though $v$ in the interacting system is smooth and has compact support. This turns out to be impossible for potentials in the Lieb class~\cite{Lieb-83b} $L^{3/2}(\R^3)+L^\ii(\R^3)$.

\begin{corollary}[Failure of non-interacting $v$-representability in Lieb class]
There exists no Kohn--Sham potential $v_{\s}\in L^{3/2}(\R^3)+L^\ii(\R^3)$ so that the density constructed in Theorem~\ref{thm:counterexample} satisfies $\rho=|\varphi_1|^2+|\varphi_2|^2$ with $(-\nabla^2/2+v_{\s})\varphi_j=\varepsilon_j\,\varphi_j$ and $\varphi_1\geq0$. 
\end{corollary}
In particular, this density is not non-interacting ground-state $v$-representable in the Lieb class, since in this case the lowest-energy orbital must then be occupied and can be chosen nonnegative.

\begin{proof}
Argue by contradiction, assuming the existence of $v_{\s}$, and $\varphi_1$ and $\varphi_2$. 
The assumption that $\varphi_1\geq0$ means it is a ground state, that is $\varepsilon_1$ is the bottom of the spectrum. In principle we do not need that it is non-degenerate. 
By unique continuation~\cite{JerKen-85}, it is known that the $\varphi_j$'s cannot vanish on a set of non-zero Lebesgue measure, but this of course does not discard a line, that has zero Lebesgue measure. Nevertheless, both $\varphi_1$ and $\varphi_2$ have to vanish on the whole $z$ axis, in the average local sense
\begin{equation}
\eps^{-3}\int_{B(\br_0,\eps)}|\varphi_j(\br)|^2\,\rd\br\leq \eps^{-3}\int_{B(\br_0,\eps)}\rho(\br)\,\rd\br\underset{\eps\to0}{\longrightarrow}0,
\end{equation}
for every $\br_0$ on the $z$ axis, by continuity of $\rho$. We can then apply a result from Ref.~\onlinecite{OrsPon-16}, which provides a stronger unique continuation result for non-negative functions. It says that for $v_{\s}\in L^{3/2}(\R^3)+L^\ii(\R^3)$, the nodal set $\{\br \in \R^3 \mid \varphi_1(\br) = 0\}$ must have vanishing $W^{2,3/2}$-capacity. This discards a line by Cor.~5.1.15 in Ref.~\onlinecite{AdaHed-96}, leading to the desired contradiction.
\end{proof}

The result is the same if we use a spin-singlet Kohn--Sham wave function with the doubly occupied orbital $\varphi_1$. 
We mention that, for interacting Coulomb electronic systems, a counterexample to the Kohn-Sham (ensemble) ground-state $v$-representability in the Lieb class $L^{3/2}(\R^3)+L^\ii(\R^3)$ has recently been provided in Ref.~\onlinecite{OpenAI-KS}. The phenomenon is there very different, since the density in question is strictly positive everywhere and it is all about its decay at infinity. Of course, failure of non-interacting ground-state $v$-representability within the Lieb class does not necessarily preclude a Kohn-Sham representation in a larger class of potentials.

\section{Example of $\cM_\rho$ not being a manifold on the torus}\label{sec:not_a_manifold}

We consider $N=1$ electron on the $d$-dimensional torus $\Omega=\bT^d$ of volume $1$ with Hilbert space $\cH:=L^2(\bT^d\times\{\uparrow,\downarrow\};\C)$. The two usual spin functions are denoted by $\alpha(\sigma):= \delta_{\uparrow,\sigma}$ and $\beta(\sigma):= \delta_{\downarrow,\sigma}$. The density is taken as the constant function $\rho_0 = 1$. Then $L^2_{\rho_0}(\bT^d)=L^2_{1/\rho_0}(\bT^d)=L^2(\bT^d)$. For $0 \le \eta \le 1/2$, we consider the path of wave functions $\Psi_\eta \in {\cal H}$ with spin state $\alpha$ defined by
\begin{equation}
\Psi_\eta(\b{r},\sigma) := \varphi_\eta(\b{r}) \alpha(\sigma),
\end{equation}
where 
\begin{equation}
\varphi_\eta(\br):=\1_{\bT^d\setminus B(0,\eta)}(\br)-\1_{B(0,\eta)}(\br).
\end{equation}
We have $\rho_{\Psi_\eta}(\b{r}) =  |\varphi_\eta(\b{r})|^2 = 1$ for all $\eta \in [0,1/2]$ and $\b{r} \in \bT^d$, and therefore $\Psi_\eta$ is in the density-constraint set ${\cal M}_{\rho_0}$
for all $\eta \in [0,1/2]$. Simple calculations give the tangent space at $\Psi_0$
\begin{align}
{\mathcal T}_{\Psi_0}\mathcal M_{\rho_0} & = \left( i L^2(\bT^d;\R) \otimes  \ket{\alpha} \right) \oplus \left( L^2(\bT^d;\C) \otimes  \ket{\beta} \right),
\end{align}
and the normal space at $\Psi_0$
\begin{align}
{\mathcal N}_{\Psi_0}\mathcal M_{\rho_0} & = L^2(\bT^d;\R)  \otimes  \ket{\alpha}.
\end{align}
In this special case, any $C^1$ function $t \mapsto \Psi(t) \in \mathcal H$ such that $\Psi(0)=\Psi_0$ and $\Psi(t) \in \mathcal M_{\rho_0}$ for all $t$ satisfies
\begin{equation}
\partial_t\Psi(0) \in {\mathcal T}_{\Psi_0}\mathcal M_{\rho_0},
\end{equation}
and conversely, for every
\begin{equation}
\delta\Psi = i \theta \otimes  \ket{\alpha} + \varphi \otimes  \ket{\beta}  \in {\mathcal T}_{\Psi_0}\mathcal M_{\rho_0},
\end{equation}
there exists a $C^1$ function $t \mapsto \Psi(t) \in \mathcal H$ such that $\Psi(0)=\Psi_0$ and $\Psi(t) \in \mathcal M_{\rho_0}$ such that $\partial_t\Psi(0) = \delta\Psi$. For $t$ sufficiently small, one can for instance take
\begin{equation}
\Psi(\br,\sigma,t) = \frac{e^{it\theta(\b{r})} \alpha(\sigma) + t\varphi(\b{r}) \beta(\sigma)}{\sqrt{1+t^2 |\varphi(\br)|^2}}.
\end{equation}
Thus, if $\mathcal M_{\rho_0}$ were a submanifold of $\mathcal H$ in the vicinity of $\Psi_0$, then its tangent space at $\Psi_0$ would indeed be ${\mathcal T}_{\Psi_0}\mathcal M_{\rho_0}$. But we can compute
\begin{align}
\Psi_\eta-\Psi_0 &= -2\, \1_{B(0,\eta)}\otimes \ket{\alpha}
\nonumber\\
& \in {\mathcal N}_{\Psi_0}\mathcal M_{\rho_0} =({\mathcal T}_{\Psi_0}\mathcal M_{\rho_0})^{\perp_\R}, 
\end{align}
and
\begin{align}
\| \Psi_\eta-\Psi_0 \|_{\mathcal H} = 2 |B(0,\eta)|^{1/2} \mathop{\longrightarrow}_{\eta \to 0^+} 0.
\end{align}
This proves that a decomposition such as~\eqref{eq:local_chart} cannot be true. In fact we have here $\tau=0$ but the normal component is not 0, hence there cannot exist a map $\nu_{\Psi_0}$ satisfying~\eqref{eq:condition_local_chart}. Intuitively, the $L^2$ norm allows large changes of $\Psi_0$ on sufficiently small regions to count as small wave-function variations. In this example, this produces wave functions in $\cM_{\rho_0}$ arbitrarily close to $\Psi_0$ in a purely normal direction, which is incompatible with a smooth manifold.

Likewise, if $d\geq 2$ and $\sqrt{\rho}\in H^1(\bT^d)$, the set $\mathcal M_{\rho}^1:=\mathcal M_\rho\cap H^1(\bT^d \times \{ \uparrow,\downarrow \};\mathbb{C})$ is not, in general, a smooth submanifold of $H^1(\bT^d \times \{ \uparrow,\downarrow \};\C)$ either. This can be shown using a regularized version of the above counterexample. Thus, in this example on the torus, the obstruction does not come from the fact that $\rho$ may vanish or decay exponentially fast. It is related to the $L^2$ topology and, for $d\geq 2$, to the $H^1$ topology being too weak to control the nonlinear density constraint. Similar topological subtleties arise in ground-state DFT~\cite{Garrigue-21,CorLae-25}. 

\section{Non-compactness and instability of the spectrum of $S_\Psi''-1$}
\label{app:noncompactness}

In Theorem~\ref{thm:compact}, we proved that the operator $S_\Psi''-1$, defined in Eq.~\eqref{eq:formula_S2''}, is compact for finite-rank wave functions $\Psi$. On the torus we can also prove compactness if $\Psi\in\cH^k$ with $k=1$ in 1D and $k=2$ in dimension $d\in\{2,3\}$, when $\rho_\Psi>0$ everywhere (see proof of Theorem~\ref{thm:gap_S_torus}). Here we provide an example where $S''_\Psi-1$ is not compact. We use the same example to show that the spectrum of $S''_\Psi-1$ can change very abruptly under arbitrarily small changes of $\Psi$ in $\cH$, even at fixed density. The example is essentially a more involved version of the example in Appendix~\ref{sec:not_a_manifold}, illustrating again the limitations of the $L^2$ topology. 

For simplicity we work in dimension $d=3$, but this is not essential.
We start with the cube $C:=(0,2^{1/3})^3$ of volume $|C|=2$ and call $A_1:=(0,1)^3$ the inside cube of volume 1 and $A_2:=C\setminus A_1$ its complement, also of volume $1$. First we place two opposite-spin electrons in the cube $C$ with the following Slater determinant $\Psi_t \in \cH$,
\begin{equation}
\Psi_t(\b{r}_1,\sigma_1,\b{r}_2,\sigma_2) := \left(\varphi_1(\b{r}_1) \alpha(\sigma_1)\right) \wedge \left( \varphi_2(\b{r}_2) \beta(\sigma_2)\right),
\end{equation}
where $\alpha(\sigma):= \delta_{\uparrow,\sigma}$ and $\beta(\sigma):= \delta_{\downarrow,\sigma}$ are the two usual spin functions, and
each orbital is a linear combination of uniform functions over $A_1$ and $A_2$
\begin{align*}
\varphi_1(\b{r}) &:= \sqrt{t}\1_{A_1}(\b{r})+\sqrt{1-t}\1_{A_2}(\b{r}),\\
\varphi_2(\b{r}) &:= \sqrt{t}\1_{A_2}(\b{r})-\sqrt{1-t}\1_{A_1}(\b{r}),
\end{align*}
for some $0\leq t\leq1$. Defining $q_1(\b{r}) :=|\varphi_1(\b{r})|^2 = t\1_{A_1}(\b{r})+(1-t)\1_{A_2}(\b{r})$ and $q_2(\b{r}) :=|\varphi_2(\b{r})|^2 = t\1_{A_2}(\b{r})+(1-t)\1_{A_1}(\b{r})$, we see that the density is simply
$\rho_{\Psi_t}(\b{r})=q_1(\b{r})+q_2(\b{r})=\1_C(\b{r})$, and the pair density is
\begin{equation}
\rho_{2,\Psi_t}(\br,\br')=q_1(\br)q_2(\br')+ q_2(\br) q_1(\br').
\label{eq:rho2_cube}
\end{equation}
Therefore, on $L^2(C)$, the operator $S''_{\Psi_t}-1$ is
\begin{equation}
S''_{\Psi_t}-1=\big|q_1\big\rangle\big\langle q_2\big|
+\big|q_2\big\rangle\big\langle q_1\big|
\label{eq:Sq1q2}
\end{equation}
and has at most rank $2$. We recover the finite-rank conclusion of Theorem~\ref{thm:compact}. In the basis $(\1_{A_1},\1_{A_2})$, the right-hand side of Eq.~\eqref{eq:Sq1q2} reduces to the matrix
\begin{equation}
\begin{pmatrix}
2t(1-t)&t^2+(1-t)^2\\
t^2+(1-t)^2&2t(1-t)
  \end{pmatrix}
\end{equation}
whose eigenvalues are $1$ and $-(1-2t)^2$. Hence for this example
\begin{equation}
\sigma(S''_{\Psi_t})=\big\{4t(1-t),1,2\big\}
\end{equation}
with the eigenvalue $1$ of infinite multiplicity and the other two eigenvalues $4t(1-t)$ and $2$ of multiplicity 1 each (for $t \neq 1/2$). When $t\in\{0,1\}$ the lowest eigenvalue touches 0 and we recover the example in Section~\ref{sec:kernelSPsi} of one electron in $A_1$ and one electron in $A_2$, so that $w=\1_{A_1}-\1_{A_2}$ is in the kernel of $S_{\Psi_t}$. In all the other cases $S_{\Psi_t}$ is invertible with a gap. The first eigenvalue covers $[0,1]$ when $t$ is varied in $[0,1]$.

Now we can use this simple example to generate a very wild one. We tile our space $\R^3$ with disjoint cubes $C_n$ of volume $|C_n|=2$ and consider the wave function
\begin{equation}
\Psi:=\sum_n\sqrt{p_n}\,\Psi^{C_n}_{t_n},\qquad \sum_n p_n=1, \qquad p_n > 0. 
\end{equation}
Here, $\Psi^{C_n}_{t_n}$ denotes our previous Slater determinant $\Psi_{t_n}$ with parameter $t_n$, translated into the cube $C_n$. The two electrons are thus simply placed together in a cube $C_n$ with probability $p_n$. The density is $\rho_\Psi=\sum_n p_n\1_{C_n}$ and, since all the $\Psi_{t_n}$ have disjoint supports,  the pair density is
\begin{equation}
\rho_{2,\Psi}(\br,\br')=\sum_n p_n\,\rho_{2,\Psi^{C_n}_{t_n}}(\br,\br'),
\end{equation}
with $\rho^{C_n}_{2,\Psi_{t_n}}$ given in Eq.~\eqref{eq:rho2_cube} translated into the corresponding cube $C_n$. Next we note that
\begin{equation}
\frac{\rho_{2,\Psi}(\br,\br')}{\sqrt{\rho_\Psi(\br)\rho_\Psi(\br')}}=\sum_n \rho_{2,\Psi_{t_n}}^{C_n}(\br,\br')
\end{equation}
since $\rho_\Psi=p_n$ in the cube $C_n$. Since the cubes are disjoint, this implies that the operator $S''_\Psi$, defined on $L^2(\R^3)$, is given by the direct-sum decomposition
\begin{equation}
S''_\Psi\simeq \bigoplus_n S''_{\Psi_{t_n}}
\end{equation}
and hence the spectrum is the closure of the union of the spectra. The coefficients $p_n$ play no role in the calculation of the spectrum of $S''_\Psi$ in this example! 
By choosing $t_n$ any sequence that is dense in $[0,1]$ we conclude that the point spectrum is dense in $[0,1]$, while the full spectrum contains the entire interval,
\begin{equation}
\sigma(S''_\Psi)=[0,1]\cup\{2\}.
\end{equation}
This shows that $S''_\Psi-1$ is not a compact operator.
What we should retain from this wild example is that $S''_\Psi$ is completely insensitive to the decay of the probabilities $p_n$, which is also the decay at infinity of $\rho_\Psi$. In particular we cannot have a bound of the form $\rho_{2,\Psi}(\br,\br')\leq C\rho_\Psi(\br)\rho_\Psi(\br')$ as we hope is true for ground states. The fact that the electrons can, with positive probability, both end up in the same box implies the occurrence of the spectrum of $S''_{\Psi_{t_n}}$. Although we expect the spectrum of $S''_\Psi$ to be nice in most physical cases, this example shows that the situation is in fact rather intricate mathematically. The example can of course be adapted to a general electron number $N$.

Finally, we can also deduce from this example that the spectral properties of $S_\Psi''$  are highly unstable with respect to the $L^2$ norm of the wave function $\Psi$, i.e. the spectrum of $S_\Psi''$ can change drastically under an arbitrarily small $L^2$ variation of $\Psi$. Consider for instance the wave function
\begin{equation}
\Psi_m:=\sum_{n=0}^{m-1}\sqrt{p_n}\Psi^{C_n}_{t_0}+\sum_{n=m}^\ii \sqrt{p_n}\Psi_{t_n}^{C_n}
\end{equation}
with $p_n$ and $\Psi^{C_n}_{t}$ as before. The sequence $t_n$ is again chosen dense in $[0,1]$, with $0<t_0<1$. Notice that 
\begin{equation}
\Psi_m\underset{m\to\ii}\longrightarrow \Psi_\ii:=\sum_{n=0}^{\ii}\sqrt{p_n}\Psi^{C_n}_{t_0},
\end{equation}
in fact we have $\norm{\Psi_m-\Psi_\ii}^2_\cH\leq 4\sum_{n=m}^\ii p_n\to0$. Note also that the density is preserved:
\begin{equation}
\rho_{\Psi_m}=\rho_{\Psi_\ii}=\sum_{n=0}^\ii p_n\1_{C_n}.
\end{equation}
Our analysis is independent of the rate of decay of the density and does not require the density to vanish on a set of non-zero measure. It is in fact strictly positive almost everywhere.
The spectrum of the limit $\Psi_\ii$ is 
\begin{equation}
\sigma(S''_{\Psi_\ii})=\{4t_0(1-t_0),1,2\},
\end{equation}
each eigenvalue having an infinite multiplicity. Note that there is a spectral gap $g_S=4t_0(1-t_0)>0$. On the other hand, from the above analysis the spectrum of $\Psi_m$ is
\begin{equation}
\sigma(S''_{\Psi_m})=[0,1]\cup\{2\}.
\end{equation}
This is because the density of the $t_n$'s in $[0,1]$ is not affected by the removal of the first ones. The limit $\Psi_m\to\Psi_\ii$ is therefore very singular. In particular, an arbitrarily small $L^2$ perturbation of the wave function $\Psi_\ii$, even at fixed density, can destroy the spectral gap of $S''_{\Psi_\ii}$.

\section{Proofs of Theorems~\ref{thm:gapKgroundstate},~\ref{thm:threegaps} and~\ref{thm:manifold}}\label{app:proofs}

\subsection{Proof of Theorem~\ref{thm:gapKgroundstate}}
\label{app:proof_thm:gapKgroundstate}
Under our assumptions on $v$ and $v_{\ee}$, we know that $\Psi\in \cH^2$ by Ref.~\onlinecite{Lewin-spectral} (Thm.~6.2). Then $\rho = \rho_\Psi \in H^2(\R^d)$ by Theorem~\ref{thm:H2}, whence it is a continuous function.
M.~and T.~Hoffmann-Ostenhof have shown\cite{Hof-77} that $\sqrt\rho\in H^1(\R^d)$ and, using $v_{\ee}\geq0$, that
\begin{equation}
-\frac{\nabla^2\sqrt\rho}{2}+(v+\Sigma-E_0)\sqrt\rho\leq 0
\label{eq:HO-inequality}
\end{equation}
where we recall that $\Sigma$ is the bottom of the essential spectrum, which coincides with the ground-state energy of $N-1$ electrons among the spin sectors compatible with removing one electron, by the Hunziker-van Winter-Zhislin (HVZ) theorem~\cite{Lewin-spectral}. Now, the general idea is that Eq.~\eqref{eq:HO-inequality} implies a lower bound on the Bohm potential $v_\text{B}$ in Eq.~\eqref{eq:formulaK''}, which itself yields a lower bound on the bottom of the essential spectrum of $K_\rho''$ through
\begin{equation}
K''_\rho=-\nabla^2+\frac{\nabla^2\sqrt\rho}{\sqrt\rho}\geq -\nabla^2 +2(\Sigma-E_0+v).
\end{equation}
Because the potential $v$ tends to $0$ at infinity by assumption, Weyl's theorem implies that the essential spectrum of $-\nabla^2 +2(\Sigma-E_0+v)$ starts at $2(\Sigma-E_0)>0$, and therefore the essential spectrum of $K''_\rho$ cannot start below this value, by comparison (see Ref.~\onlinecite{Lewin-spectral}, Thm.~5.17).
For the reader worried about domains, we provide a more detailed proof of these claims. We start by choosing a sequence $f_n\in C^\ii_c(\R^d)$ such that $\sqrt{\rho} f_n$ weakly converges to $0$ in $L^2(\R^d)$, $\int_{\R^d}\rho|f_n|^2=1$ and $\int_{\R^d}\rho|\nabla f_n|^2\to I:=\inf\sigma_{\rm ess}(K_\rho'')$. We use here that $C^\ii_c(\R^d)$ is dense in $H^1_\rho(\R^d)$. Integrations by parts using $f_n\in C^\ii_c(\R^d)$ and the previously mentioned properties of $\rho$ imply
\begin{align}
&\int_{\R^d}|\nabla f_n(\br)|^2\rho(\br)\,\dr\nonumber\\
&\qquad =\int_{\R^d}|\nabla (\sqrt\rho f_n)(\br)|^2\,\dr\nonumber\\
&\qquad \qquad +\int_{\R^d}\sqrt{\rho(\br)}\,\nabla^2\sqrt{\rho(\br)}\, |f_n(\br)|^2\dr\nonumber\\
&\qquad \geq 2\int_{\R^d}(\Sigma-E_0+v(\br))|f_n(\br)|^2\rho(\br)\dr\nonumber\\
&\qquad \geq 2(\Sigma-E_0)-2\eps_R+2\int_{B(0,R)}v(\br)|f_n(\br)|^2\rho(\br)\dr
\end{align}
where $\eps_R=\sup_{|\br|\geq R}|v(\br)|$, which tends to 0 by assumption. From our assumption that $\rho$ is continuous and does not vanish, we know that $(f_n)$ is bounded in $H^1(B(0,R))$ and, since $\sqrt{\rho} f_n$ weakly converges to $0$ in $L^2(\R^d)$, it follows that $f_n$ weakly converges to $0$ in $H^1(B(0,R))$. This implies that the last term tends to 0 by Lem.~1.10 in Ref.~\onlinecite{Lewin-spectral}. Taking $R\to\ii$ at the end provides the claimed inequality, by Thm.~5.17 in Ref.~\onlinecite{Lewin-spectral}.\qed 

\subsection{Proof of Theorem~\ref{thm:threegaps}}
\label{app:proof_thm:threegaps}
Let $f\in H^1_\rho(\R^d)$ be so that $\int_{\R^d}f(\br)\rho(\br)\,\dr=0$. Consider the excited wave function $\tilde\Psi:=\left(\sum_{j=1}^Nf(\br_j)\right)\Psi$, which belongs to the same spin sector as $\Psi$. Then
\begin{equation}
\braket{\tilde\Psi}{\Psi}=\int_{\R^d}f(\br)\rho(\br)\,\dr=0
\end{equation}
so that $\tilde\Psi$ is orthogonal to $\Psi$. The assumption that $\Psi$ is non-degenerate implies
\begin{equation}
\bra{\tilde\Psi}\hat{H}-E_0\ket{\tilde\Psi}\geq g_H\|\tilde\Psi\|^2_{\cal H}=g_H\pscal{f,S_\Psi f}
\label{eq:gap_H}
\end{equation}
from the definition of $S_\Psi$ in Eq.~\eqref{eq:formula_S}. On the other hand, using $(\hat{H}-E_0)\Psi=0$ we can write
\begin{equation}
\bra{\tilde\Psi}\hat{H}-E_0\ket{\tilde\Psi}=\frac12 \bra{\Psi} [\hat{F},[\hat{H},\hat{F}]]\ket{\Psi}
\end{equation}
with $\hat{F}=\sum_{j=1}^Nf(\br_j)$. An explicit computation of the double commutator yields
\begin{equation}
[\hat{F},[\hat{H},\hat{F}]]=\sum_{j=1}^N|\nabla f(\br_j)|^2,
\end{equation}
which is called the Ismagilov-Morgan-Simon-Sigal (IMS) formula in the literature~\cite{CycFroKirSim-87}. We have thus shown that
\begin{equation}
\bra{\tilde\Psi}\hat{H}-E_0\ket{\tilde\Psi}=\frac12\int_{\R^d}|\nabla f(\br)|^2\rho(\br)\,\dr
\end{equation}
and the result follows after insertion in Eq.~\eqref{eq:gap_H}. To justify all the previous calculations we can take $f\in C^\ii_c(\R^d)$ and then pass to the limit in the $H_\rho^1$ norm, using that $C^\ii_c(\R^d)$ is dense in $H_\rho^1(\R^d)$, established in Theorem~\ref{thm:K}.\qed 

\subsection{Proof of Theorem~\ref{thm:manifold}}
\label{app:proof_thm:manifold}
Any $\rho\in H^k(\bT^d)$ satisfying $\rho>0$ and $\int_{\bT^d}\rho(\br)\d\br=N$ arises from at least one $\Psi\in\cH^k$. This is because $H^k(\bT^d)$ is an algebra, hence we have automatically $\sqrt{\rho}\in H^k(\bT^d)$ as well. In 1D, this is because $\nabla \sqrt{\rho} = (\nabla \rho)/(2\sqrt{\rho})\in L^2(\bT)$
as $\rho\in H^1(\bT)$ and $\rho\geq c>0$ by continuity. The argument is similar in 2D and 3D. From the smooth Hobby--Rice theorem \cite{LazLie-13,Rutherfoord-13,FriSup-22} we can find any number $N$ of smooth phases $\theta_j$ so that the orbitals $\varphi_j=\sqrt{\rho/N}e^{i\theta_j}$ are orthonormal. The corresponding spin-polarized Slater determinant then has the desired density. This proves that $\cM_\rho^k$ is non-empty. Of course, similar functions can be constructed in other spin sectors. 

Let us then consider any reference point $\Psi\in\cM_\rho^k$ and prove that $\cM_\rho^k$ is a manifold in its neighborhood. From Theorem~\ref{thm:gap_S_torus}, we know that $S''_\Psi$ is  invertible on $L^2(\bT^d)$ with a bounded inverse. Let us prove that the same holds on the smaller space $H^k(\bT^d)$. Of course, $S''_\Psi$ is also injective on  $H^k(\bT^d)$. We wish to prove that the inverse $(S''_\Psi)^{-1}$ \emph{a priori} taking its values in $L^2(\bT^d)$ in fact sends $H^k(\bT^d)$ to itself and is a bounded operator for the $H^k(\bT^d)$ norm. To show this, we first notice that $S''_\Psi$ is a bounded operator from $H^k(\bT^d)$ to itself. 
For instance, we have
\begin{equation}
\nabla^k_\br S''_\Psi \, w=\nabla^k_\br w+\nabla^k_{\br}\int_{\bT^d}\frac{\rho_{2,\Psi}(\br,\br')}{\sqrt{\rho(\br)\rho(\br')}}w(\br')\,\rd\br'
\end{equation}
and thus obtain 
an estimate of the form
\begin{equation}
    \norm{S''_\Psi w -w}_{H^k}\leq C \norm{\rho^{-1/2}}^2_{H^k}\norm{\Psi}^2_{\cH^k}\norm{w}_{H^k}
    \label{eq:estim_S_Psi_Sobolev}
\end{equation}
for every $w\in H^k(\bT^d)$. This proves that $S''_\Psi$ is a bounded operator on $H^k(\bT^d)$. If $\Psi\in C^\ii((\bT^{d}\times\{\uparrow,\downarrow\})^N;\C)$ it is easy to see that $S''_\Psi-1$ is compact in $H^k(\bT^d)$, because $\nabla_\br^k(\rho_{2,\Psi}(\br,\br') / \sqrt{\rho_\Psi(\br) \rho_{\Psi}(\br')})$ is a bounded function, hence is the kernel of a Hilbert-Schmidt operator. Using a density argument, with smooth wave functions $\Psi_n$ converging to $\Psi$ in $\cH^k$, and the convergence of $S_{\Psi_n}''-1\to S_\Psi''-1$ in the $H^k$-operator norm by Eq.~\eqref{eq:estim_S_Psi_Sobolev}, we conclude that $S''_\Psi-1$ is also a compact operator on $H^k(\bT^d)$. 
Since $S_\Psi''$ is injective, there is also a gap for the $H^k$-norm and $S''_\Psi$ admits a bounded inverse on $H^k(\bT^d)$ by the Fredholm alternative. We have thus proved that for every source $g\in H^k(\bT^d)$ we can find a unique $w\in H^k(\bT^d)$ such that $S''_\Psi w =g$, with $\|w\|_{H^k}\leq C\|g\|_{H^k}$. Using the invertibility of $S''_\Psi$ on $L^2(\bT^d)$ we would have only obtained $w\in L^2(\bT^d)$, which is not sufficient to ensure that the normal vector $\nu=\sum_{j=1}^N w(\br_j)\Psi\in \cH^k$. Here we have $w\in H^k(\bT^d)$, which implies $w\in L^\ii(\bT^d)$ and $\nu\in\cH^k$. The invertibility of $S''_\Psi$ with a gap in $H^k(\bT^d)$ is transferred to $S'_\Psi$ and $S_\Psi$. We can here multiply or divide by $\rho$ and $\sqrt\rho$ as we wish and still stay in $H^k(\bT^d)$.

Let us now discuss the normal and tangent spaces. Let $\eta$ be any vector in $\cH^k$ and write $\eta=\tau+\nu$ with $\tau\in\cT_\Psi\cM_\rho$ and $\nu\in\cN_\Psi\cM_\rho$, using the orthogonal projections associated with the real $L^2$ inner product. By Eqs.~\eqref{eq:NPsiMrhofinal} and~\eqref{eq:w_for_any_deltaPsi} we know that $\nu=\sum_{j=1}^N w(\br_j)\Psi$ with the unique $w=S_\Psi^{-1} \rho_{\Psi,\eta}$. But $\rho_{\Psi,\eta}\in H^k(\bT^d)$ because $\Psi,\eta\in\cH^k$. This implies that $w\in H^k(\bT^d)$, from the invertibility of $S_\Psi$ in that space explained in the previous paragraph and hence $\nu\in \cH^k$. By difference we conclude that $\tau\in \cT_\Psi\cM_\rho \cap \cH^k$. Hence the orthogonal decomposition in $\cH$ has its projections in $\cH^k$, with a $w\in H^k(\bT^d)$. This is what was claimed in the second part of the statement.

At this point we have everything we need to apply the Banach implicit function theorem. Recalling Eq.~\eqref{eq:manifold_eq_to_be_solved}, we introduce the function 
\begin{equation}
\cF(\tau,w):=2S_\Psi w +\rho_{\tau+\nu}
\label{eq:implicit_F}
\end{equation}
from $\cT_\Psi\cM^k_{\rho_\Psi}\times H^k(\bT^d)$ to $H^k(\bT^d)$ with, of course, $\nu=\sum_{j=1}^Nw(\br_j)\Psi$. The function $\cF$ is quadratic in $\tau$ and $w$, hence it is jointly $C^\ii$ in these two variables, within the specified space. We have $\cF(0,0)=0$ and the derivative in $w$ at $(0,0)$ is 
\begin{equation}
D_w\cF(0,0)=2S_\Psi.
\end{equation}
The invertibility of $S_\Psi$ in $H^k(\bT^d)$ is exactly what we need to apply the implicit function theorem and conclude the existence of a $C^\ii$ map $w_\Psi(\tau)$ defined in a neighborhood of $0$, such that the equation $\cF(\tau,w)=0$ admits a unique solution $w=w_\Psi(\tau)\in H^k(\bT^d)$ near $w=0$, for every $\tau$ in this region. Defining $\nu_\Psi(\tau):=\sum_{j=1}^N w_\Psi(\tau)(\br_j) \Psi$, we obtain a $C^\ii$ normal correction satisfying $\nu_\Psi(0)=0$ and $D\nu_\Psi(0)=0$, the latter following from $D_\tau \cF(0,0)=0$. Thus, the map $\tau \mapsto \Psi + \tau + \nu_\Psi(\tau)$ provides the desired local parametrization of $\cM_\rho^k$. This concludes the proof of the theorem. \qed


%
\end{document}